**CHALMERS UNIVERSITY OF TECHNOLOGY**
Gothenburg, Sweden
**www.chalmers.se**

For centuries, the existence of planets orbiting stars other than the Sun remained a matter of speculation. The first confirmed detection of such planets in the early 1990s marked the beginning of the modern era of exoplanetary science. This thesis introduces the principal methods used to detect and characterise exoplanets before examining the physical and orbital properties of the observed planet population, with particular emphasis on multi-planetary systems, their architectures, and the similarities and differences among planets within the same system. Future exoplanet studies will ultimately expand and refine our knowledge of the intrinsic exoplanet population, including the processes governing planet formation, evolution, and diversity.

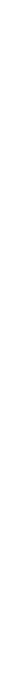

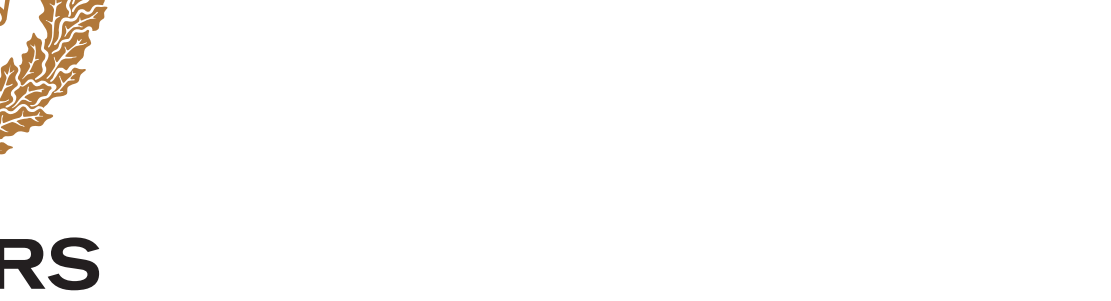

CHALMERS
UNIVERSITY OF TECHNOLOGY

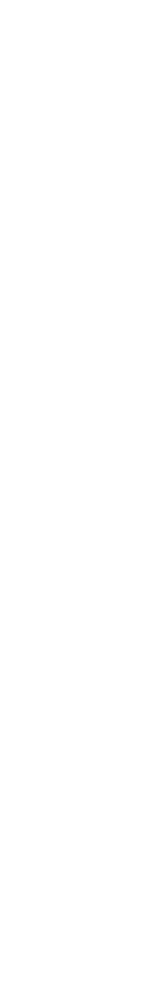

ALEXANDRA MURESAN

Space Journey Beyond Earth

2026

CHALMERS
UNIVERSITY OF TECHNOLOGY

**PHD THESIS**

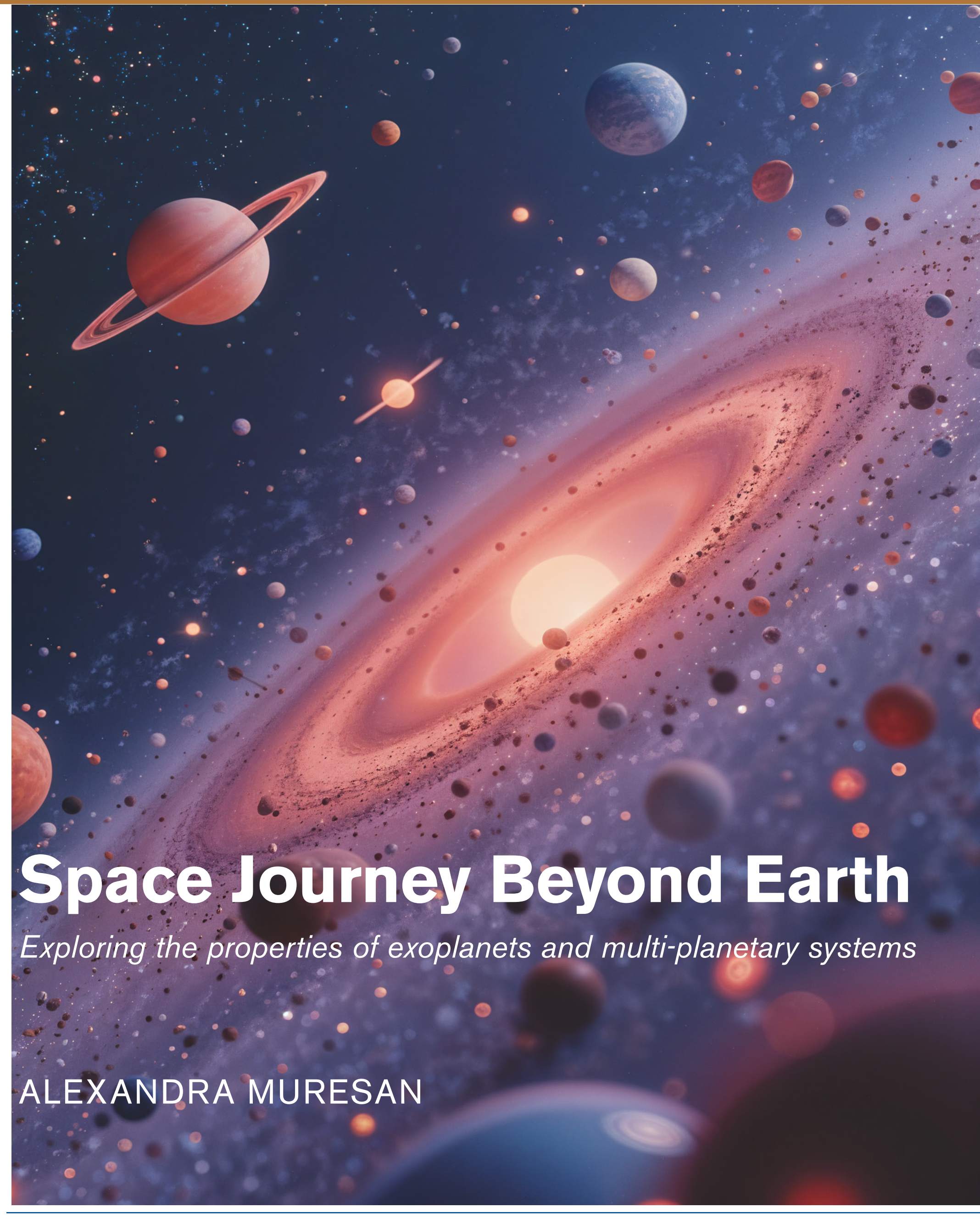

# Space Journey Beyond Earth

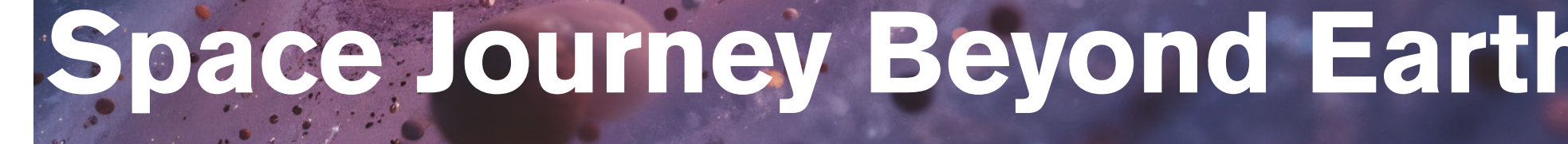

*Exploring the properties of exoplanets and multi-planetary systems*

ALEXANDRA MURESAN

**DEPARTMENT OF PHYSICS AND ASTRONOMY**
CHALMERS UNIVERSITY OF TECHNOLOGY
Gothenburg, Sweden 2026
www.chalmers.se

THESIS FOR THE DEGREE OF DOCTOR OF PHILOSOPHY

# Space Journey Beyond Earth

## Exploring the properties of exoplanets and multi-planetary systems

ALEXANDRA MURESAN

Department of Physics and Astronomy

CHALMERS UNIVERSITY OF TECHNOLOGY

Gothenburg, Sweden 2026

**Space Journey Beyond Earth**

Exploring the properties of exoplanets and multi-planetary systems

ALEXANDRA MURESAN

Acknowledgements, dedications, and similar personal statements in this thesis, reflect the author's own views.

Department of Physics and Astronomy
Chalmers University of Technology
SE-412 96 Gothenburg, Sweden
Telephone + 46 (0)31-772 1000

Cover illustration: AI-generated illustration of exoplanets and stars from www.freepik.com, accessed on 2026-09-19.

This thesis has been prepared using LaTeX.

**Space Journey Beyond Earth**
Exploring the properties of exoplanets and multi-planetary systems
ALEXANDRA MURESAN
Department of Physics and Astronomy
Chalmers University of Technology

# Abstract

The discovery of more than 6300 planets beyond the Solar System has revealed a broad range in the physical and orbital properties of exoplanets. In contrast to this observed diversity, planets within many multi-planetary systems exhibit pronounced intra-system similarities in their radii, masses, and orbital spacings. This thesis investigates the planetary properties and system architectures, focusing on the observed intra-system similarities, the inferred planetary interior compositions, and the differences between single- and multi-planetary systems.

Paper A shows that planets within the same system tend to have similar orbital spacings even when their sizes or masses differ substantially. The spacing similarity is more pronounced in orbital period ratios than in mutual Hill separations, while the period ratios are more strongly correlated with the masses than with the radii of adjacent planets. These results indicate that uniform orbital spacings represent a common architectural property and are not necessarily associated with similarities in planetary size or mass.

Paper B compares transiting planets in observed single- and multi-planetary systems (singles and multis, respectively). After excluding hot Jupiters, the distributions of planet types and radii of singles multis orbiting FGK stars show no significant differences, suggesting a common underlying planet population. Around FGK host stars, however, an unexpected overabundance of multis is identified within a specific planet radius range. In contrast, singles and multis hosted by M dwarfs exhibit differences in their planet-type and radius distributions, although the small sample sizes render these results less conclusive.

Paper C finds that non-gas-giant planets in systems hosting detected gas giants are, on average, larger, more massive, and more irregularly spaced than planets in systems without detected gas giants. Adjacent planets exhibit significant similarities in radius, mass, and bulk density, whereas their inferred core and water mass fractions show substantial scatter and no strong correlations. These results indicate that similarities in bulk properties do not necessarily translate into similarities in interior compositions.

Together, these studies place new observational constraints on the processes governing planet formation and dynamical evolution, while also demonstrating that the observed planet multiplicity and intra-system similarities in mass, radius, and orbital spacing provide only a partial representation of planetary systems.

# List of publications

This thesis includes the following papers:

[A] **Alexandra Muresan**, Carina Persson, Malcolm Fridlund, "Diversities and similarities exhibited by multi-planetary systems and their architectures: I. Orbital spacings". December 2024, *A&A*, 692, A122, doi: 10.1051/0004-6361/202451353.

[B] **Alexandra Muresan**, Carina Persson, Malcolm Fridlund, "Diversities and similarities exhibited by multi-planetary systems and their architectures: II. Radii of singles and multis". March 2026, *A&A*, 707, A355, doi: 10.1051/0004-6361/202557468.

[C] **Alexandra Muresan**, Carina Persson, Lorena Acuña-Aguirre, Malcolm Fridlund, "Diversities and similarities exhibited by multi-planetary systems and their architectures: III. Trends in planetary interior compositions". Submitted to *A&A* in August 2026.

Other publications by the author, not included in this thesis, are:

[D] C. Persson, ..., **A. Muresan**, et al., "TOI-1438: A rare system with two short-period sub-Neptunes and a tentative long-period Jupiter-like planet orbiting a K0V star". *A&A*, 2025, 702, A69.

[E] J. Korth, D. Gandolfi, ..., **A. Muresan**, et al., "TOI-1130: A photodynamical analysis of a hot Jupiter in resonance with an inner low-mass planet". *A&A*, 2023, 675, A115.

[F] J. Cabrera, D. Gandolfi, ..., **A. Muresan**, et al., "The planetary system around HD 190622 (TOI-1054)". *A&A*, 2023, 675, A183.

[G] J. Orell-Miquel, ..., **A. Muresan**, et al., "HD 191939 revisited: New and refined planet mass determinations, and a new planet in the habitable zone". *A&A*, 2023, 669, A40.

# Acknowledgments

First and foremost, I want to convey my profound thanks and gratitude to my supervisor, Carina Persson, and my co-supervisor, Malcolm Fridlund, for their immense help and guidance. I am extremely thankful for their great support and constructive feedback, and I deeply appreciate their good and thoughtful advice.
I am also colossally grateful to my parents and brother for their enormous help and unwavering support. I greatly thank them for their huge encouragement and trust during my journey in life and for always providing me with love and care.

Alexandra Muresan
Gothenburg, 23 September 2026

# Abbreviations

The following list consists of the abbreviations that are mentioned more than once throughout this thesis.

- **Ariel** : Atmospheric Remote-sensing Infrared Exoplanet Large-survey
- **CHEOPS** : CHaracterising ExOPlanet Satellite
- **ESA** : European Space Agency
- **HARPS** : High Accuracy Radial Velocity Planet Searcher
- **HATNet** : Hungarian Automated Telescope Network
- **IAU** : International Astronomical Union
- **JWST** or **Webb** : James Webb Space Telescope
- **NASA** : National Aeronautics and Space Administration
- **PLATO** : PLAnetary Transits and Oscillations
- **Roman** : Nancy Grace Roman Space Telescope
- **RV** : Radial Velocity
- **TESS** : Transiting Exoplanet Survey Satellite
- **TRAPPIST** : TRAnsiting Planets and PlanetesImals Small Telescope
- **TTV** : Transit Timing Variation
- **WASP** : Wide Angle Search for Planets

# Symbols and constants

The following list comprises the symbols and constants in SI units used throughout this thesis (Prša et al. 2016).

$M_\star$ : Stellar mass

$R_\star$ : Stellar radius

$T_{\rm eff}$ : (Stellar) Effective temperature

$M_\odot$ : Solar mass $= 1.988 \times 10^{30}$ kg

$R_\odot$ : Solar radius $= 6.957 \times 10^{8}$ m

$M_{\rm p}$ : Planet mass

$R_{\rm p}$ : Planet radius

$T_{\rm eq}$ : (Planetary) Equilibrium temperature

$M_\oplus$ : Earth mass $= 5.971 \times 10^{24}$ kg

$R_\oplus$ : Earth radius $= 6.3781 \times 10^{6}$ m

$F_{\odot,\rm in}$ : Average solar flux incident on Earth $= 1361\ {\rm W\,m^{-2}}$

$M_{\rm J}$ : Jupiter mass $= 1.898 \times 10^{27}\,{\rm kg} \approx 318\,M_\oplus$

$R_{\rm J}$ : Jupiter radius $= 7.1492 \times 10^{7}\,{\rm m} \approx 11\,R_\oplus$

$G$ : gravitational constant $= 6.6743 \times 10^{-11}\,{\rm m^3\,kg^{-1}\,s^{-2}}$

au : 1 astronomical unit $= 1.496 \times 10^{11}$ m

pc : 1 parsec $= 3.086 \times 10^{16}$ m

# Contents

# CHAPTER 1

## Introduction

Prior to the early 1990s, no planet beyond the Solar System had been robustly confirmed. Subsequent advances in observational instrumentation, data acquisition, and analysis techniques have revealed that such extrasolar planets, referred to as exoplanets, are ubiquitous in the local Milky Way Galaxy and display a rich diversity in their physical and orbital properties (e.g. Deeg & Belmonte 2018; Zhu & Dong 2021).

The first confirmed detections were made in 1992, when two planets were discovered orbiting a pulsar (Wolszczan & Frail 1992), followed in 1995 by the discovery of a planet orbiting a Sun-like star (Mayor & Queloz 1995). Ground- and space-based facilities have enabled the confirmation of more than 6300 exoplanets and the identification of thousands of additional planet candidates[1]. The rapidly growing number of known exoplanets has enabled a shift from the characterisation of individual planets toward population-level studies of exoplanet demographics and planetary system architectures.

The observed planets span a wide range of sizes, from radii of approximately 0.3 Earth radii (Barclay et al. 2013), slightly larger than the Moon, to inflated gas giants with radii exceeding two Jupiter radii ($> 22\,R_\oplus$; Galazutdinov et al. 2023). Observations have also uncovered remarkable and unexpected worlds, including lava planets with orbital periods of mere hours (e.g. Rappaport et al. 2013), as well as extremely low-density, inflated planets (e.g. Masuda 2014), and young planets revolving around their host stars once every several hundred years (e.g. Marois et al. 2008).

[1]Based on data from the NASA Exoplanet Archive accessed on 2026-09-03.

Notably, nearly half of the confirmed exoplanets to date reside in multi-planetary systems containing between two and eight detected planets[2]. A large fraction of these systems host planets in compact configurations with orbital periods shorter than 100 days. However, the observed architecture of a planetary system does not necessarily represent its intrinsic architecture. Current observational biases and limitations, particularly the incomplete sensitivity to low-mass planets and planets with long orbital periods, hinder a robust characterisation of the intrinsic planet population and cause the observed planet multiplicity to underestimate the true number of planets in a system. Consequently, despite the discovery of more than 1060 multi-planetary systems to date, none of them exhibits an architecture closely analogous to that of the Solar System, comprising multiple inner terrestrial planets and several outer, long-period giant planets.

In contrast to the observed diversity among exoplanets, many multi-planetary systems exhibit substantial intra-system similarities. Planets within the same system often have similar radii and masses, and are regularly spaced in nearly circular and coplanar orbits (e.g. Millholland et al. 2017; Weiss et al. 2018a; Van Eylen & Albrecht 2015; Otegi et al. 2022; Muresan et al. 2024). These planetary properties and system architectures provide important constraints on the processes that govern planet formation and subsequent dynamical evolution.

Another fundamental question concerns whether the observed differences between single- and multi-planetary systems reflect two intrinsically distinct planetary populations. A subset of systems in which only one planet has been detected may in reality contain additional planets that remain undetected due to their small sizes, long orbital periods, large eccentricities, high mutual inclinations, or unfavourable viewing geometry (e.g. Weiss et al. 2018c; Millholland et al. 2021; He et al. 2020; Muresan et al. 2026). An important open question in exoplanetary science is therefore whether the observed planets in single- and multi-planetary systems originate from a common underlying population or represent two different populations.

[2]Based on data from the NASA Exoplanet Archive accessed on 2026-09-03.

Despite the observed intra-system similarities in planetary bulk properties, it has not yet been established on a population level whether planets within the same system also possess similar interior compositions. Planets with comparable radii, masses, or bulk densities may nonetheless possess different interior structures and relative fractions of iron, silicates, volatiles, and gaseous envelopes. Investigating whether the observed intra-system similarities extend from bulk properties to interior compositions provides further insight into the degree of uniformity among planets within the same system.

## 1.1 Exoplanets and their host stars

### 1.1.1 Definition of an exoplanet

An exoplanet is defined as a planet beyond the Solar System, and it is therefore first necessary to review the definition of a planet. In 2006, the International Astronomical Union (IAU) adopted a formal definition of a planet within the Solar System[3]. According to this definition, a planet is a celestial body that satisfies the following three criteria:

**1**) It orbits the Sun,

**2**) It has sufficient mass for its self-gravity to overcome rigid-body forces and attain a nearly round shape through hydrostatic equilibrium, and

**3**) It has cleared the neighbourhood around its orbit.

The IAU definition was formulated specifically for bodies in the Solar System and does not constitute a formal definition of an exoplanet. In studies of extrasolar objects, the first criterion must instead stipulate that an exoplanet orbits one or multiple host stars outside the Solar

[3]The corresponding IAU resolution is accessible at https://www.iau.org/static/resolutions/Resolution_GA26-5-6.pdf.

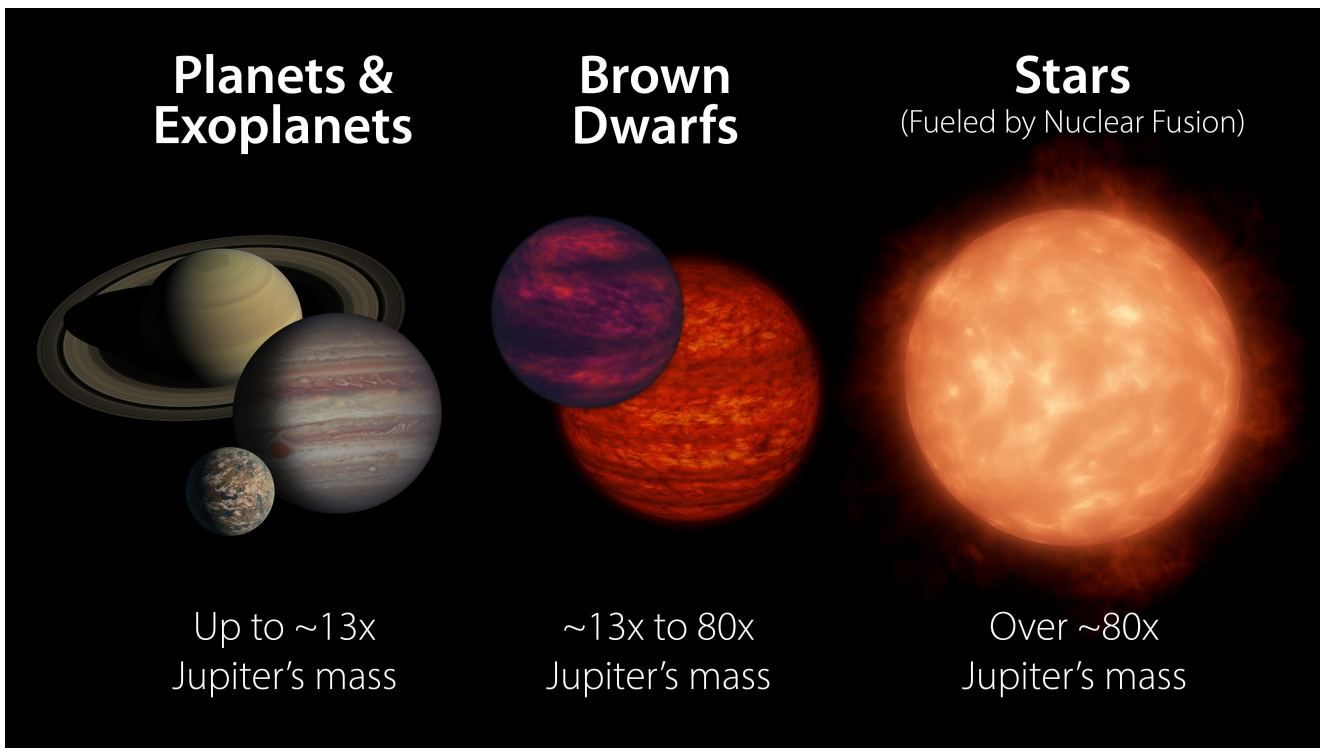

**Figure 1.1:** Approximate nominal mass ranges typically used to distinguish between planets, brown dwarfs, and stars. The frequently adopted upper planetary mass limit of 13 $M_\mathrm{J}$ corresponds approximately to the onset of deuterium burning. Brown dwarfs are commonly classified as sub-stellar bodies with masses of approximately $13 - 80\,M_\mathrm{J}$, while stars represent more massive objects in which hydrogen fusion occurs, depending on composition. These boundaries should be regarded as sharp physical divisions. Illustration credit: NASA/JPL-Caltech (2020).

System. Furthermore, observations have also revealed a population of planetary-mass objects that are not gravitationally bound to any star, commonly referred to as free-floating planets (e.g. Coleman 2024). This type of objects are, however, not considered in this thesis.

Additional practical criteria are frequently adopted, in particular an upper mass boundary separating planets from brown dwarfs. This thesis adopts the commonly used upper limit of 13 $M_\mathrm{J}$ for planetary masses, as illustrated in Fig. 1.1. This value corresponds approximately to the onset of deuterium burning for objects of roughly solar composition, as formulated in the IAU statement from 2003 (Lecavelier des Etangs & Lissauer 2022). Under this mass-based classification, sub-stellar objects above the nominal 13 $M_\mathrm{J}$ limit are classified as brown dwarfs and are estimated to reach up to 80 $M_\mathrm{J}$, which is the approximate mass at which an object can begin to fuse hydrogen into helium in its core (Burrows et al. 2001; Chabrier & Baraffe 2000). This value marks the frequently adopted lower mass boundary of main-sequence stars and corresponds approximately to the lowest mass of an M-type star.

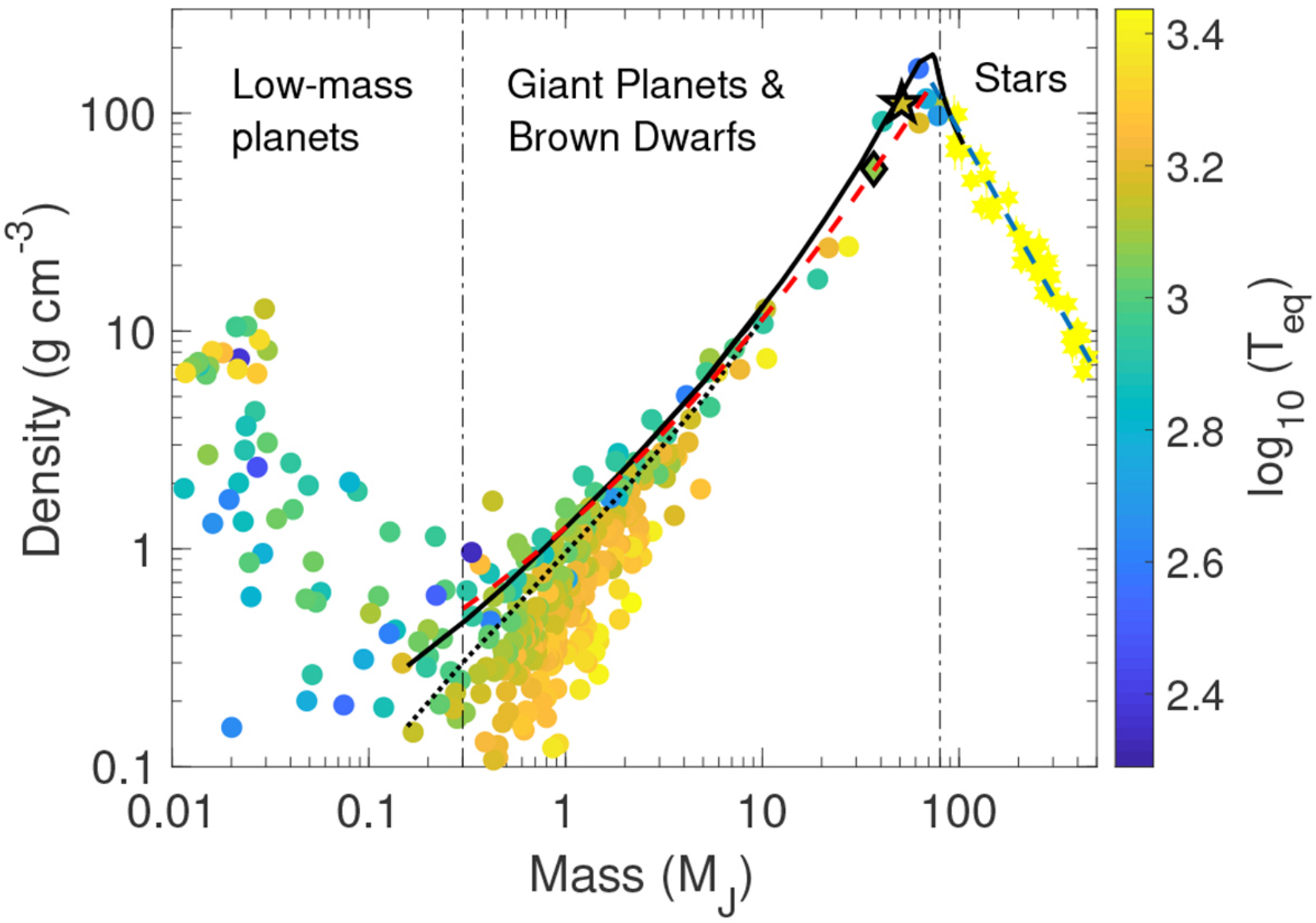

**Figure 1.2:** Mass–density diagram from Persson et al. (2019) for planets, brown dwarfs, and low-mass stars in eclipsing binaries with relative uncertainties in the measured masses and densities below 20%. The colours of the planets and brown dwarfs indicate their equilibrium temperatures, assuming isotropic re-radiation and a Bond albedo of zero. The vertical dashed lines at 0.3 and 80 $M_\mathrm{J}$ separate the three regions labelled low-mass planets, giant planets and brown dwarfs, and low-mass stars. The red and blue lines show fits to the data, while the black solid and dotted curves represent theoretical models.

The 13 $M_\mathrm{J}$ limit should not be interpreted as a unique physical division between planets and brown dwarfs, and the use of a fixed mass boundary remains debated. For example, Chabrier et al. (2014) argued that the distinction should instead be based on formation mechanisms, for which the corresponding mass ranges may overlap. Empirical mass–density relations likewise do not reveal a sharp transition at 13 $M_\mathrm{J}$ (Hatzes & Rauer 2015; Persson et al. 2019), as illustrated in Fig. 1.2. The upper planetary mass limit adopted in this thesis should therefore be regarded as a practical classification criterion rather than a fundamental physical boundary.

|  | Spectral type | Mass [$M_\odot$] | Radius [$R_\odot$] | Effective temp. [K] | Fraction of stars in the Solar neighbourhood |
|---|---|---|---|---|---|
| Early | O | ≥ 18 | ≥ 7.2 | ≥ 31 500 | < 0.01 % |
| ↓ | B | 2.3 – 18 | 2.2 – 7.2 | 10 000 – 31 500 | 1 % |
|  | A | 1.7 – 2.3 | 1.75 – 2.2 | 7 300 – 10 000 | 1 % |
|  | F | 1.07 – 1.7 | 1.13 – 1.75 | 6 000 – 7 300 | 3 % |
|  | G | 0.9 – 1.07 | 0.83 – 1.13 | 5 300 – 6 000 | 8 % |
|  | K | 0.58 – 0.9 | 0.6 – 0.83 | 3 900 – 5 300 | 12 % |
| Late | M | 0.08 – 0.58 | ≤ 0.6 | 2 300 – 3 900 | 75 % |

**Table 1.1:** Conventional spectral classification of main-sequence stars. The values are approximate and compiled from Pecaut & Mamajek (2013), Mamajek (2022), and Ledrew (2001).

### 1.1.2 Nomenclature of exoplanets

Exoplanets are generally designated by the name of their host star followed by a lower-case letter. The first planet discovered in a system is typically assigned the letter b, followed by c, d, and so forth. These letters indicate the order of discovery and do not necessarily correspond to the planets' orbital ordering. When several planets are discovered simultaneously, the letters are usually assigned in order of increasing orbital distance.

A star is often listed in several catalogue and can, therefore, undertake multiple designations. Star names often include the initials of their respective catalogues followed by specific numbers, as exemplified by the star GJ 9827. The most frequently encountered star catalogues are the GJ, HD, HIP, and HR catalogues (e.g. Perryman 2018, Ch. 1). Certain stars are also named after the constellation in which they reside, as for example the star 51 Pegasi in the constellation Pegasus, while few others are designated based on their coordinates. Additionally, many host stars receive survey-based names such as *Kepler*, K2, WASP, and TESS (Sect. 3.3.3).

### 1.1.3 Host-star characterisation

Accurate host-star characterisation is essential because the inferred planetary radii and masses depend directly on the adopted stellar radius and mass. The semi-major axis of a planet's orbit, which indicates the size of the orbit, inferred through Kepler's third law (Eq. 2.1) also depends on the stellar mass. Consequently, uncertainties in the stellar parameters propagate directly into the derived planetary properties and contribute to their associated uncertainties (Kjeldsen 2018; Persson 2024).

Stars are studied and characterised through different techniques, such as photometry, astrometry, spectroscopy, asteroseismology, interferometry, and comparisons with stellar evolutionary models. Accurate determination of host-star parameters is essential for characterising planets and investigating their influence on the planetary physical, chemical, and orbital properties. Reliable planet and host-star parameters are particularly important for the population analyses conducted in the three papers presented in this thesis, which compare planetary radii and masses across different systems and stellar types.

Table 1.1 summarises the conventional spectral classification of main-sequence stars, including the characteristic mass, radius, and effective temperature ranges for the seven spectral types presented (Pecaut & Mamajek 2013; Mamajek 2022). Each class is further sub-divided into several categories, which are however not listed. The table also presents the approximate fractions of main-sequence stars in the Solar neighbourhood belonging to each spectral type.

Of particular interest are M-type stars, also referred to as M dwarfs, which constitute approximately 75% of this local main-sequence stellar population (Henry & Jao 2024). Their low stellar masses and radii also make M dwarfs favourable targets for the detection and characterisation of small planets, since a planet of a given radius produces a deeper transit around a smaller star, while a planet of a given mass and orbital period induces a larger radial-velocity signal around a lower-mass star (Ch. 3). It is therefore important to distinguish planets orbiting M dwarfs from those hosted by FGK stars, as implemented in paper B presented in this thesis. Furthermore, the lowest-mass M dwarfs have the longest main-sequence lifetimes of all stars, reaching up to trillions of years, vastly exceeding the current age of the Universe.

## 1.2 Scientific aims

The rapidly growing number of exoplanet detections has enabled planetary systems to be studied both through detailed characterisations of individual planets and through comparative analyses of planetary properties, system architectures, and population-level trends. A significant finding is that the rich diversity of exoplanets is accompanied by a substantial degree of uniformity observed within individual multi-planetary systems, as described in the beginning of this chapter.

Nevertheless, several fundamental questions remain unresolved. For instance, it is yet unclear how prevalent the observed intra-system similarities in planetary bulk properties are across planetary systems, and whether they extend to planetary interior compositions. Another open question is whether the apparent distinction between single- and multi-planetary systems reflects intrinsically different populations or is largely driven by observational biases and limitations.

The overall aim of this thesis is to investigate differences and similarities both within and across the observed planetary systems, with particular emphasis on the architectures of multi-planetary systems and the properties of their exoplanets. The three papers included in this thesis address complementary aspects of this topic.

Paper A investigates the orbital architectures of confirmed multi-planetary systems, focusing on the similarities of orbital spacings between adjacent planets and their relationships with planetary radii and masses.

Paper B examines whether planets in observed single- and multi-planetary systems are consistent with originating from the same underlying population based primarily on the distributions of their types and radii across different host-star classes.

Paper C investigates whether the observed intra-system similarities in planetary radius, mass, and bulk density also extend to the corresponding inferred interior compositions, and whether the presence of detected gas giants is associated with differences in the properties and architectures of the non-gas-giant planets within the same systems.

The main research questions addressed in this thesis are therefore:

- How similar are the physical properties and orbital spacings of planets within observed multi-planetary systems?
- Are planets in observed single- and multi-planetary systems consistent with belonging to the same underlying planet population based on the distributions of their types and radii?
- Do intra-system similarities in observable planetary bulk properties extend to the inferred interior compositions?
- Are the physical and orbital properties of non-gas-giant planets different in systems with and without detected gas giants?

## 1.3 Thesis outline

The remainder of this thesis provides the theoretical and observational background required for the studies presented in Papers A–C. Chapter 2 introduces the orbital elements commonly used to describe planetary systems. Chapter 3 reviews the principal methods for detecting and characterising exoplanets, with particular emphasis on transit photometry, radial-velocity measurements. The observed exoplanet population and its physical properties, including demographic studies, mass–radius relations, and planetary interior compositions are presented in Ch. 4. Following, Ch. 5 focuses on planetary system architectures, including observed planet multiplicities and intra-system similarities. Finally, Ch. 6 summarises the main analyses and results of Papers A–C, prior to discussing future observational opportunities and remaining questions in Ch. 7.

CHAPTER 2

# Orbital parameters of planetary systems

This chapter introduces the orbital parameters frequently employed in studies of exoplanets and planetary systems. Planetary motion is commonly described using the Keplerian two-body approximation, in which a planet and its host star are treated as point masses interacting only through their mutual Newtonian gravitational attraction. The solution to the Kepler problem gives the positions and velocities of the two bodies as functions of time (Murray & Correia 2010). In multi-planetary systems, mutual gravitational perturbations between the planets cause deviations from exact Keplerian motion and can lead to variations in their orbital elements over time. Nevertheless, Keplerian orbits provide the fundamental framework for describing the orbital parameters of exoplanets.

In the two-body approximation, the planet and its host star orbit their common centre of mass, referred to as the barycentre. The distances of the star and planet from the barycentre depend on their mass ratio and orbital separation. For $M_{\rm p} \ll M_{\star}$, the barycentre is usually located inside the star, as in the Sun–Earth system. However, for a sufficiently high companion-to-star mass ratio and/or a sufficiently large orbital separation, the barycentre may lie outside the stellar surface, as in the Sun–Jupiter system.

For a bound two-body system, both the planet and the star follow elliptical orbits around the barycentre, which is located at one focus of each ellipse. Equivalently, the relative orbit of the planet with respect to the star forms an ellipse with the star located at one focus. These are referred to as Keplerian orbits, for which Kepler's three laws of planetary motion apply (e.g. Murray & Correia 2010).

A Keplerian orbit and the position of the orbiting body along it at a given reference epoch are commonly described by six orbital elements. Five of these specify the size, shape, and orientation of the orbit: the semi-major axis $a$, eccentricity $e$, inclination $i$, longitude of the ascending node $\Omega$, and argument of periapsis $\omega$. The sixth element specifies the orbital phase at the reference epoch and is commonly expressed as the mean anomaly $M_0$. Equivalent phase parameters, such as the true anomaly $\nu_0$ at a specified epoch, may also be used. In the following description, the instantaneous orbital phase is specified by the true anomaly $\nu$ (Murray & Correia 2010; Perryman 2018, Ch. 2).

The six aforementioned orbital elements are presented below, with four of them also illustrated in Fig. 2.1. Additional quantities, such as the orbital period, mutual inclination, and orbital spacing between planets are introduced as well.

- Eccentricity, $e$:
  It defines the shape of the orbit. A bound elliptical orbit has $0 \leq e < 1$, with $e = 0$ representing a circular orbit. Increasing eccentricity corresponds to an increasingly elongated ellipse, while parabolic and hyperbolic trajectories are unbound and have $e = 1$ and $e > 1$, respectively.

- Semi-major axis, $a$:
  It describes the size of the orbit and equals half of the major axis of the ellipse. In this thesis, $a$ denotes the semi-major axis of the relative planet–star orbit. It can be expressed in terms of the periapsis and apoapsis distances as follows:

$$a = \frac{r_{\rm pe} + r_{\rm ap}}{2} \ .$$

The periapsis distance, $r_{\rm pe}$, denotes the minimum separation between the planet and host star during an orbit:

$$r_{\rm pe} = a\,(1 - e) \ ,$$

while the apoapsis distance, $r_{\rm ap}$, represents their maximum separation:

$$r_{\rm ap} = a\,(1 + e) \ .$$

For a circular orbit ($e = 0$), $a$ corresponds to the planet–star separation, which is constant and equal to the radius of the orbit.

- Orbital period, $P$:
  It is the time required for a planet to complete one orbital revolution and is related to the semi-major axis through Kepler's third law:

$$P^2 = \frac{4\pi^2}{G\,(M_\star + M_{\rm p})}\,a^3 \ . \tag{2.1}$$

  Two planets are close to a mean-motion commensurability when their orbital periods are close to a ratio of small integers, for example 2:1, 3:2, or 4:3. A true mean-motion resonance additionally requires at least one associated resonant angle to librate rather than circulate. In such configurations, the planets' orbital motions are gravitationally coupled, causing the planets to experience periodic gravitational perturbations that can significantly influence their long-term dynamical evolution.

- Orbital spacing:
  In multi-planetary systems, the orbital spacing between two adjacent planets is frequently expressed as their orbital period ratio, $P_{i+1}/P_i$ , where the planets are indexed by increasing semi-major axis. Orbital period ratios constitute one of several quantities used throughout this thesis to characterise planetary system architectures.

  An alternative measure of the orbital spacing is the separation in units of the mutual Hill radius. For two adjacent planets with masses $M_i$ and $M_{i+1}$ and semi-major axes $a_i$ and $a_{i+1}$, the mutual Hill radius is

given by:

$$R_H = \left( \frac{M_i + M_{i+1}}{3M_\star} \right)^{1/3} \frac{a_i + a_{i+1}}{2} \ .$$

Their dimensionless separation is therefore:

$$\Delta = \frac{a_{i+1} - a_i}{R_H} \ . \tag{2.2}$$

In contrast to the period ratio, $\Delta$ also depends on the planetary masses and is therefore useful for characterising the dynamical spacing of planetary systems.

- Inclination, $i$:
  It denotes the angle between the orbital plane and the reference plane, which is usually specified to be the sky plane, as illustrated in Fig. 2.2. Equivalently, $i$ is the angle between the orbital angular-momentum axis and the observer's line of sight. The inclination is defined in the range $0° \leq i \leq 180°$, where orbits with $i = 0°$ and $i = 180°$ are seen face-on, whereas an orbit with $i = 90°$ is viewed edge-on, as shown in Fig. 2.2. Transiting planets therefore have orbital inclinations close to 90°, as discussed in Sect. 3.3.

- Longitude of the ascending node, $\Omega$:
  The intersection between the orbital plane and the reference plane defines the line of nodes. The ascending node is the point at which the orbiting body crosses from the negative to the positive side of the reference plane. The longitude of the ascending node, $\Omega$, denotes the angle measured in the reference plane from a predefined reference direction to the ascending node, with $0° \leq \Omega < 360°$. For most exoplanets, $\Omega$ cannot be determined from transit photometry or radial-velocity observations alone.

- Mutual inclination:
  In multi-planetary systems, the mutual inclination between two planets denotes the angle between their orbital planes. Small mutual inclinations, therefore, correspond to nearly coplanar orbits. Since transit photometry generally does not constrain the longitudes of the ascending nodes, the true mutual inclinations cannot usually be determined from transits alone.

- Argument of periapsis, $\omega$:
  It denotes the angle measured in the orbital plane from the ascending node to the periapsis, with $0° \leq \omega < 360°$. For a circular orbit, the periapsis is undefined, and $\omega$ therefore has no unique physical value. When required in a numerical parametrisation of a circular orbit, $\omega$ may be fixed to an arbitrary value, with $\omega = 90°$ conventionally adopted.

- True anomaly, $\nu$:
  It is the angle measured in the orbital plane, at the focus of the ellipse, from the direction of periapsis to the instantaneous position of the orbiting body. It varies with time and specifies the instantaneous orbital phase, with $0° \leq \nu < 360°$. Its value at the reference epoch, $\nu_0$, may be used as the sixth orbital element instead of the mean anomaly $M_0$.

The orbital parameters introduced in this chapter provide the basis for the detection methods discussed in Ch. 3 and for the characterisation of the observed planetary population and system architectures in Chs. 4 and 5.

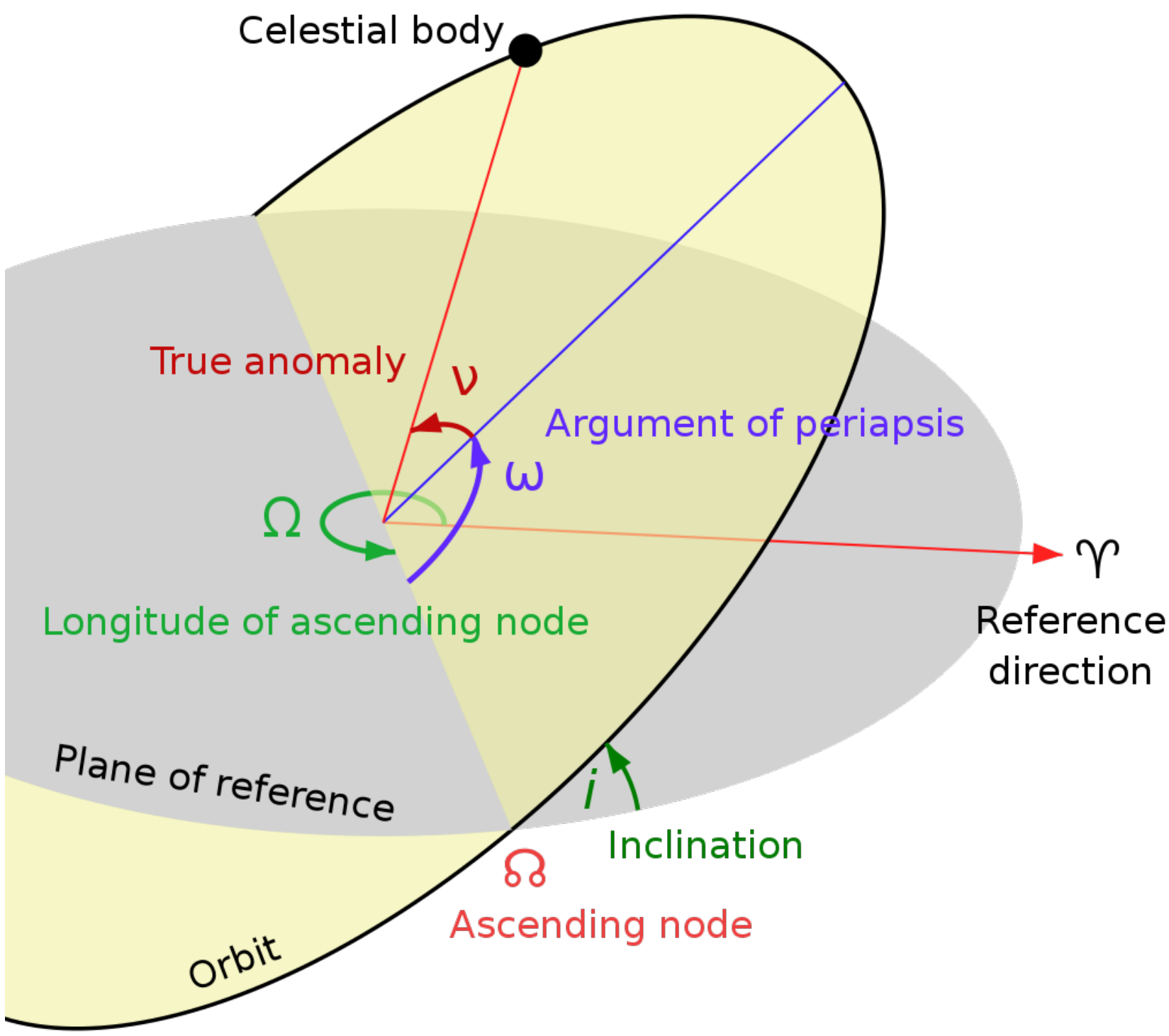

**Figure 2.1:** Geometry of a Keplerian orbit with the sky plane as the plane of reference. The three orientation angles $i$, $\Omega$, and $\omega$ are shown together with the instantaneous true anomaly $\nu$. The semi-major axis $a$ and eccentricity $e$, which describe the size and shape of the orbit, respectively, are not indicated. Illustration credit: Snyder (2007).

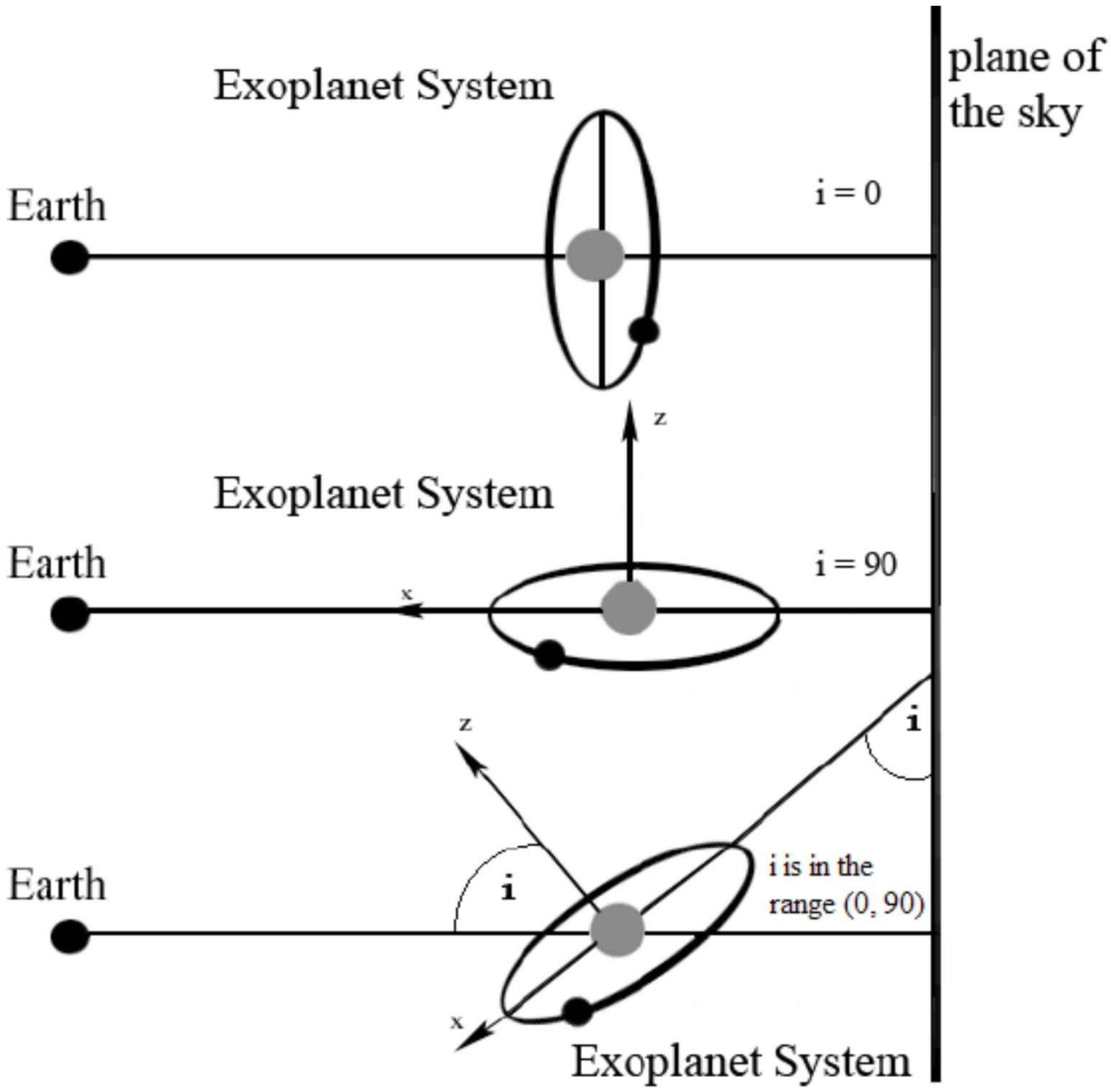

**Figure 2.2:** Orbital inclination $i$ of an exoplanet as viewed from Earth. $i$ denotes the angle between the orbital plane and the sky plane, and is equivalently defined as the angle between the orbital angular-momentum axis and the observer's line of sight. The top and middle panels show planetary systems viewed face-on ($i = 0°$) and edge-on ($i = 90°$), respectively, while the bottom panel illustrates an intermediate inclination. Illustration credit: Todorov (2008).

CHAPTER 3

# Exoplanet detection methods

## 3.1 Overview

This chapter provides an overview of the principal techniques currently employed for the detection of exoplanets. As of 3 September 2026, the number of known exoplanets has reached 6360, while more than 8000 additional planetary candidates are awaiting validation (NASA Exoplanet Archive). Direct observations of exoplanets are challenging due to the large contrast in brightness between planets and their host stars and their small angular separations on the sky. Most known exoplanets have therefore been discovered indirectly, through the effect of the planet on the motion or observed brightness of its host star, or through its gravitational effect on the light from a more distant background source. Direct imaging constitutes an important exception because this method detects light from the planet itself.

Figure 3.1 shows the cumulative number of confirmed exoplanets discovered with eleven different techniques as a function of time. For a planet detected with more than one technique, the figure indicates the method responsible for the initial discovery. For instance, a planet discovered through transit photometry may later have its mass measured using radial-velocity (RV) observations or transit-timing variation (TTV) measurements, while a planet discovered with RVs may subsequently be found to transit. Different techniques therefore provide complementary information about the same planet.

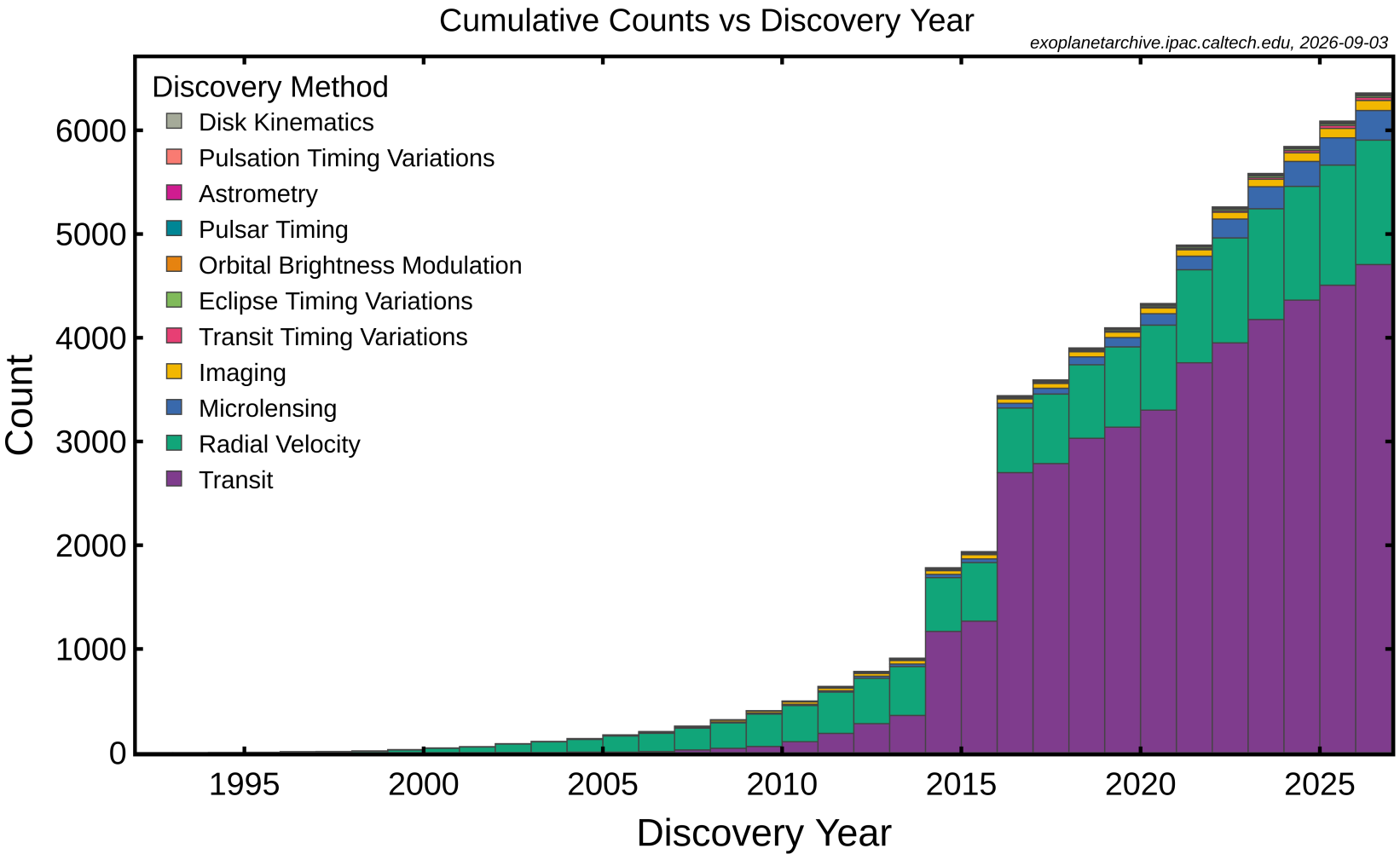

**Figure 3.1:** Cumulative number of confirmed exoplanets discovered over time using different detection techniques. For planets detected through multiple techniques, only the method responsible for the initial discovery is shown. Data from the NASA Exoplanet Archive, as of 2026-09-03.

The vast majority of exoplanets have been discovered using in total six of the techniques displayed in Fig. 3.1, comprising one direct and five indirect methods. These are presented in the remainder of this chapter in the following order: radial-velocity measurements, transit photometry, transit-timing variations, astrometry, gravitational microlensing, and direct imaging. The numbers and fractions of planets detected using each of these techniques are enumerated in Table 3.1.

These principal methods are sensitive to different planetary and stellar properties and therefore detect different subsets of the planets that exist. Transit photometry primarily measures the decrease in stellar brightness caused by a transiting planet and is therefore strongly sen-sitive to planetary size and orbital geometry. The RV method detects the motion of the host star and is primarily sensitive to the planetary mass and orbital period. TTVs measure changes in transit times caused by gravitational interactions between planets within the same system. Astrometry measures the motion of the host star across the sky, while microlensing detects the gravitational deflection of light from a background star. In contrast, direct imaging is employed to capture light from the planet itself.

| Detection method | Number of discovered planets (% of all 6360 planets) | Number of detected planets (% of all 6360 planets) |
|---|---|---|
| **Transit photometry** | 4706 (≈ 74 %) | 4736 (≈ 74 %) |
| **Radial velocity** | 1200 (≈ 19 %) | 2516 (≈ 40 %) |
| **Gravitational microlensing** | 285 (≈ 4 %) | 285 (≈ 4 %) |
| **Direct imaging** | 97 (≈ 2 %) | 102 (≈ 2 %) |
| **Transit Timing Variations** | 29 (≈ 0.5 %) | 495 (≈ 8 %) |
| **Astrometry** | 6 (≈ 0.1 %) | 175 (≈ 3 %) |
| **Total** | **6323 (≈ 99 %)** | |

**Table 3.1:** The number and percentage of the 6360 confirmed exoplanets, as of 2026-09-03, discovered (second column) and detected (third column) using each of the six main detection techniques. Transit photometry is the most successful technique and has, together with the radial-velocity method, led to the discovery of approximately 93% of all the confirmed exoplanets to date. Data collected from the NASA Exoplanet Archive.

Differences among these detection methods influence the observed distribution of planetary radii, masses, orbital periods, and multiplicities. A sample of detected planets therefore does not constitute a random sample of the underlying planet population. To interpret exoplanet demographics and planetary system architectures, it is necessary to understand both what each method measures and which planets are most likely to be detected.

This chapter first provides a somewhat detailed description of the radial-velocity and transit methods, which together account for the majority of planetary mass and radius measurements used in current exoplanet studies. TTVs are then introduced as an additional technique for measuring planetary masses in multi-planetary systems. Astrometry, gravitational microlensing, and direct imaging are subsequently de-scribed because they extend sensitivity to different orbital separations and stellar environments. The final section compares the primary observational biases and explains their consequences for studies of planetary populations and system architectures.

## 3.2 Radial velocity

The radial-velocity method was used to discover 51 Pegasi b in 1995, the first confirmed exoplanet orbiting a main-sequence star (Mayor & Queloz 1995). The planet is a gas giant with a short orbital period of 4.2 days, a configuration not represented in the Solar System. Its discovery provided early evidence that planetary system architectures can differ substantially from that of the Solar System. RV measurements remain one of the primary techniques both for detecting planets and for measuring the masses of planets discovered by other methods.

### 3.2.1 Radial-velocity parameters

As described in Ch. 2 for a two-body system, a planet and its host star revolve around their common barycentre in elliptical orbits. The RV method measures the radial component of a host star's velocity, i.e. its component along the observer's line of sight (e.g. Lovis & Fischer 2010; Hatzes 2016, Ch. 8). In contrast, the planet is usually too faint for its orbital motion to be measured directly.

The stellar radial velocity is determined spectroscopically from the Doppler shifts of absorption lines in the stellar spectrum, while the rest wavelengths of these lines are known from laboratory measurements. When the star moves towards the observer, the lines are shifted towards shorter wavelengths, producing a blueshift and a negative radial velocity, as illustrated in Fig. 3.2. Conversely, when the star moves away from the observer, they are shifted towards longer wavelengths, inducing a redshift and a positive radial velocity. A planet therefore induces a periodic pattern of alternating blueshifts and redshifts as the star moves around the system barycentre.

The size of the periodic RV variation is described by the RV semi-amplitude $K$. For a circular orbit, $K$ is half the difference between the maximum and minimum radial velocities, as indicated in Fig. 3.3. For a planet with $M_\mathrm{p} \ll M_\star$,

$$K = \left(\frac{2\pi G}{P}\right)^{1/3} \frac{M_\mathrm{p} \sin i}{M_\star^{2/3}} \frac{1}{\sqrt{1-e^2}} , \tag{3.1}$$

where $G$ is the gravitational constant, $P$ represents the orbital period, $M_\mathrm{p}$ and $M_\star$ denote the planetary and stellar masses, respectively, while $i$ and $e$ denote the orbital inclination and eccentricity, respectively. Notably, the planet and its host star have the same $P$, $i$, and $e$ because they orbit their common barycentre within the same orbital plane.

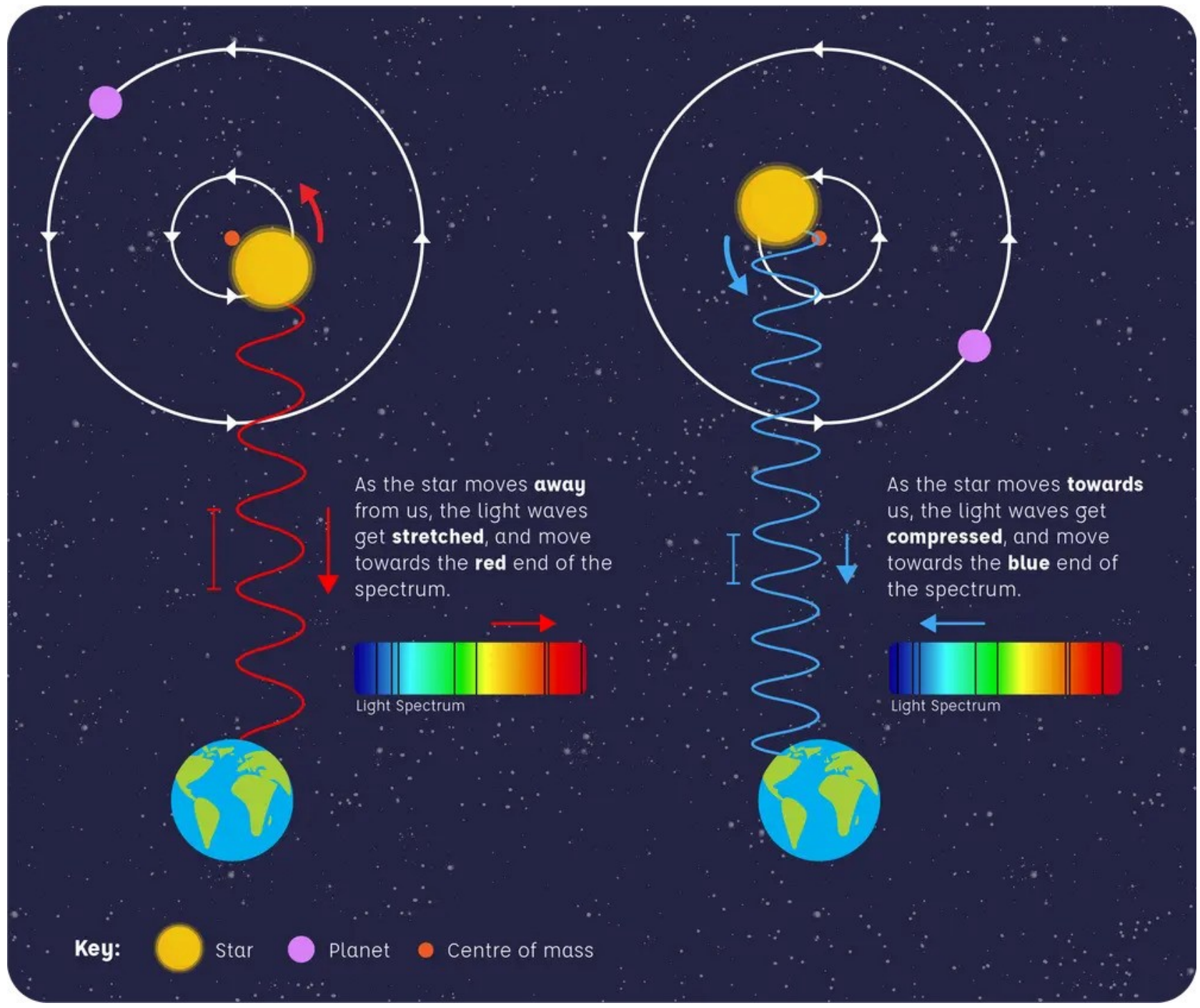

**Figure 3.2:** Visualisation of the radial-velocity method for detecting exoplanets. As the star and planet orbit their common barycentre, the resulting radial motion of the host star can be measured through Doppler shifts in the stellar spectral lines. Motion towards the observer blueshifts the spectral lines, whereas motion away from the observer produces a redshift. Illustration credit: Hopkinson & Barton, LCO.

Equation 3.1 reveals several of the most important properties of the RV method. More massive planets produce larger RV signals, while for a given planetary mass the signal decreases for longer orbital periods and more massive host stars. Consequently, RV surveys are most sensitive to massive planets on relatively short-period orbits, particularly around lower-mass host stars.

For a circular orbit, the stellar radial velocity varies sinusoidally, as illustrated in Fig. 3.3. The orbital period is determined from the repetition time of the RV signal, while $K$ represents the amplitude of the signal. In contrast, an eccentric orbit produces a non-sinusoidal RV curve whose shape contains information about both the eccentricity and the orientation of the elliptical orbit relative to the observer. With sufficiently well-sampled data, an RV time series can therefore constrain the orbital period, semi-amplitude, and eccentricity.

An important limitation follows from the factor $\sin i$ in Eq. 3.1. The RV method measures only the line-of-sight component of the stellar velocity, and can therefore only determine the minimum planetary mass $M_\mathrm{p} \sin i$ unless the inclination $i$ is known. A nearly face-on orbit has a small $\sin i$ and produces little radial motion even if the planet is massive. A transiting planet, however, has an orbital inclination close to 90°, which can be constrained from the transit geometry (Sect. 3.3). Combining transit and RV measurements therefore allows the planet mass $M_\mathrm{p}$ to be determined.

The inferred planetary mass also depends on the adopted mass of the host star. Stellar characterisation is therefore an essential component of RV-based exoplanet studies because uncertainties in $M_\star$ propagate into the measured planet mass. The same general principle applies to the planetary radii obtained from transit photometry, which depend on the measured radii of the host stars, as discussed in Sect. 3.3.

**Figure 3.3:** Schematic of a radial-velocity curve for a circular orbit ($e = 0$). The orbital period $P$ and radial-velocity semi-amplitude $K$ (Eq. 3.1) are indicated. Illustration credit: Altamura et al. (2022).

The scale of the expected RV signals demonstrates why the early RV discoveries around Sun-like stars were dominated by giant planets. As expressed in Eq. 3.1, the RV signal induced by a planet depends on both the orbital period and the mass ratio between the planet and host star. For example, a Jupiter-mass planet orbiting a Sun-like star at a semi-major axis of 0.1 au would induce an RV semi-amplitude of approximately 90 $\mathrm{m\,s^{-1}}$, whereas at 5 au, comparable to Jupiter's orbital semi-major axis in the Solar System, the produced semi-amplitude would be only approximately 12 $\mathrm{m\,s^{-1}}$. These values are much greater than those generated by a planet of 1 Earth mass at a distance of 0.1 au and 1 au, corresponding to $\approx 0.3\,\mathrm{m\,s^{-1}}$ and $0.09\,\mathrm{m\,s^{-1}}$, respectively. The radial-velocity method is therefore most sensitive to massive and/or short-period planets, and detecting low-mass planets requires both very precise spectroscopy and careful treatment of variations originating from the star itself.

### 3.2.2 Observational limitations and stellar variability

Stellar variability is an important limitation on the precision of RV measurements. Oscillations, convection, magnetic activity, and surface features, such as starspots, can change the shapes or apparent wavelengths of stellar absorption lines and thereby induce apparent RV variations (Hatzes 2016, Chs. 9, 10). These variations can add noise to a planetary signal, obscure it, or in some cases produce a periodic signal that resembles the effect of an orbiting planet.

As an example, dark starspots remove light from part of the rotating stellar surface and alter the observed spectral-line profiles as the spot moves across the visible disc. Distinguishing such activity signals from planetary motion can therefore require information from both the RV measurements and independent indicators of stellar activity (e.g. Aigrain & Foreman-Mackey 2023; Barragán et al. 2022).

The attainable RV precision also depends strongly on the stellar spectrum. Velocities are easier to measure when the spectrum contains many deep and narrow absorption lines whose positions can be determined precisely. Hotter stars generally have fewer and broader absorption lines, and rapid stellar rotation broadens the lines further. These properties make precise RV measurements more difficult for hot or rapidly rotating stars (Hatzes 2016, Ch. 3).

Cool, low-mass stars present a different combination of advantages and disadvantages. Equation 3.1 shows that a planet of a given mass and orbital period produces a larger RV signal around a lower-mass star. These properties make K and M dwarfs favourable targets in searches for low-mass planets. Many M dwarfs are nevertheless faint at optical wavelengths and can display strong magnetic activity. High-precision RV measurements therefore depend not only on the expected signal amplitude but also on stellar brightness, spectral type, rotation, activity, and the wavelength range of the observations.

Additionally, both the total time span and the frequency of the observations are important factors. A planet can only be securely identified if the observations cover the stellar orbital variation sufficiently well and over a long enough interval. Long-period planets require long observing baselines, while sparse or irregular sampling can produce ambiguities in the recovered parameters. In multi-planetary systems, several planetary signals can be superimposed on one another together with stellar variability. RV analyses therefore often involve modelling several planetary signals simultaneously, together with stellar variability when necessary.

### 3.2.3 Radial-velocity instrumentation

Precise radial-velocity work requires high-resolution spectrographs with extremely stable wavelength calibration. The position of a stellar absorption line must be measured to a tiny fraction of the line width, and instrumental drifts must be distinguished from true stellar velocity variations. Stabilised spectrographs, for instance HARPS (Mayor et al. 2003) and HARPS-N (Cosentino et al. 2012), can reach metre-per-second precision, while newer instruments, such as ESPRESSO (Pepe et al. 2010, 2021) and EXPRES (Jurgenson et al. 2016), were designed to attain higher precision.

At the lowest amplitudes, however, improved instrumental precision does not automatically translate into equally precise planetary masses because stellar variability can become the dominant source of uncertainty. The practical detectability of an RV planet is therefore determined by a combination of planetary signal amplitude, stellar properties, the frequency of observations, the total observing baseline, and instrumental precision. These selection effects are important when interpreting samples of planets with measured masses.

## 3.3 Transit photometry

Transit photometry detects a planet through the decrease in observed stellar flux that occurs when the planet passes in front of its host star, as viewed from an observer's line of sight. In contrast to the RV method, which mainly probes the planetary mass, the transit signal is primarily sensitive to the planetary size relative to the host star. Transit photometry has become the most effective exoplanet discovery method because a single survey can monitor the brightness of many thousands of stars simultaneously.

A transit is only observed for a favourable orbital orientation. The planet must pass across the apparent stellar disc as viewed by the observer, which requires the orbital plane to be sufficiently close to edge-on. Transit surveys therefore detect only the planetary systems whose orbital orientations allow the planets to pass in front of their host stars as seen by the telescope. This selection effect is central to the interpretation of both planet occurrence rates and observed planet multiplicities.

### 3.3.1 Transit geometry and observables

A transit light curve represents the stellar flux measured as a function of time during the passage of a planet across the stellar disc. A schematic of an idealised and unbiased transit light curve is provided in Fig. 3.4. The basic shape of the transit contains information about the size of the planet and the geometry of its orbit. The most important quantities are introduced below.

#### Impact parameter

A conjunction occurs when the sky-projected separation between the centres of the star and planet reaches a minimum, as viewed by an observer. In the case of a transit, this configuration occurs at the midpoint of the transit. The impact parameter $b$ is defined as the sky-projected distance between the centres of the star and planet at conjunction, in units of stellar radii (Winn 2010):

$$b = \frac{a \cos(i)}{R_\star} \left( \frac{1 - e^2}{1 + e \, \sin(\omega)} \right) \; , \tag{3.2}$$

where $a$ and $i$ correspond to the orbital semi-major axis and inclination, respectively, while $R_\star$ denotes the stellar radius. In the conventional approach, only the absolute value $|b|$ is regarded. For a circular orbit ($e = 0$), the impact parameter is simplified as follows:

$$b = \frac{a \cos(i)}{R_\star} \tag{3.3}$$

Equation 3.3 reveals that, for a circular orbit, the impact parameter depends on the orbital inclination and the orbital distance relative to the stellar radius. As illustrated in Fig. 2.2, an orbit seen edge-on has to $i = 90°$, for which the impact parameter is $b = 0$, and the corresponding transit is referred to as a central transit. Increasing $b$ moves the path of the planet across the stellar disc, referred to as the transit chord, closer to the stellar limb (the apparent edge of the visible stellar disc). Sufficiently large values produce grazing transits in which the planetary disc never lies completely inside the projected stellar disc.

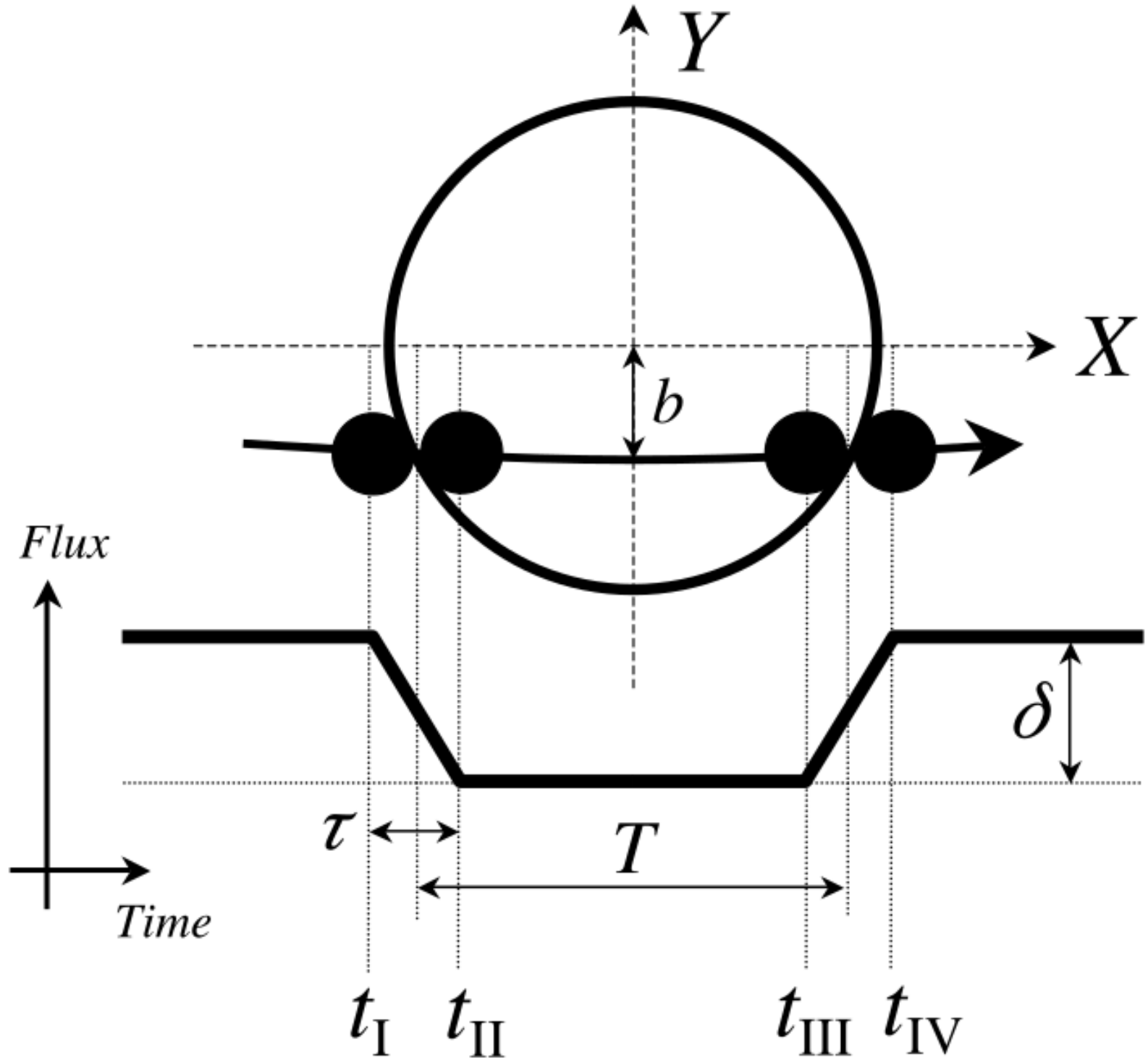

**Figure 3.4:** Schematic of a planet (black circle) transiting its host star from left to right, as viewed by an observer. The included idealised transit light curve shows the relative stellar flux as a function of time. The impact parameter $b$, transit depth $\delta$, ingress and egress intervals, and the four contact points from $t_1$ to $t_4$ are indicated. Illustration credit: Winn (2010).

### Contact points and transit duration

As illustrated in Fig. 3.4, the disc of a non-grazing transiting planet intersects the stellar circumference at four contact points, which correspond to four specific times ranging from $t_1$ to $t_4$. The first contact, $t_1$, occurs when the planetary and stellar discs first intersect, as viewed by an observer. The ingress denotes the interval from $t_1$ to $t_2$, during which the planetary disk progressively. The planet is fully projected against the stellar disc between $t_2$ and $t_3$, and the egress occurs between $t_3$ and $t_4$. In the case of an eccentric orbit, the durations $\tau_{\text{in}}$ and $\tau_{\text{eg}}$ generally differ, whereas for a nearly circular orbit, they may be approximated as equal and collectively denoted by $\tau$ (Winn 2010).

The total transit duration, $T_{\text{tot}}$, which is henceforth referred to as the transit duration, denotes the time elapsed between the first and fourth contact points:

$$T_{\text{tot}} = t_4 - t_1 \,.$$

During a non-grazing transit, the full transit occurs when the entire planetary disk is projected onto the stellar disk, thereby producing the flat-bottomed part of the transit light curve shown in Fig. 3.4. Its duration is given by

$$T_{\text{f}} = t_3 - t_2 \,.$$

The total transit duration depends on the length of the transit chord and on the planet's projected orbital velocity across the stellar disc. It therefore contains information about the orbital period $P$, the orbital separation expressed in units of the stellar radius, $a/R_\star$, the impact parameter $b$, and, for eccentric orbits, the orbital speed at the time of transit. The duration and the shape of ingress and egress are consequently important for determining the transit geometry and distinguishing cen-tral from grazing events.

### Transit depth

As illustrated in Fig. 3.4, the planet blocks part of the stellar disc during a transit. For a non-grazing transit, assuming a uniform stellar surface brightness and negligible contribution from the planetary flux, the fractional transit depth $\delta$ can be approximated as follows:

$$\delta = \left(\frac{R_\mathrm{p}}{R_\star}\right)^2 , \tag{3.4}$$

where $R_\mathrm{p}$ and $R_\star$ denote the radii of the planet and its host star, respectively. The transit depth therefore directly constrains the radius ratio $R_\mathrm{p}/R_\star$, and not the absolute planetary radius. An independent estimate of $R_\star$ is required to obtain $R_\mathrm{p}$, and uncertainties in the stellar radius consequently propagate into the inferred planetary radius.

Equation 3.4 also reveals that a planet of a given size produces a deeper transit around a smaller star. The dependence of the transit depth on the host-star radius is illustrated in Fig. 3.5. For example, an Earth-sized planet transiting a Sun-sized star blocks only about 0.01% of the stellar light, whereas a Jupiter-sized planet produces a flux decrease of approximately 1%. For the same planets orbiting host stars with half the Solar radius, the corresponding transit depths are four times larger.

### Parameters obtained from transit light curves

The following three quantities, commonly referred to as observables, can be directly measured from a transit light curve: transit depth $\delta$ (Eq. 3.4), the total transit duration $T_\mathrm{tot}$, and the full transit duration $T_\mathrm{f}$ (Winn 2010). A fourth observable, the orbital period $P$, can additionally be determined from a photometric time series containing at least two planetary transits. Moreover, the semi-major axis $a$ is related to the orbital period via Kepler's third law (Eq. 2.1.

The following four combinations of physical parameters can be inferred directly from the four observables (Seager & Mallén-Ornelas 2003): the planet-to-star radius ratio $R_\mathrm{p}/R_\star$ (from Eq. 3.4), the impact parameter $b$ (Eq. 3.3), the ratio $a/R_\star$ (from the total transit duration), and the mean density of the host star $\rho_\star$ (from the total transit duration and Eq. 2.1). Specifically, for a circular orbit and $M_\mathrm{p} \ll M_\star$, the orbital period and scaled semi-major axis provide an estimate of the mean stellar density,

$$\rho_\star \simeq \frac{3\pi}{GP^2}\left(\frac{a}{R_\star}\right)^3 .$$

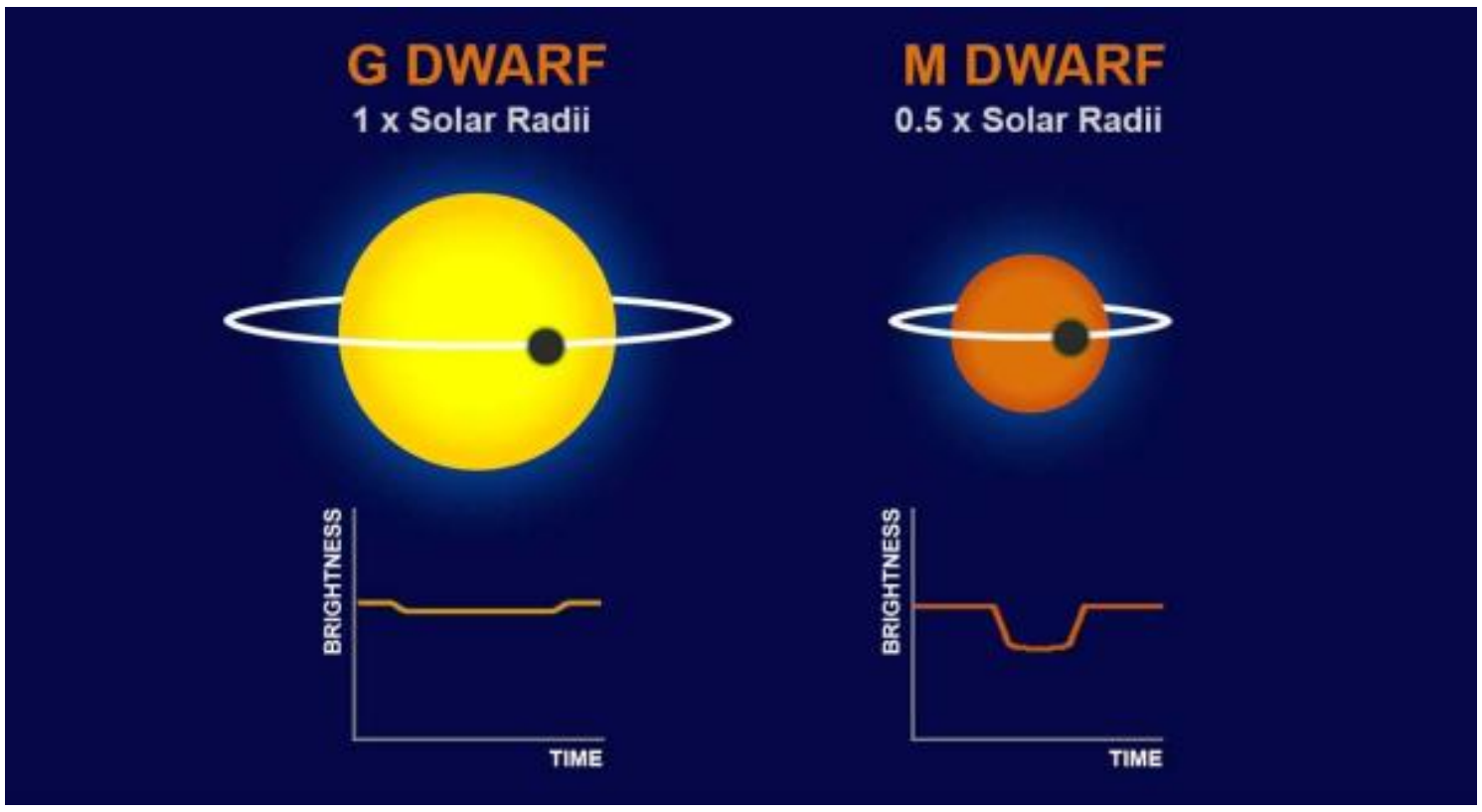

**Figure 3.5:** Illustration of identical planets transiting stars of different radii. Left: The planet transits a star with a radius of $1\,R_\odot$, producing a shallow transit depth, as shown in the transit light curve. Right: For a smaller star with a radius of $0.5\,R_\odot$, the same planet blocks a larger fraction of the stellar flux. An analogous principle applies to host stars of identical radii transited by planets of different sizes, as the larger planet produces a deeper transit. Illustration credit: Eisner (2024).

This quantity is commonly referred to as the transit-derived stellar density. It is useful because the stellar density is constrained by the transit geometry rather than by the transit depth. In practice, the accuracy of the inferred density depends on the adopted orbital model, particularly when the eccentricity is unknown.

Additionally, the orbital inclination $i$ can be inferred from the transit geometry through the impact parameter and the scaled semi-major axis. For a planet that also has radial-velocity measurements, the transit inclination removes the $\sin i$ ambiguity of the RV mass. Transit photometry and RV measurements therefore provide complementary informa-tion: transits yield the planet-to-star radius ratio and the orbital geometry, while RVs provide the mass scale. Combining these measurements yields the true planetary mass, which, together with the planetary radius, allows the bulk density to be determined and subsequently used to constrain the planetary composition.

### 3.3.2 Stellar effects on transit light curves

The simple relation in Eq. 3.4 assumes that the stellar disc has uniform and constant surface brightness. However, real stars violate both of these assumptions. Limb darkening modifies the brightness across the stellar disc, while starspots, faculae, oscillations, and other forms of variability change the brightness with time. These effects must be included when precise planetary radii and orbital parameters are inferred.

#### Limb darkening

In reality, stars exhibit limb darkening (LD), whereby the stellar disk is brightest at its centre and becomes progressively fainter towards its limb (the apparent edge of the stellar disc). This phenomenon occurs due to the radial decrease in photospheric temperature (e.g. Mandel & Agol 2002; Winn 2010). Near the disk centre, the line of sight penetrates into deeper, hotter layers, whereas near the limb it probes shallower and cooler layers. Consequently, a planet blocks regions with different surface brightnesses as it transits the star, and limb darkening therefore changes the transit shape and affects the inferred impact parameter and planet-to-star radius ratio. The variation in stellar intensity across the disc can be incorporated into the analysis of transit light curves by adopting appropriate limb-darkening models (e.g. Mandel & Agol 2002; Claret 2004; Southworth 2008).

#### Starspots and stellar variability

As for radial-velocity measurements, magnetic activity can affect transit observations as well. For example, prominent dark starspots primarily produce the following three features:

- An out-of-transit flux modulation caused by stellar rotation.

- An overestimation of the transit depth, and hence the planetary radius, due to unocculted starspots during a transit.
- A temporary increase in the observed flux during a transit due to starspots occulted by the planet, potentially leading to an underestimated planetary radius.

These effects must be taken into account when analysing observations of active host stars.

### 3.3.3 Transit missions and observing strategies

The properties of a transit survey determine the types of planets it can detect. The number of monitored stars, their brightness and spectral types, the photometric precision, the frequency of observations, and the total observing time affect the resulting planet sample. These differences are important when samples from several surveys are combined or compared.

The *Kepler* mission monitored more than 150,000 stars in a single field continuously for approximately four years (Borucki et al. 2010). Its long, nearly continuous baseline enabled the detection of many small planets and allowed multiple transits to be observed even for orbital periods of months. The same long baseline also made *Kepler* particularly powerful for measuring transit-timing variations (Sect. 3.4 in multi-planetary systems. A disadvantage for follow-up characterisation is that many of the host stars are relatively faint, making precise RV measurements difficult.

After the original *Kepler* observing programme ended, the spacecraft continued as the K2 mission and observed a sequence of different regions of the sky (Howell et al. 2014). Each field was surveyed for a much shorter time than the original *Kepler* field, which reduced the sensitivity to long-period planets and produced a different selection of detected planets.

The Transiting Exoplanets Survey Satellite (TESS) uses a different strategy, monitoring a much larger fraction of the sky and targeting many brighter nearby stars (Ricker et al. 2014). Most sky regions are observed for substantially shorter intervals than the original *Kepler* field, although overlapping sectors provide longer coverage in some regions. TESS is therefore particularly effective at discovering short-period tran-siting planets around stars that are bright enough for detailed follow-up, including RV mass measurements.

The CHaracterising ExOPlanet Satellite (CHEOPS) differs from these discovery surveys because it primarily performs targeted high-precision observations of selected stars rather than searching large samples of stars for previously unknown planets (Benz et al. 2021). It is designed to refine radii and transit parameters for known planets. This distinction is relevant for population studies: measurements from a targeted follow-up mission can improve the precision of individual planetary properties, although its target sample cannot be used directly to determine how common different types of planets are.

The *Kepler*, K2, TESS, and targeted follow-up samples are therefore affected by different observational selection effects and should not be assumed to represent identical underlying stellar or planetary populations simply because they are all observed with transit photometry.

### 3.3.4 Observational biases and limitations

Transit photometry has several strong selection effects. Some arise from orbital geometry, while others occur because a transit must be sufficiently deep and repeated often enough to be recognised in noisy photometric data. Additionally, transit-like signals can be produced by astrophysical systems that do not contain a transiting planet.

#### Transit probability

A planet transits only when its orbital orientation places it across the stellar disc from the observer's point of view. For a circular orbit with $R_\mathrm{p} \ll R_\star$, the geometric transit probability is approximately

$$p_\mathrm{tr} \simeq \frac{R_\star}{a}. \tag{3.5}$$

Equation 3.5 shows that the probability decreases with orbital distance. Eccentricity can modify this probability because the star–planet distance at the time of transit can differ from the semi-major axis. Viewed from a random direction, the probability that Earth would transit the Sun is only about 0.47%, whereas a planet in a ten-day orbit around a Sun-sized star has a transit probability of roughly 5%. Transit surveys are consequently strongly biased towards short-period planets even before instrumental detectability is considered.

In multi-planetary systems, the same geometry affects the number of planets that are seen to transit. If the planetary orbits are not perfectly coplanar, one planet may transit while another does not. An intrinsically multi-planetary system can therefore be observed as a lower-multiplicity system or even as an apparent single-planet system. This distinction between observed and intrinsic multiplicity is discussed further in Ch. 5.

### Signal strength, observing baseline, and completeness

A planet that transits geometrically is not necessarily detected because the signal must also be large enough relative to the photometric noise. The transit depth increases as $(R_\mathrm{p}/R_\star)^2$, so large planets and planets orbiting small stars are easier to detect. The signal-to-noise ratio also improves when several transits are observed and combined. Short-period planets therefore benefit twice: they have a higher geometric transit probability and produce more transit events during a fixed observing interval.

The observing baseline sets an additional limit. A survey lasting only a few weeks cannot normally establish a long orbital period from repeated transits, whereas a multi-year baseline can detect planets with much longer periods. Data gaps and irregular sampling can further reduce the probability that the required number of transit events is observed. The probability that a transiting planet present in the data is recovered by the automated search is commonly described as the detection completeness. It depends on quantities such as planetary radius and orbital period, stellar noise, data gaps, the observing baseline, and the properties of the detection pipeline.

Detection completeness is often estimated by inserting artificial transit signals into real or simulated light curves and testing how many are recovered by the same automated detection procedure used for the survey. This procedure is commonly called an injection–recovery test. The fraction that is recovered provides an empirical estimate of the detection efficiency as a function of different quantities, including planet radius, orbital period, and host-star properties (e.g. Thompson et al. 2018). In order to infer intrinsic planet occurrence rates, the resulting detection completeness must be combined with the geometric transit probability and the survey target selection.

### False positives and validation

A periodic decrease in stellar flux is not sufficient on its own to demonstrate the presence of a planet. Eclipsing binary stars, grazing stellar eclipses, and blended background systems can all produce transit-like signals (Cameron 2012). Planet candidates are therefore checked for features, such as inconsistent odd and even transit depths, secondary eclipses, nearby contaminating sources, and stellar or orbital parameters that are inconsistent with a planetary interpretation (e.g. Morton 2012; Thompson et al. 2018; Lafarga et al. 2026). RV measurements can provide independent confirmation when the host star is sufficiently bright and the expected signal is measurable. Otherwise, statistical validation can be used to test whether false-positive explanations are sufficiently unlikely.

The concepts of completeness and reliability should be kept distinct. Completeness indicates what fraction of real transiting planets with specified properties are recovered, whereas reliability describes what fraction of the reported planet candidates correspond to genuine planets rather than astrophysical false positives or false alarms.

## 3.4 Transit-timing variations

If a transiting planet revolves around its star with an exactly constant orbital period and without significant gravitational perturbations from other planets, its successive transit times follow

$$t_n = t_0 + nP,$$

where $t_0$ is a reference transit time, $P$ is the orbital period, and $n$ is the transit number. In a multi-planetary system, however, the planets gravitationally perturb one another and slightly speed up or slow down each other's orbital motion. Individual transits can therefore occur earlier or later than predicted from a constant orbital period (e.g. Steffen et al.

2013; Perryman 2018, Ch. 6).

The differences between the observed and predicted transit times form a pattern that depends on the planetary masses and orbital configuration. By modelling these gravitational interactions, TTV analyses can constrain planet-to-star mass ratios and, together with the stellar mass, determine planetary masses.

TTVs can also reveal additional planets that do not transit. A non-transiting planet can gravitationally perturb a transiting neighbour and thereby leave a detectable timing signature. This method is therefore based on gravitational interactions between planets rather than on the motion of the host star, as measured with RVs.

TTV-derived masses do not carry the same $\sin i$ projection degeneracy as RV minimum masses. They are nevertheless model dependent: the inferred masses can depend on the assumed orbital configuration, eccentricities, and the number of interacting planets included in the model. In some systems, different combinations of planetary mass and orbital eccentricity can produce similar timing variations, which can make the individual parameters difficult to determine uniquely. Sufficient time coverage is also required to sample the timing pattern.

TTV signals are often larger when neighbouring planets have orbital period ratios close to simple integer ratios, such as 2:1 or 3:2, because the gravitational effects can build up over many orbits. A period ratio close to such a value does not by itself demonstrate that the planets are dynamically resonant. Mean-motion resonances and resonant chains are discussed further in Sect. 5.3.

The TTV method is particularly important for closely spaced planets in multi-planetary systems discovered by *Kepler*, many of whose host stars are too faint for precise RV measurements. TRAPPIST-1 provides a well-known example in which the planetary masses are constrained through TTVs (e.g. Agol et al. 2021). Consequently, TTV masses represent a different subset of planetary systems than RV masses, which is important when samples containing masses measured with both methods are analysed.

## 3.5 Other detection methods

The three detection techniques discussed above are the most relevant to the planetary radii, masses, and multi-planetary systems analysed later in this thesis. Astrometry, gravitational microlensing, and direct imaging nevertheless provide important complementary detections because they are sensitive to different orbital separations and planet types. Their main principles and selection effects are summarised below.

### 3.5.1 Astrometry

Astrometry detects planets by measuring small changes in the position of the host star on the sky. In addition to the star's proper motion and annual parallax, an orbiting planet causes the star to revolve around the system barycentre (Malbet & Sozzetti 2018; Perryman 2018, Ch. 3). The planetary signal is larger for more massive planets, wider orbits, and nearby systems, although very long orbital periods can be difficult to measure during a finite observing programme. Astrometry also constrains the orbital inclination and can therefore determine the true planetary mass rather than the projected quantity $M_{\mathrm{p}} \sin i$. It is consequently complementary to RV measurements, particularly for massive planets on relatively wide orbits.

### 3.5.2 Gravitational microlensing

Gravitational microlensing occurs when a foreground star passes close to the line of sight to a more distant background star, and its gravity magnifies the background light. A planet orbiting the foreground star can produce an additional perturbation in this magnification signal (Rektsini & Batista 2024; Perryman 2018, Ch. 5). This method is particularly sensitive to planets at projected separations of a few astronomical units and can detect systems at much larger distances than typical transit and RV surveys.

Microlensing events are transient and generally non-repeating, thereby rendering detailed physical characterisation of individual planets more difficult. The resulting planet samples therefore have selection effects that differ substantially from those of transit and RV surveys.

### 3.5.3 Direct imaging

Direct imaging detects light from the planet itself, spatially separated from the much brighter host star. The detected light can arise from thermal emission, particularly at infrared wavelengths, or from reflected stellar light (Claudi 2016; Perryman 2018, Ch. 7). The main challenges are the large brightness contrast between the star and planet as well as the small angular separation between them. This detection method is therefore most sensitive to young, massive giant planets on wide orbits, which are both comparatively bright and well separated from their host stars on the sky.

The HR 8799 system provides an important example and contains four directly imaged giant planets on wide orbits (Marois et al. 2008, 2010). Direct imaging therefore probes a region of parameter space that is poorly sampled by transit surveys. However, the resulting planetary sample is subject to strong selection effects and is not representative of the overall exoplanet population.

## 3.6 Comparing detection methods and observational biases

The six detection techniques described in this chapter do not sample the underlying exoplanet population uniformly. Their sensitivities depend on planetary mass and radius, orbital period and geometry, host-star properties, distance, observing strategy, and measurement precision. The observed sample is therefore shaped by both the underlying exoplanet population and the selection effects of the surveys used to detect and characterise the planets.

A useful concept is the *selection function*: the probability that an object with specified properties enters the observed sample. For a transit survey, for example, this probability includes both the geometric probability that the planet transits and the probability that the resulting signal is detected by the survey pipeline. Different missions and detection methods have different selection functions. A sample constructed from several surveys therefore generally has a more complicated selection function than any one survey alone.

| Detection method | Measured quantity | Main planetary information obtained | Particularly sensitive to |
|---|---|---|---|
| **Radial velocity** | Doppler shifts of host-star absorption lines | $M_p \sin i$, $P$, and $e$ | More massive and shorter-period planets around stars suitable for precise spectroscopy |
| **Transit photometry** | Periodic decrease in host-star flux | $R_p/R_\star$, $P$, and transit geometry | Larger and shorter-period planets around smaller stars, and with orbital orientations that produce transits |
| **Transit-timing variations** | Changes in the times of successive transits | Planet-to-star mass ratio and gravitational interactions | Interacting planets in multi-planetary systems, particularly neighbouring planets with period ratios close to integer ratios |
| **Astrometry** | Motion of the host star across the sky | Planetary mass and orbital geometry | Massive planets on relatively wide orbits around nearby stars |
| **Gravitational microlensing** | Gravitational magnification of a background star | Planet-to-star mass ratio and projected separation | Planets at projected separations of a few au, including systems at large Galactic distances |
| **Direct imaging** | Planetary light separated from the host star | Planetary brightness or spectrum and projected orbital separation | Young, luminous giant planets on wide orbits |

**Table 3.2:** Summary of the six principal exoplanet detection methods discussed in this chapter. The sensitivities are qualitative and do not represent sharp detection boundaries.

Table 3.2 summarises the quantities measured by the six principal detection techniques and the types of planets to which they are particularly sensitive. The entries describe broad trends rather than strict limits, and the actual sensitivity also depends on the properties of the host star, the instrument, and the observing strategy.

This distinction is central to population- and system-level studies. An apparent excess or deficit of planets within some range of planetary masses, radii, or orbital periods can represent a genuine astrophysical feature, but it can also arise because those planets are easier or more difficult to detect. Analogously, a system that appears to contain only one planet may in reality contain several additional planets below the detection threshold or on non-transiting orbits.

### 3.6.1 Sensitivity to planetary properties and orbital separations

Transit photometry is primarily sensitive to planetary size through Eq. 3.4. Its geometric probability decreases with orbital separation according to Eq. 3.5, and short-period planets also produce a larger number of repeated transits during a finite observing interval. Transit samples are therefore strongly weighted towards short-period planets and, for a given host star, towards larger planets.

The RV signal depends directly on planetary mass through Eq. 3.1. Massive and short-period planets induce the largest amplitudes. RV surveys therefore provide a large fraction of the known planetary masses, although this sample is biased towards planets that produce sufficiently strong signals around stars suitable for precise spectroscopy.

Transit and RV surveys therefore both preferentially detect planets that produce large signals, but their efficiencies for detecting large samples of small planets differ substantially. Transit surveys can monitor large numbers of stars simultaneously, whereas high-precision RV surveys generally require repeated spectroscopic observations of individual targets. This difference, together with the high photometric precision achievable from space, has enabled transit surveys to build much larger samples of small, generally low-mass planets. *Kepler*, for example, revealed large numbers of super-Earths and sub-Neptunes that are much less well represented among RV discoveries.

The two methods also differ when the orbital period approaches or exceeds the observing baseline. A transit can only be detected if at least one transit occurs during the observations, and a single transit does not normally provide a direct measurement of the orbital period. RV observations, in contrast, can reveal a long-period companion through a long-term trend or curvature even when only a fraction of its orbit has been observed, although its orbital parameters may remain poorly constrained.

TTVs are sensitive to the gravitational effects that planets exert on one another rather than directly to planetary radius or to the motion of the host star. Their detectability depends strongly on the architecture of the system and is often enhanced in systems with closely spaced planets whose period ratios are near simple integer ratios. TTVs therefore often probe a different subset of planets than the RV method, even though both techniques ultimately provide a value of $M_{\mathrm{p}}$.

Astrometric sensitivity increases with planetary mass and orbital separation, while direct imaging favours young, luminous giant planets on wide separations. Microlensing is sensitive to planet-to-star mass ratio and lensing geometry, commonly probing planets at projected separations of a few astronomical units. These methods extend the observed planet population far beyond the short-period region where transit detections predominate.

### 3.6.2 Dependence on host-star properties and distance

Host-star properties affect every major detection method. For instance, for a planet of fixed radius, the transit depth is larger around a smaller star, as shown in Fig. 3.5, while for a planet of fixed mass and orbital period, the RV amplitude is larger around a lower-mass star. At the same time, high-precision RV spectroscopy requires sufficient stellar photon counts and well-resolved, relatively narrow stellar absorption lines. Consequently, very faint stars, rapid rotators, and magnetically active stars are generally less amenable to precise RV measurements.

The distance to the system introduces another selection effect. The received flux decreases with the inverse square of distance, making precise photometric and spectroscopic measurements more difficult for faint, distant stars. The angular astrometric signal decreases with distance as well, and direct imaging becomes more difficult at large distances because both the planetary flux and the angular planet–star separation become less favourable. Transit depth itself does not depend on distance, but the photometric precision obtainable for the host star usually decreases as the target becomes fainter. Microlensing differs from these methods because it can access planetary systems at much larger Galactic distances.

These dependencies imply that different stellar populations are sampled with different efficiencies. Observed distinctions between planets orbiting, for example, FGK stars and M dwarfs must therefore be interpreted together with the different detectability of planets around those stars.

### 3.6.3 Combining data from different surveys

Exoplanet samples used in population studies often combine planets detected by several surveys. These missions may differ in target selection, observing baseline, wavelength range, photometric or spectroscopic precision, detection algorithms, and the methods used to derive stellar parameters. A sample assembled from various surveys therefore contains planets that were selected and characterised in different ways, even if they satisfy the same final selection criteria.

As discussed in Sect. 3.3.3, *Kepler*, K2, TESS, and CHEOPS differ in observing strategy, target selection, and observing baseline. Samples combining planets characterised by these missions therefore inherit different observational biases.

A further selection is introduced when a population study requires a measured planetary property. A radius sample is dominated by transiting planets because transit photometry directly provides $R_{\mathrm{p}}/R_{\star}$. A mass sample requires successful RV, TTV, astrometric, or other dynamical measurements and is therefore more selective. A sample requiring both mass and radius is more restricted, and such well-characterised planets need not be representative of the observed planet population.

The source of the mass measurement may also be important. RVs and TTVs have different observational requirements and therefore different selection effects. Combining them increases the available sample, but the resulting sample contains planets affected by different observational biases. This distinction should be kept in mind when interpreting population-level trends (e.g. Leleu et al. 2024).

### 3.6.4 Consequences for observed multi-planetary systems

Observational biases have additional consequences for studies of planetary system architectures. The observed planet multiplicity is the number of detected planets in a system, whereas the intrinsic multiplicity denotes the true number of planets residing in the system. Several factors can lead to discrepancies between the observed and intrinsic planet multiplicities, including non-transiting planets, planets with radii or orbital periods that render them difficult to detect, and planets whose RV or TTV signals fall below the available measurement precision.

In addition to underestimating the intrinsic planet multiplicity, undetected planets in a system affect the apparent adjacency. Two detected planets are usually called adjacent when no other detected planet lies between them in orbital period. If an intermediate planet is missed, two observed neighbouring planets are not intrinsically adjacent. Their period ratio spans two or more true interplanetary spacings and can therefore be much larger than the individual spacings in the underlying system. Missing planets can consequently alter the observed distributions of orbital spacings, estimates of spacing regularity, and correlations between spacing and planetary mass or radius.

Transit geometry introduces a particularly important multiplicity bias. In a multi-planetary system, mutual orbital inclinations can cause only a subset of planets to transit from a given viewing direction. A system in which only one planet is observed to transit is therefore not necessarily an intrinsically single-planet system. This distinction is central when observed singles and multis are compared (e.g. Weiss et al. 2018c; Muresan et al. 2026).

The absence of a detected planet of a particular type should likewise not be equated automatically with physical absence. A system classified as having no detected gas giant planet may still contain an undiscovered long-period giant. These considerations are particularly relevant when classifying planetary systems based on the presence or absence of a giant planet (e.g. paper C).

The probability of identifying additional planets in a system can also depend on the survey's automated detection procedure. Particularly, the detection efficiency of the *Kepler* pipeline decreases for additional planets in a light curve that already contains one or more transiting planets (Zink et al. 2019). Population-level analyses of single- and multi-planetary systems therefore require accounting for both the transit geometry and detection completeness.

### 3.6.5 Combining complementary measurements

Although the various detection methods introduce different biases, their complementarity is one of the main strengths of exoplanet observations. Transit photometry yields the planet-to-star radius ratio and constrains the orbital geometry. RV observations provide $M_\mathrm{p} \sin i$, and the transit inclination allows the true mass to be obtained. Combining planetary mass and radius further yields the mean bulk density, which constitutes an important observational constraint on planetary interior composition, as discussed in Ch. 4.

TTVs provide an alternative route to planetary masses in interacting multi-planetary systems, especially for faint host stars which are difficult to observe with RVs. Astrometry can determine orbital inclinations and true masses for non-transiting RV planets. Direct imaging and microlensing extend detections to wider orbital separations. A more complete picture of planetary systems therefore requires complementary information from several methods rather than treating any one technique as a complete census.

Figures 3.6 and 3.7 illustrate the capability of the detection methods to populate different regions of the observed period–radius and period–mass planes, respectively. Transit discoveries dominate the sample of planets with measured radii at relatively short periods. RV discoveries make a substantial contribution to the sample of planets with measured masses, particularly at long orbital periods and high masses. The colours indicate the discovery method rather than any additional technique that may have been employed for subsequent detection. For example, a planet discovered through transit photometry may subsequently have its mass measured using RV or TTV observations.

The complementary sensitivities illustrated in these figures reveal that the observed exoplanet sample cannot be interpreted as an unbiased representation of the underlying population of planetary systems. The following chapter considers the physical properties and demographics of the observed planets, while the consequences of incomplete detection for multi-planetary system architectures are discussed further in Ch. 5.

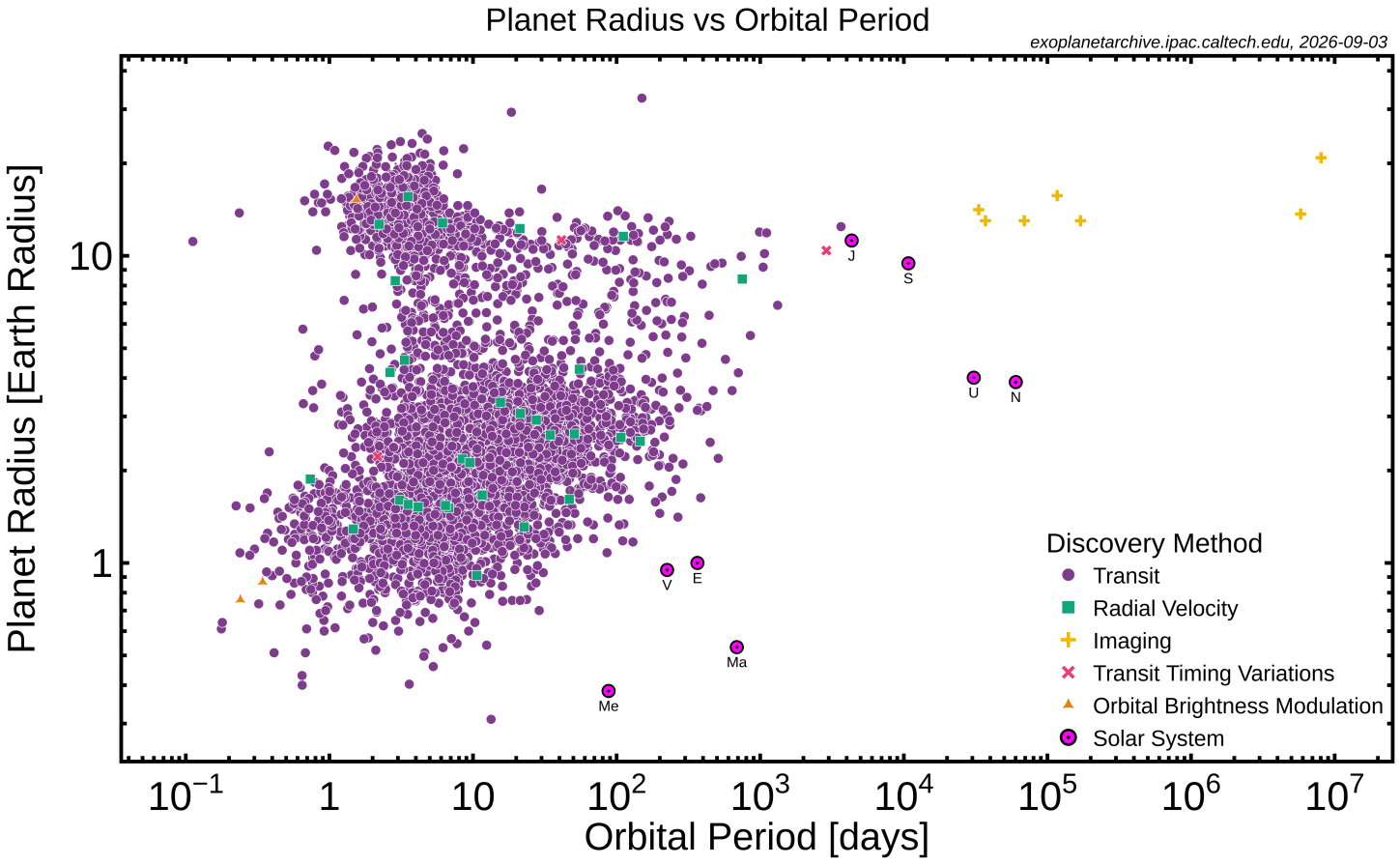

**Figure 3.6:** Planetary radius in units of Earth radii as a function of orbital period for confirmed planets with measured radii, including the Solar System planets. The points are colour-coded according to the discovery method. Transit photometry is the predominant method in this sample, particularly at relatively short orbital periods. Data from the NASA Exoplanet Archive, as of 2026-09-03.

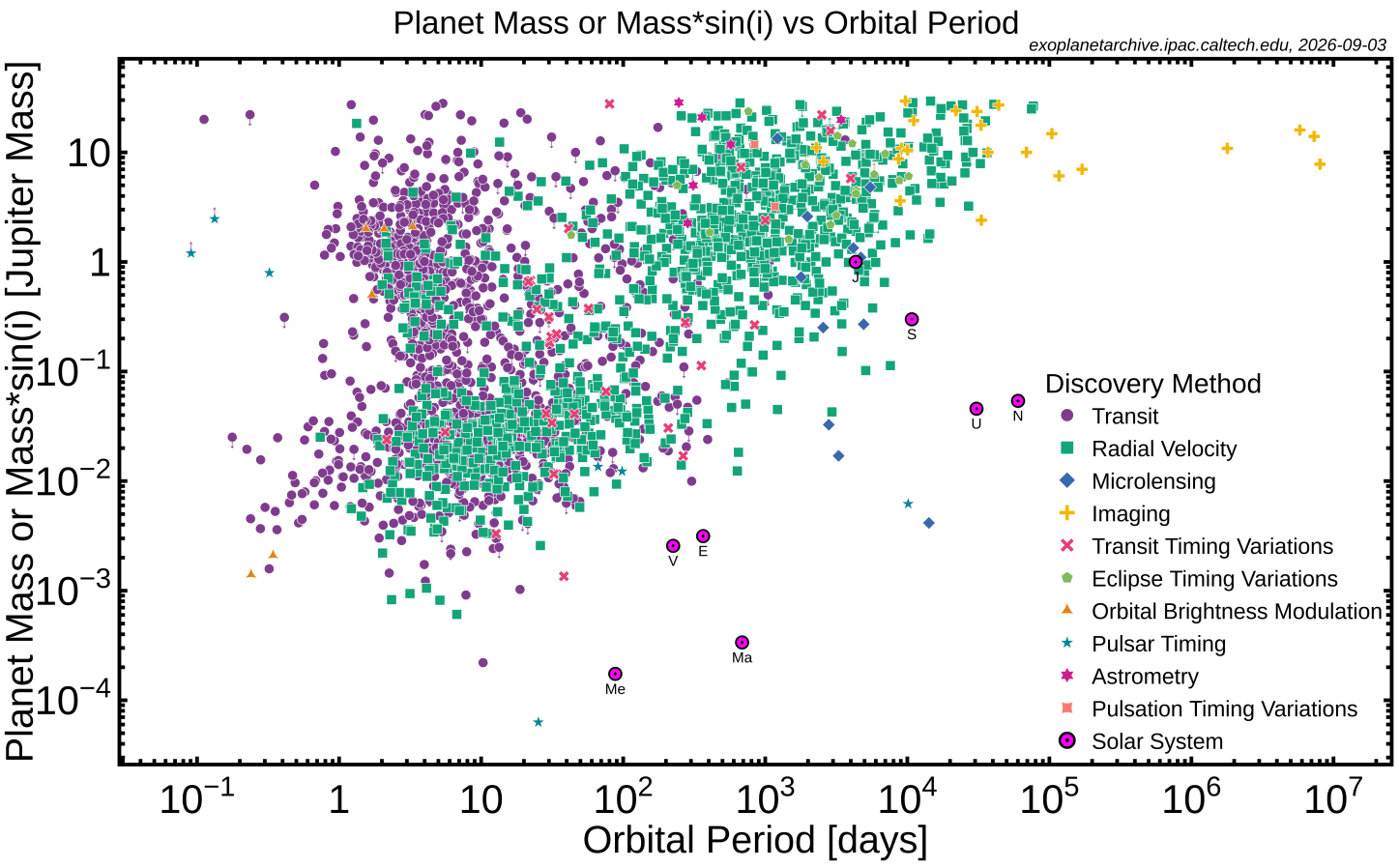

**Figure 3.7:** Planetary mass $M_\mathrm{p}$, or minimum mass $M_\mathrm{p} \sin i$, in units of Jupiter masses as a function of orbital period for confirmed planets with mass measurements, including the Solar System planets. The colour of each planet indicates the discovery method, which is not necessarily the technique subsequently used to measure the planetary mass. RV discoveries make a substantial contribution at longer orbital periods and higher planetary masses. Data from the NASA Exoplanet Archive, as of 2026-09-03.

CHAPTER 4

# Exoplanet properties and demographics

This chapter describes the main orbital and physical properties and the demographics of the observed planet population. Exoplanet demographics comprises the statistical distributions and occurrence of planets as functions of different parameters, such as planetary radius, mass, orbital period, and host-star properties. These distributions provide observational constraints, for instance, on planet formation and evolution, although their interpretation requires consideration of the observational selection effects presented in Ch. 3. This chapter first introduces commonly adopted planet types and mass–radius relations, followed by the inference of planetary interior structures from observed bulk properties. Finally, the distributions of planets, their dependence on host-star properties, and the statistical methods used to analyse them are discussed.

## 4.1 Planet types

One of the most notable discoveries in exoplanet research is the remarkable diversity of exoplanets and planetary systems (e.g. Winn & Fabrycky 2015; Zhu & Dong 2021; Biazzo et al. 2022). The observed pop-ulation spans a broad range of planetary masses, radii, temperatures, orbital periods, and system configurations. Exoplanets can therefore be classified in several complementary ways, including by the stellar environment in which they reside, the number of detected planets in the system, as well as their physical, orbital, and irradiation properties.

Notably, although more than 50% of the stars in the Solar neighbourhood are estimated to reside in binary- or higher-order multiple-star systems (Duchêne & Kraus 2013; Reylé et al. 2021), approximately 91% of all the confirmed exoplanets have been found in orbits around single stars. The remaining 8% and 1% of the planets have been detected in systems with two and three stars, respectively, while only one known exoplanet resides in a quadruple-star system[1]. In binary-star systems, planets are commonly classified as circumstellar, or S-type, when they orbit one of the stars, and circumbinary, or P-type, when they orbit both stars (Perryman 2018).

Detecting and characterising planets in multiple-star systems may, however, be subject to observational biases. First, current exoplanet detection techniques are predominantly sensitive to planets orbiting single stars, potentially resulting in an incomplete census of planets in multiple-star systems. Second, stellar companions are often difficult to detect, particularly when they have very long orbital periods and/or substantially smaller masses than the primary stars in their respective systems.

A further classification of exoplanets distinguishes between those residing in single- and multi-planetary systems, and these planets are denoted throughout this thesis as singles and multis, respectively. This terminology refers to the observed planet multiplicity in the system and does not necessarily correspond to the intrinsic multiplicity, as discussed in Sect. 5.2.

Most importantly, planets are classified according to their observed physical properties, particularly radius and mass (e.g. Borucki et al. 2011; Winn & Petigura 2024). However, a key consideration is that a planet's radius or mass alone does not uniquely determine its interior composition. The following radius ranges, dividing exoplanets into different types, provide an approximate descriptive classification, with the terminology and adopted boundaries varying between studies.

[1] Based on data from the NASA Exoplanet Archive accessed on 2026-09-03.

- Sub-Earths and Earth-sized planets: approximately $R_p < 1.1\,R_\oplus$. Planets in this size range are generally expected to be predominantly rocky, although their relative fractions of iron and silicate material can vary considerably.

- Super-Earths: approximately $R_p \simeq 1.1 - 1.8\,R_\oplus$. This term refers primarily to planetary size rather than to similarity with Earth in composition, atmosphere, surface conditions, or habitability.

- Sub-Neptunes: approximately $R_p \simeq 1.8{-}3.5\,R_\oplus$. Their larger radii generally require lower-density material in addition to a rocky interior, such as water-rich material or a gaseous envelope. Different combinations of these components can produce similar planetary radii.

- Neptune-sized planets: approximately $R_p \simeq 3.5 - 6\,R_\oplus$. These planets generally possess substantial volatile-rich or gaseous envelopes.

- Gas giants: approximately $R_p \gtrsim 6\,R_\oplus$. They possess thick H/He-rich envelopes surrounding interiors enriched in heavier elements.

These radius intervals provide a useful descriptive classification but do not constitute unique physical boundaries between planet types. The approximate planet classes are also apparent in the period–radius and period–mass distributions shown in Figs. 3.6 and 3.7, respectively. However, planets of similar size can have substantially different masses and compositions, as further discussed in Sect. 4.3.

Planets are also commonly described as hot, warm or temperate, or cold according to the stellar irradiation they receive, although the boundaries between these categories are not uniquely defined. They may additionally be classified according to their orbital periods. For instance, ultra-short-period planets have orbital periods of approximately one day or less and are predominantly small planets (e.g. Sanchis-Ojeda et al. 2014), while hot Jupiters are gas giants on short-period orbits, commonly defined as having $P < 10$ days (e.g. Dawson & Johnson 2018).

A key concept related to stellar irradiation is the habitable zone, conventionally defined as the range of orbital distances from a host star where a rocky planet with a suitable atmosphere could maintain liquid water on its surface (e.g. Zsom et al. 2013; Kopparapu et al. 2013; Kane 2021). A planet located in the habitable zone is not necessarily habitable, since the surface conditions also depend on properties such as the planetary mass, atmospheric composition and pressure, orbital properties, and the characteristics and evolution of the host star. The habitable zone should therefore be regarded as a first-order constraint on the conditions that may allow surface liquid water rather than as a classification of an Earth-like planet.

## 4.2 Mass–radius relations

The approximate planet classes introduced in Sect. 4.1 provide a useful description of the observed population, but planetary mass or radius alone does not uniquely determine the composition. Considering mass and radius together provides substantially more information about a planet's bulk properties.

The mean bulk density of a planet is given by:

$$\rho_{\rm p} = \rho_{\oplus} \frac{M_{\rm p}}{R_{\rm p}^3} \ , \qquad (4.1)$$

where $M_{\rm p}$ and $R_{\rm p}$ are the planetary mass and radius, respectively, expressed in Earth units, while $\rho_{\oplus}$ denotes the bulk density of Earth where $\rho_{\oplus} = 5.514\,{\rm g\,cm^{-3}}$ (Prša et al. 2016). Equation 4.1 indicates that planets with the same mass can have very different radii and densities, while planets of similar radius can possess substantially different masses. Since $\rho_{\rm p} \propto M_{\rm p} R_{\rm p}^{-3}$, uncertainties in both mass and radius propagate into the bulk density. To first order, the fractional uncertainty in radius enters the propagated density uncertainty with a factor of three. Therefore, a well-constrained planetary radius is particularly important for obtaining a precise bulk density.

A comparatively small amount of low-density material can have a large influence on planetary radius. For example, a rocky planet surrounded by a low-mass gaseous envelope can have a substantially larger radius than a bare rocky planet of similar total mass. Consequently, the fractional contribution of an outer envelope to a planet's radius generally

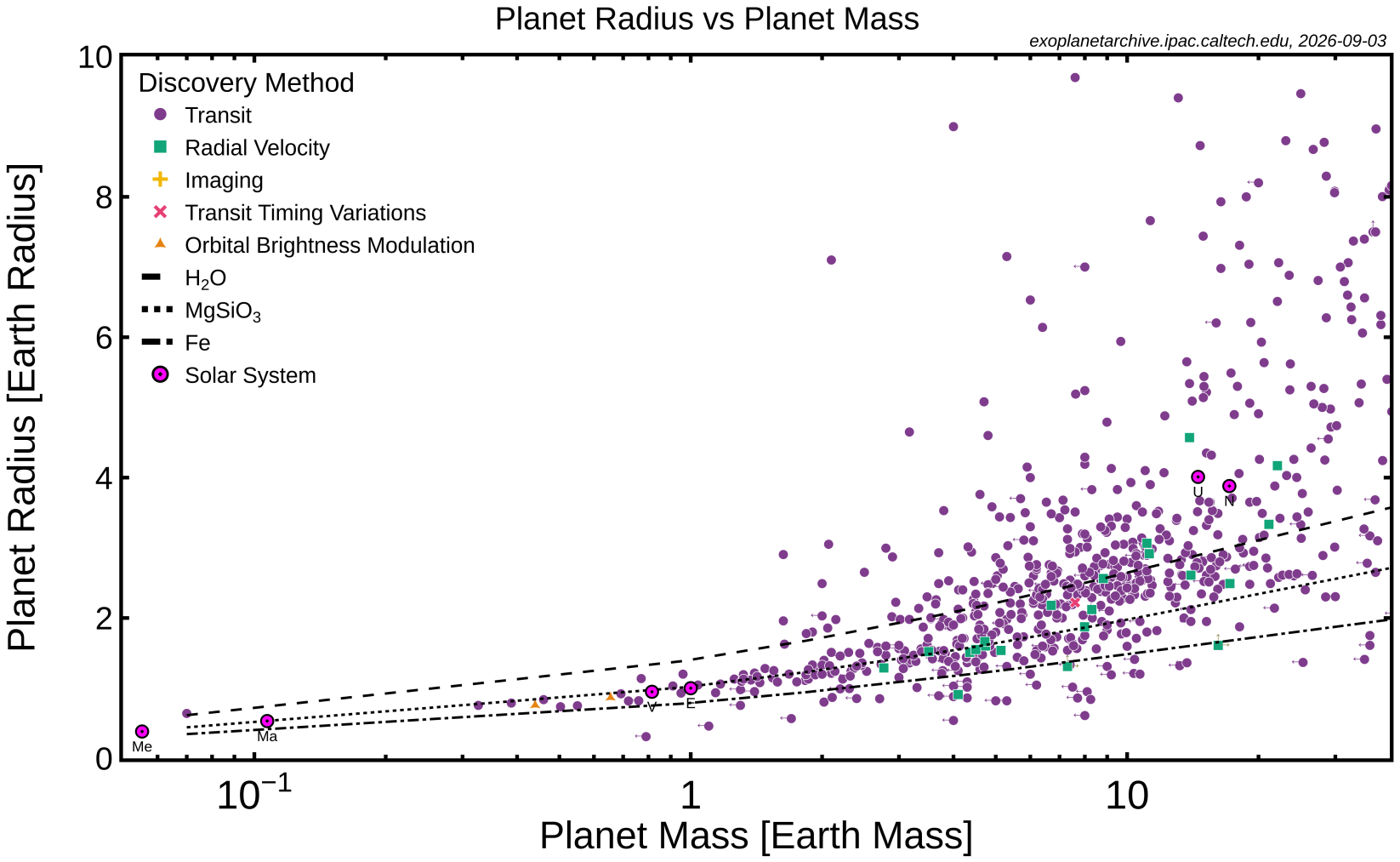

**Figure 4.1:** Radius as a function of mass for confirmed exoplanets with measured radii and masses, including the four terrestrial planets of the Solar System. The data are colour-coded by discovery method and shown for planetary masses up to $40\,M_\oplus$ and radii up to $10\,R_\oplus$. The black curves indicate theoretical mass–radius relations for idealised $H_2O$, $MgSiO_3$, and Fe compositions from Zeng et al. (2016). Figure from the NASA Exoplanet Archive, as of 2026-09-03.

does not correspond to a comparable fraction of the total planetary mass.

Figure 4.1 shows the radius as a function of mass for a subset of confirmed exoplanets with measured radii and masses, illustrating the large diversity in planetary bulk properties. For a given mass, a comparatively small radius and high bulk density favour a larger contribution from dense materials such as iron and silicates. A larger radius and lower bulk density indicate a greater contribution from lower-density material.

The mass–radius plane can therefore exclude some compositions without uniquely identifying the correct one. Even a well-measured combination of $M_\mathrm{p}$ and $R_\mathrm{p}$ can be reproduced by different combinations of an iron-rich core, silicate mantle, water-rich material, and a gaseous envelope. This compositional degeneracy becomes particularly important for sub-Neptunes, for which different combinations of rocky material, water-rich material, and H/He can reproduce similar bulk properties (e.g. Zeng et al. 2019; Acuña et al. 2024).

Mass–radius relations can broadly be divided into theoretical and empirical relations. A theoretical relation predicts the radius expected for a planet of a given mass and assumed composition. Such relations are calculated using interior-structure models and equations of state, which describe how materials such as iron, silicates, and water behave under the high pressures and temperatures inside planets. Material compression becomes increasingly important at high pressure, and realistic planets do not follow simple constant-density relations over large mass ranges.

On the other hand, an empirical mass–radius relation describes the statistical relationship between the measured masses and radii of the observed planet population. It can, for example, be used to estimate the typical mass associated with a measured radius, or vice versa, when a direct measurement of one quantity is unavailable. Obtaining precise planetary masses is generally more observationally demanding than measuring radii for transiting planets, and many planets with measured radii therefore lack precise mass measurements.

Since empirical mass–radius relations incorporate both measurement uncertainties and the intrinsic diversity of planetary properties, they should not be interpreted as equivalent to direct measurements. Several such relations have been derived using planets with measurements of both mass and radius, with several examples shown in Fig. 4.2 (e.g. Weiss et al. 2013; Chen & Kipping 2017; Bashi et al. 2017; Otegi et al. 2020a; Müller et al. 2024).

The scatter around the empirical mass–radius relations presented in Fig. 4.2 reflects the intrinsic diversity of planetary structures and compositions as well as the uncertainties in the measured masses and radii. Planets of similar size do not necessarily have the same mass or interior composition. Empirical mass–radius relations therefore provide useful population-level estimates, but physical interior models are required when investigating the possible interior compositions of individual planets.

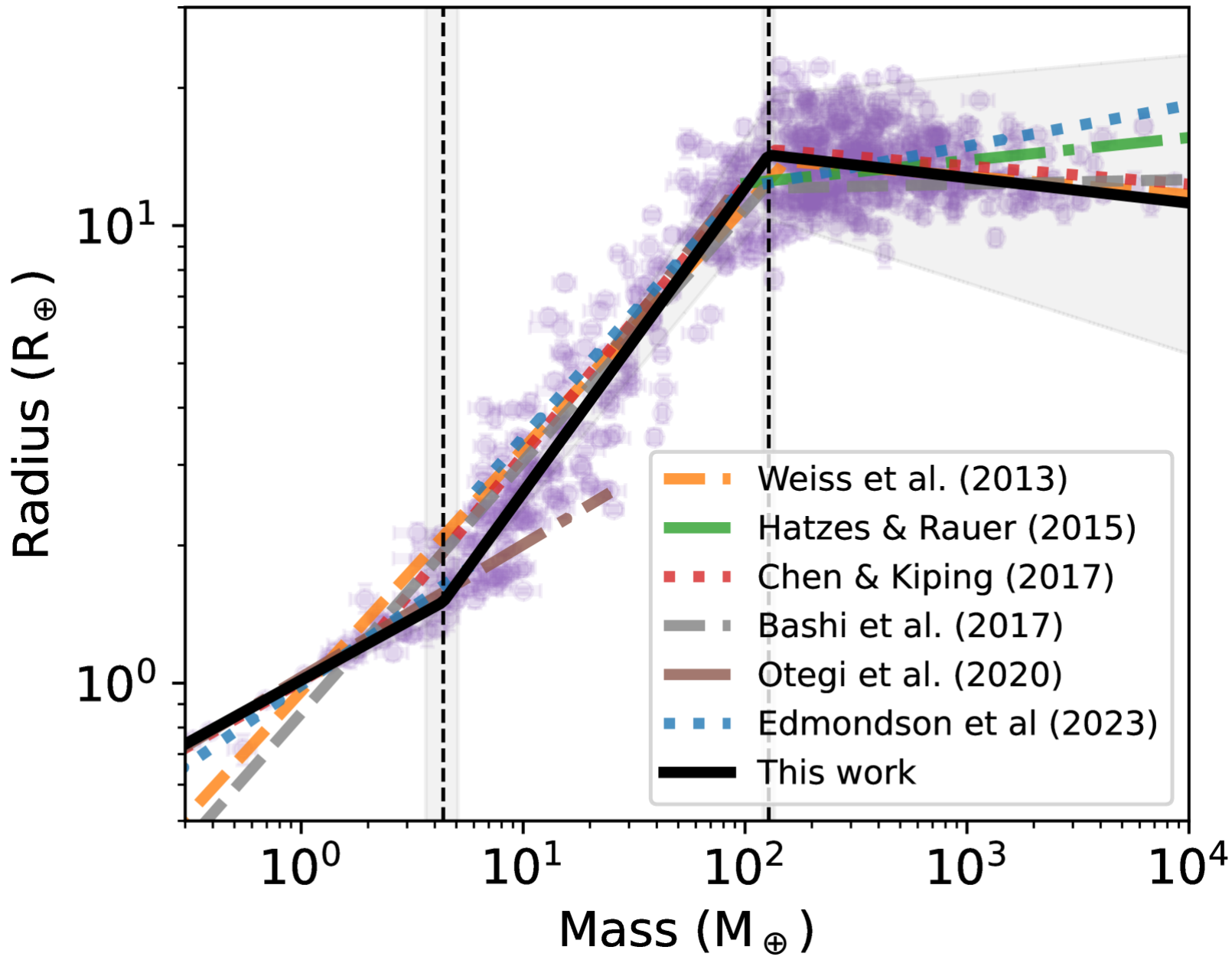

**Figure 4.2:** Radius versus mass for confirmed exoplanets with measurements of both quantities, together with mass–radius relations from the studies indicated in the figure legend. The black solid line represents the relations derived by Müller et al. (2024), who divided the population into small, intermediate, and giant planets, separated by the vertical dashed lines. The planetary sample shown in the figure contains measurements available as of July 2023. Figure from Müller et al. (2024).

## 4.3 Interior structures and compositions of exoplanets

To translate the bulk properties discussed in Sect. 4.2 into constraints on planetary composition, models of planetary interiors are required. A planet's interior structure is shaped by its bulk composition, formation history, differentiation, and subsequent thermal and dynamical evolution (e.g. Baumeister et al. 2025).

For a specified planetary mass and composition, an interior model calculates how pressure, density, and enclosed mass change from the centre towards the outer layers of the planet. Equations of state describe how different materials behave under the relevant pressures and temperatures. Together with the equations of hydrostatic equilibrium and mass conservation, they allow the planetary radius corresponding to a particular interior structure to be calculated (e.g. Zeng et al. 2019; Aguichine et al. 2021). Repeating these calculations for different masses and compositions produces theoretical mass–radius relations such as those displayed in Fig. 4.2.

For rocky and moderately volatile-rich planets, a common simplified representation consists of chemically differentiated concentric layers. These may include an iron-rich core surrounded by a silicate mantle and, for volatile-rich planets, an outer water-rich layer. Planets can additionally possess gaseous envelopes, including H/He-rich envelopes. Giant planets require different interior descriptions because large fractions of their masses and volumes consist of hydrogen and helium surrounding interiors of heavier elements (e.g. Müller & Helled 2023).

A three-layer model consisting of an iron-rich core, silicate mantle, and water-rich outer layer can be described using the core mass fraction (CMF), mantle mass fraction (MMF), and water mass fraction (WMF):

$$\mathrm{CMF} = \frac{M_{\mathrm{core}}}{M_{\mathrm{p}}} \,, \qquad \mathrm{MMF} = \frac{M_{\mathrm{mantle}}}{M_{\mathrm{p}}} \,, \qquad \mathrm{WMF} = \frac{M_{\mathrm{water}}}{M_{\mathrm{p}}} \,,$$

where $M_{\mathrm{p}}$ is the total planetary mass and $M_{\mathrm{core}}$, $M_{\mathrm{mantle}}$, and $M_{\mathrm{water}}$ are the masses assigned to the three respective layers. Within this adopted model,

$$\mathrm{CMF} + \mathrm{MMF} + \mathrm{WMF} = 1 \,.$$

Increasing the CMF generally increases the planetary mean density because iron-rich material is denser than silicates or water-rich material. Conversely, increasing the fraction of lower-density material generally produces a larger radius for a given mass. These relationships are non-linear because the different materials are compressed to different degrees under planetary interior conditions.

The interpretation of a water-rich layer requires particular care. The WMF represents the total amount of material assigned to water within the adopted model. At the pressures and temperatures encountered inside many planets, this material need not resemble a liquid surface ocean. Water can instead occur as vapour, supercritical water, or high-pressure phases.

There is also an important degeneracy between water-rich material and H/He. Since hydrogen and helium have very low densities, an H/He envelope containing only a small fraction of the total planetary mass can still produce a substantial increase in radius. A model without a separate H/He layer may consequently reproduce part of this radius increase by assigning a larger fraction of the planetary mass to a water-rich layer. An inferred WMF should therefore be interpreted within the assumptions of the adopted model rather than as a direct measurement of a planet's actual water inventory.

The prevalence of genuinely water-rich planets within the observed population remains uncertain, partly because bulk mass and radius alone cannot uniquely distinguish water-rich interiors from planets containing low-mass H/He envelopes (Luque & Pallé 2022; Rogers et al. 2023; Parviainen et al. 2024).

Conclusively, different combinations of interior mass fractions may reproduce similar planetary masses and radii, resulting in compositional degeneracies that cannot generally be resolved from mass–radius measurements alone.

### 4.3.1 Stellar irradiation and equilibrium temperature

Stellar irradiation can affect the thermal state and radius of a planet, particularly when the planet contains volatile-rich outer layers or a gaseous envelope. The instellation denotes the stellar irradiance incident on the planet. Within the Solar System, the corresponding solar irradiation is commonly referred to as insolation.

For a circular orbit, the instellation $F_{\star,\mathrm{in}}$ can be expressed relative to the solar flux received by Earth as follows:

$$\frac{F_{\star,\mathrm{in}}}{F_{\odot,\mathrm{in}}} = \left(\frac{R_\star}{R_\odot}\right)^2 \left(\frac{T_\mathrm{eff}}{T_\odot}\right)^4 \left(\frac{1\ \mathrm{au}}{a}\right)^2 , \tag{4.2}$$

where $R_\star$ and $T_\mathrm{eff}$ denote the stellar radius and effective temperature, respectively, $a$ is the planetary orbital semi-major axis, and $F_{\odot,\mathrm{in}}$ denotes the solar flux received at Earth's orbital distance, with $F_{\odot,\mathrm{in}} = 1361\ \mathrm{W\,m^{-2}}$ (Del Genio et al. 2019).

Closely related to the instellation is the planetary equilibrium temperature $T_\mathrm{eq}$. It is a simplified estimate of the planetary temperature obtained by balancing the stellar radiation absorbed by the planet with the thermal radiation it emits, considering no additional sources of external or internal heating. For a circular orbit and complete redistribution of the absorbed stellar energy over the planet, $T_\mathrm{eq}$ can be expressed as follows:

$$T_\mathrm{eq} = T_\mathrm{eff}\ (1 - A_\mathrm{B})^{1/4} \left(\frac{R_\star}{2a}\right)^{1/2} , \tag{4.3}$$

where the Bond albedo $A_\mathrm{B}$ represents the fraction of the total incident stellar energy that is reflected by the planet. Since the Bond albedo is generally not directly measured for exoplanets, calculations of $T_\mathrm{eq}$ commonly require an assumed value. This is often predefined as $A_\mathrm{B} = 0.3$, similar to Earth's Bond albedo, or $A_\mathrm{B} = 0$ (Del Genio et al. 2019).

The equilibrium temperature should not be interpreted as the actual temperature at a particular location on a planet. An airless planet can exhibit large temperature differences between its illuminated and dark hemispheres. If an atmosphere is present, atmospheric heat transport and greenhouse effects can further modify the atmospheric and surface temperatures. Additionally, internal heat can potentially contribute to the planet's thermal state.

The irradiation level is particularly important for planets with volatile-rich outer layers or gaseous envelopes because temperature affects their density and radial extent. Two planets with identical masses and bulk compositions can therefore have different radii if their thermal states differ. Irradiation and equilibrium temperature must consequently be considered when theoretical mass–radius relations and interior models include temperature-sensitive outer layers.

### 4.3.2 Bayesian inference of planetary interior compositions

The relationship between a planet's interior composition and its observable properties can be investigated using forward and inverse models. Forward models begin with assumed physical parameters, for instance the planetary mass and composition, and predict observable quantities such as the planetary radius (e.g. Lopez & Fortney 2014; Zeng et al. 2019; Aguichine et al. 2021). In contrast, inverse models begin with measured properties and determine which combinations of model parameters are consistent with those observations (e.g. Dorn et al. 2015; Baumeister & Tosi 2023; Acuña et al. 2024).

Because several different interior structures can reproduce the same measured bulk properties, the inverse problem generally has no unique solution. Bayesian inference provides a way of quantifying the range of parameter values supported by the observations. Bayes' theorem can be written schematically as

$$p(\boldsymbol{\theta} \mid D) \propto p(D \mid \boldsymbol{\theta})\, p(\boldsymbol{\theta}) \ ,$$

where $D$ represents the observed data and $\boldsymbol{\theta}$ the model parameters. The likelihood, $p(D \mid \boldsymbol{\theta})$, describes how well the observables predicted by a particular model reproduce the measurements, while the prior, $p(\boldsymbol{\theta})$, contains the adopted information or assumptions about the parameters before the data are considered. Their combination gives the posterior probability distribution $p(\boldsymbol{\theta} \mid D)$.

A parameter combination whose predicted observables are close to the measured values will generally have a higher likelihood than one that predicts values far from the measurements. Evaluating many possible combinations of interior parameters therefore determines which regions of the parameter space are consistent with the observations.

Markov Chain Monte Carlo (MCMC) methods provide an efficient numerical technique for sampling posterior probability distributions. Rather than evaluating every possible combination of parameters in a multi-dimensional parameter space, an MCMC algorithm generates a sequence of parameter combinations whose distribution approaches the posterior probability distribution. The resulting samples can be used to determine posterior medians, credible intervals, and correlations between model parameters.

The posterior distributions can reveal compositional degeneracies, whereby different combinations of low- and high-density components can produce similar planetary masses or radii. For example, an increase in a low-density component may be compensated by an increase in denser material while maintaining a similar total planetary radius. Broad or strongly correlated posterior distributions therefore provide information about which aspects of a planetary interior are poorly constrained. When the observations provide only weak constraints on a parameter, the posterior distribution can also depend appreciably on the adopted prior.

The inferred compositions remain conditional on the adopted interior model, equations of state, layer structure, thermal assumptions, and priors. Posterior uncertainties therefore quantify the range allowed within the adopted model framework rather than all physically possible planetary interiors.

## 4.4 Distributions of planet types and properties

The preceding sections considered the physical properties of individual planets. With thousands of known exoplanets, these properties can also be examined statistically across planetary types. Such demographic studies investigate the occurrence rates of different planet types, the distributions of their properties, and the dependence of these distributions on orbital and stellar properties.

The detected exoplanet population is not a direct representation of the intrinsic planet population. As discussed in Ch. 3, the probability of detecting and characterising a planet depends on various properties, including the planet's radius, mass, orbital period, host star, and the observing strategy. Demographic studies must therefore account for these selection effects to reliably infer the underlying planetary population.

### 4.4.1 Planet occurrence rates

An important distinction is between how common a planet type is among detected planets, and how frequently that type of planet actually occurs around stars. The fraction among detected planets can be strongly affected by observational biases. A planet type that is easy to detect may therefore be overrepresented among the discovered planets even if it is intrinsically uncommon. Hot Jupiters, for example, are comparatively easy to detect, making them more prevalent in detected samples than their intrinsic occurrence around stars would imply.

Planet occurrence rates, however, attempt to infer the underlying frequency of planets after accounting for observational selection effects such as survey completeness and, for transit photometry, geometric transit probability. Depending on the study, an occurrence rate may be expressed as the average number of planets per star within a specified range of planetary and orbital properties or as the fraction of stars hosting at least one such planet. The planetary and stellar parameter ranges must therefore be specified when occurrence rates from different studies are compared.

One of the major results of exoplanet surveys is that planets are ubiquitous. Completeness-corrected studies indicate that the average number of planets per star is of order unity or higher when broad ranges of planetary sizes and orbital periods are considered (e.g. Winn & Fabrycky 2015; Zhu & Dong 2021). Stars hosting planets are therefore the rule rather than the exception in the Solar neighbourhood, and planets are expected to be common throughout the Galaxy.

The ubiquity of planets is evident even among nearby stars. Planets have been identified orbiting Proxima Centauri, the nearest star to the Sun and a member of the Alpha Centauri triple system (Anglada-Escudé et al. 2016; Suárez Mascareño et al. 2025), and also around Barnard's star, the nearest single star (González Hernández et al. 2024; Basant et al. 2025).

Occurrence-rate studies indicate that small planets are particularly common, as shown in Fig. 4.3: Super-Earths and sub-Neptunes with orbital periods shorter than approximately 100 days occur around a substantial fraction of stars, with inferred rates of several tens of per cent, depending on the adopted planet definitions and stellar sample (e.g.

Winn & Fabrycky 2015; Zhu & Dong 2021; Winn & Petigura 2024).

The occurrence rates of super-Earths and sub-Neptunes exhibit a strong dependence on orbital period, as indicated in Fig. 4.3. Both planet types are relatively uncommon at periods of approximately one day but become increasingly prevalent towards periods of about 10 days. At longer orbital periods, their occurrence rates remain approximately constant over the period range considered. Super-Earths are more common at orbital periods below 10 days, whereas sub-Neptunes become more prevalent beyond approximately 10 days.

In contrast, giant planets have a different occurrence distribution. Hot Jupiters are intrinsically rare despite being comparatively easy to detect, occurring around only approximately $0.5 - 1\,\%$ of Sun-like stars (e.g. Winn & Fabrycky 2015; Winn & Petigura 2024). The occurrence of giant planets increases towards longer orbital periods over the range probed by RV surveys, while microlensing and direct-imaging surveys extend demographic studies to even wider orbital separations (e.g. Winn & Fabrycky 2015; Zhu & Dong 2021).

Microlensing is particularly valuable for probing colder planets beyond the water snow line, which denotes the distance from the host star beyond which water can condense into ice. This region is difficult to probe with short-duration transit surveys. Microlensing observations, which predominantly probe relatively low-mass Galactic host stars, indicate that Neptune-mass planets beyond the snow line are common (e.g. Winn & Fabrycky 2015; Zhu & Dong 2021).

In addition to broad occurrence trends, demographic distributions contain local deficits and excesses that can provide clues to planetary evolution. One example is the relative scarcity of intermediate-sized planets on very short-period orbits, commonly referred to as the hot-Neptune (or sub-Jovian) desert, as illustrated in Fig. 4.4. Several mechanisms may contribute to this feature, including atmospheric loss and the formation and migration histories of close-in planets (e.g. Winn 2018; Burn & Mordasini 2024). Another notable characteristic of the observed population is the radius valley, discussed in the following section.

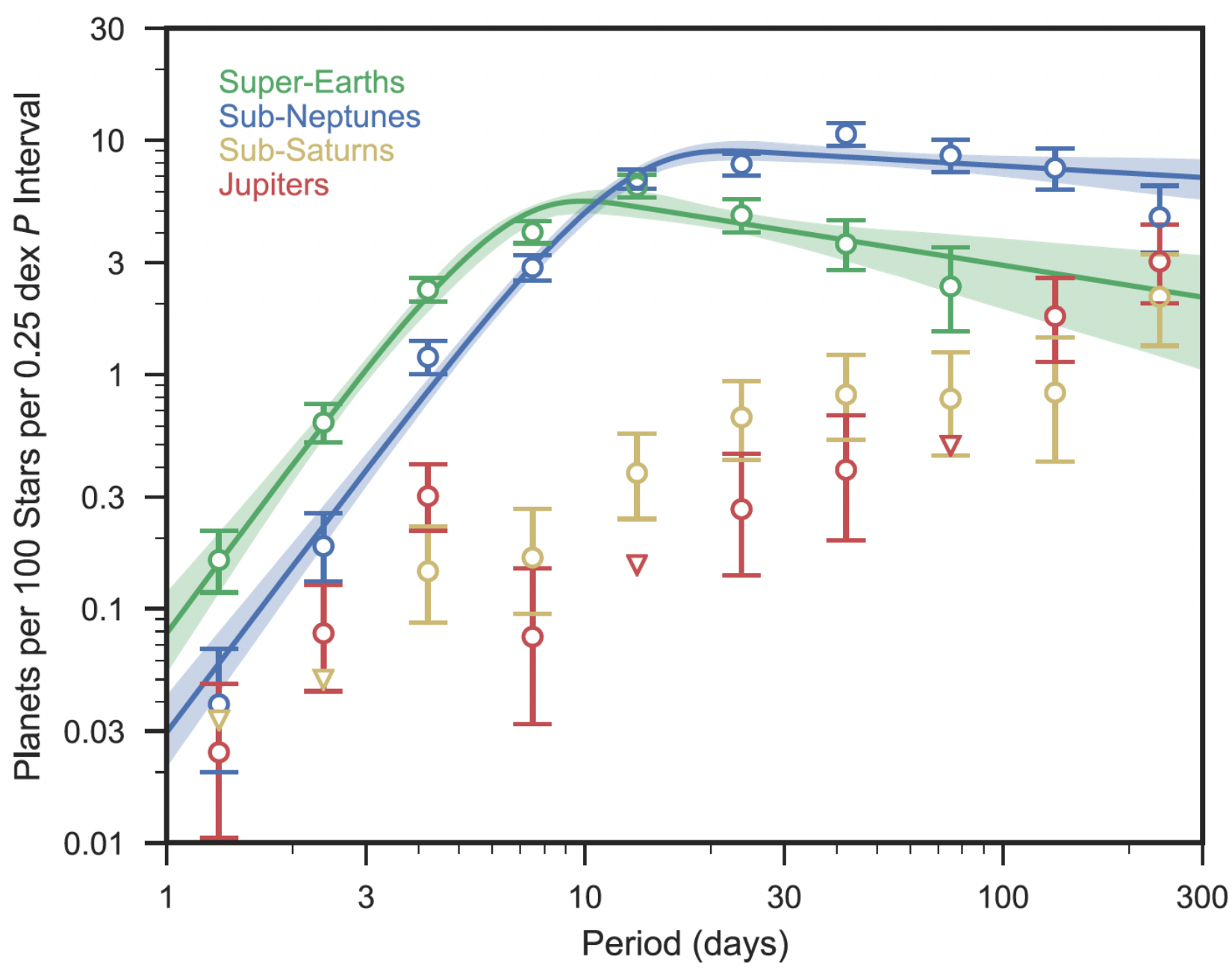

**Figure 4.3:** Planet occurrence rates per 100 stars as a function of orbital period after correcting for the detection biases of the *Kepler* survey, shown for four different planet types following the definitions adopted by Petigura et al. (2018). The occurrences of super-Earths and sub-Neptunes increase strongly between orbital periods of approximately 1 and 10 days, while becoming much flatter at longer periods. Sub-Saturns and Jupiters display a different period dependence. The solid curves and shaded regions represent the best-fit models and their $1\sigma$ confidence intervals. Figure from Petigura et al. (2018).

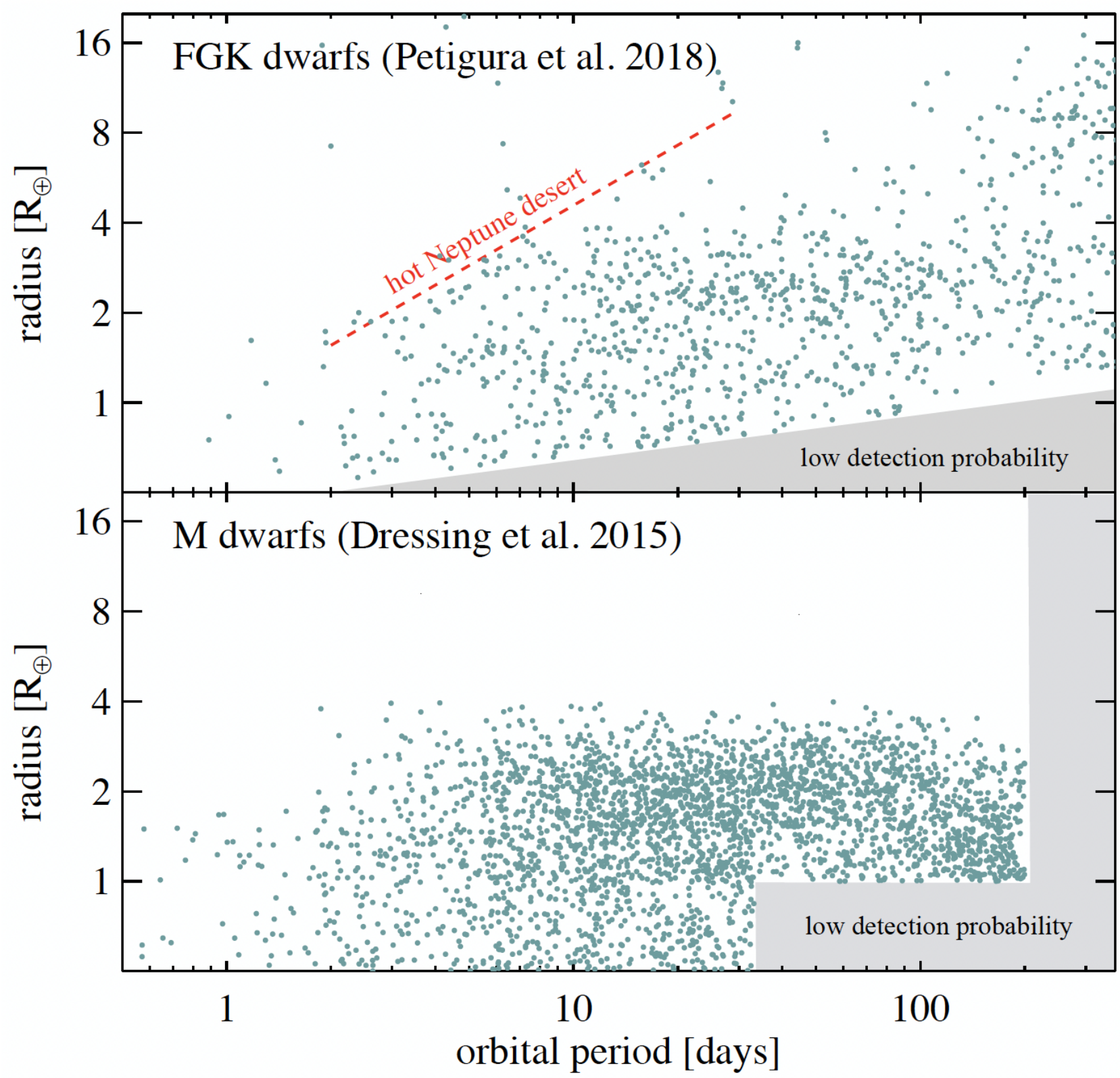

**Figure 4.4:** Radius versus orbital period for planets with $P < 100$ days based on occurrence-rate distributions for planets orbiting FGK stars from Petigura et al. (2018) (upper panel) and M dwarfs from Dressing & Charbonneau (2015) (lower panel). The figure illustrates the high occurrence of small planets around M dwarfs and the hot-Neptune desert in the FGK population. The two panels are based on different surveys and underlying samples and should therefore be interpreted as illustrations of broad population trends. Planets with $R_\mathrm{p} > 4\,R_\oplus$ are not included in the lower panel because only four planet candidates in this size range were present in the underlying sample. Figure from Winn (2018).

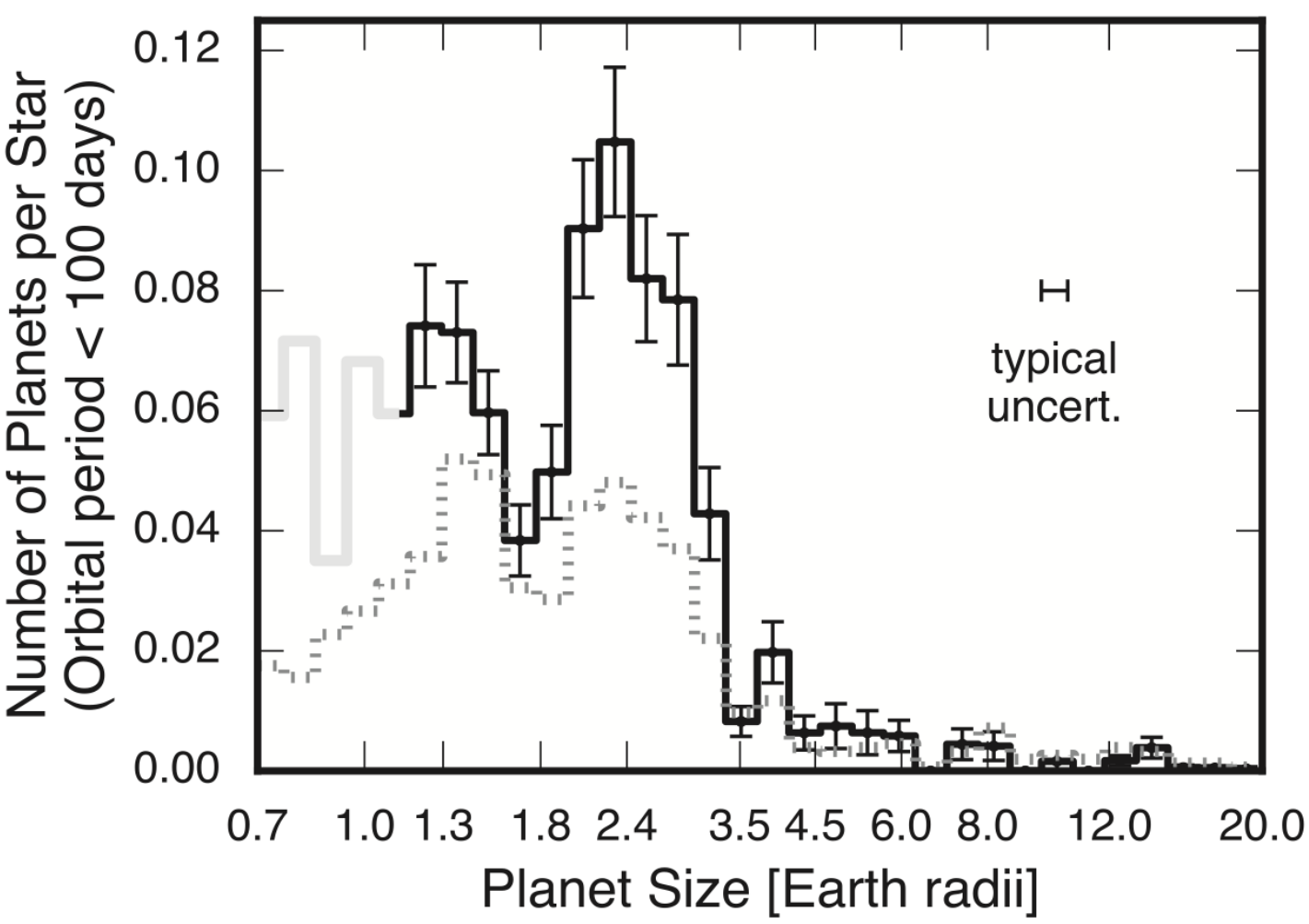

**Figure 4.5:** Radius distribution of planets with orbital periods shorter than 100 days based on *Kepler* observations. The solid line indicates the number of planets per star as a function of planetary radius after completeness corrections from Fulton & Petigura (2018), while the dotted line shows the distribution prior to these corrections. The deficit around $1.7 - 2\,R_\oplus$ is referred to as the radius valley. Figure from Fulton & Petigura (2018).

### 4.4.2 The radius valley

One of the most prominent features in the radius distribution of close-in planets is the radius valley, illustrated in Fig. 4.5: a deficit of planets with radii of $\simeq 1.7 - 2\,R_\oplus$ at orbital periods shorter than approximately 100 days (Fulton et al. 2017; Fulton & Petigura 2018; Van Eylen et al. 2018). The radius valley separates two abundant planet classes, with planets smaller than $\simeq 1.7\,R_\oplus$ generally interpreted as predominantly rocky, whereas sub-Neptunes larger than $\simeq 2\,R_\oplus$ require a greater contribution from low-density material.

The quoted radius range for the valley is only approximate. The position and depth of the radius valley also depend on orbital period and host-star properties (e.g. Van Eylen et al. 2018; Ho & Van Eylen 2023; Kamulali et al. 2026), and it therefore does not represent a fixed physical boundary between super-Earths and sub-Neptunes.

Two of the leading explanations for the radius valley involve the removal of primordial H/He envelopes. In the photoevaporation scenario, high-energy ultraviolet and X-ray radiation from the young host star heats the planetary atmosphere and can drive substantial atmospheric escape (e.g. Lammer et al. 2003; Jin et al. 2014). Some planets can consequently lose essentially all of their primordial H/He envelopes and become part of the smaller-radius rocky population, while planets that retain their envelopes maintain larger radii. This evolution can produce a deficit of planets at intermediate radii.

Core-powered mass loss represents another leading mechanism proposed to explain the radius valley, whereby energy released during the cooling of the planetary interior heats the overlying atmosphere and promotes atmospheric escape (e.g. Ginzburg et al. 2018; Gupta & Schlichting 2019). In this scenario, the energy driving the escape originates primarily from the cooling planet rather than from the high-energy radiation of the host star.

A complementary class of explanations involves differences established during planet formation. Planets forming beyond the water snow line can accrete larger amounts of water-rich material and subsequently migrate towards the host star. Such planets may contribute to the larger-radius population, while planets that form inside the snow line are expected to have more rock-dominated compositions (e.g. Zeng et al. 2019; Venturini et al. 2020; Burn et al. 2024).

Formation location, migration, and subsequent atmospheric loss may therefore all contribute to shaping the observed radius distribution. These proposed mechanisms are however not necessarily mutually exclusive. Additionally, the larger-radius population need not have a unique composition, since both water-rich material and relatively small H/He envelopes can increase planetary radii, as discussed in Sect. 4.3.

### 4.4.3 Correlations with host-star properties

The properties and occurrence rates of planets depend in several ways on the properties of their host stars (e.g. Mulders 2018; Winn 2018; Burn & Mordasini 2024). These relationships provide constraints on planet formation because stellar mass and composition are linked to the mass and chemical inventory of the protoplanetary disc. At the same time, observational selection effects also depend on stellar properties, and intrinsic population differences must therefore be distinguished from detection biases.

One of the best-established host-star correlations is the dependence of giant-planet occurrence on stellar metallicity. The latter describes the abundance of elements heavier than helium and is commonly represented in exoplanet studies by the logarithmic iron abundance relative to the Sun, denoted by [Fe/H]. Gas giants are considerably more common around metal-rich stars than around metal-poor stars (e.g. Burn & Mordasini 2024). This trend is generally interpreted within the core-accretion framework: a larger amount of solid material in the protoplanetary disc facilitates the formation of sufficiently massive planetary cores before the gaseous disc disperses.

In contrast, the occurrence of small and low-mass planets exhibits a much weaker dependence on stellar metallicity. These planets are found around stars spanning a broad metallicity range, indicating that the strong metallicity dependence observed for gas giants does not extend to the small-planet population.

Planet classes also depend on stellar mass and spectral type. M dwarfs host a large population of small planets, while gas giants are comparatively uncommon around these low-mass stars. The relative occurrence and characteristic sizes of close-in planets therefore differ between planets orbiting M dwarfs and those orbiting FGK stars.

Some observed differences also arise from detection biases. As discussed in Ch. 3, planets of a given radius or mass are not detected with equal efficiency around stars of different sizes and masses. Population comparisons between different stellar types must therefore distinguish intrinsic differences from observational selection effects. Figure 4.4 illustrates both the high occurrence of small planets around M dwarfs and the hot-Neptune desert in the FGK-star population.

## 4.5 Population-analysis methods

Statistical methods are required to determine whether apparent patterns in a planet population are likely to represent underlying relationships or can arise from sampling variability. Several complementary techniques can be used to measure correlations, compare distributions, estimate uncertainties, and test the robustness of population-level results.

A statistical hypothesis test is generally formulated in terms of a null hypothesis. For a correlation test, the null hypothesis commonly states that no association exists between the two variables. For a two-sample test, it generally states that the samples are consistent with being drawn from the same underlying population or distribution.

The null hypothesis assumes that there is no real difference between the populations being compared. The $p$-value indicates how likely it would be to obtain a difference at least as large as the observed one if this assumption were correct. A small $p$-value therefore provides evidence against the null hypothesis, and throughout this thesis, $p < 0.05$ is adopted as a conventional threshold for statistical significance.

Statistical significance should not be confused with the strength or physical importance of an effect. The statistical significance of a correlation depends not only on its magnitude but also on the sample size. For instance, a weak correlation can yield a small $p$-value when the sample size is sufficiently large, whereas a stronger correlation may fail to reach statistical significance when the sample size is small. Measures of effect strength and the distribution of the data should therefore be considered together with the $p$-value. When many related statistical tests are performed, some small $p$-values may also occur by chance, and individual results should therefore be interpreted together with effect sizes, consistency across related analyses, and the physical context.

### 4.5.1 Correlation tests

Correlation tests are used to determine whether two quantities tend to vary together. A positive correlation means that one quantity generally increases as the other increases, whereas a negative correlation means that one generally decreases as the other increases. Two main tests are introduced below.

The Pearson correlation coefficient, $r$, measures the strength and direction of a linear relationship, i.e. how closely the data follow a straight-line trend. It ranges from $-1$ to $+1$, where values close to $+1$ indicate a strong positive linear correlation, values close to $-1$ a strong negative linear correlation, and values close to zero little or no correlation. Pear-son's $r$ can be strongly affected by individual extreme values and may give a weak correlation even when two variables are clearly related in a non-linear way.

The Spearman rank correlation coefficient instead measures whether two quantities generally increase or decrease together, without requiring the relation to follow a straight line. Rather than using the measured values directly, the observations are ordered from the smallest to the largest, and the correlation between these rankings is calculated. Spearman's coefficient is therefore useful for monotonic relationships, where one variable generally increases or decreases as the other changes, and it is less sensitive to extreme values than Pearson's $r$.

These correlation tests describe relationships across a population of planets or planetary systems. Measures of similarities between planets within individual systems are introduced separately in Sect. 5.2.3.

### 4.5.2 Empirical cumulative distributions and two-sample tests

Statistical tests can also be used to determine whether two samples have different distributions. Before applying such tests, the distributions can be visualised using an empirical cumulative distribution function (CDF). For any chosen value $x$, the CDF gives the fraction of observations that are smaller than or equal to $x$. For example, if the CDF of planetary radius has a value of 0.7 at $2\,R_\oplus$, then 70 % of the planets in that sample have radii of $2\,R_\oplus$ or smaller. Two samples can therefore be compared by examining how their CDFs differ across the full range of values, without dividing the data into histogram bins.

The Kolmogorov–Smirnov (K–S) test compares the CDFs of two samples. It identifies the point at which the separation between the two CDFs is greatest and uses this difference to test whether the samples are consistent with being drawn from the same underlying distribution.

The Anderson–Darling (A–D) two-sample test addresses the same general question, but gives greater importance to differences near the ends, or tails, of the distributions. It can therefore be more sensitive than the K–S test when the samples differ mainly among their unusually small or large values.

The Mann–Whitney (M–W) test asks a slightly different question. It compares the ordering, or ranks, of the values in two independent samples and tests whether the values in one sample tend to be systematically larger or smaller than those in the other. When the two distributions have similar shapes, this can be interpreted as a comparison of their typical values. More generally, the M–W test should not be described simply as a test of whether the two medians are different.

Employing several complementary tests can be useful because the tests are sensitive to different types of differences between the samples.

### 4.5.3 Uncertainties on counts and fractions

Population studies often compare either the number of planets in different categories or the fraction of a sample belonging to each category. These two quantities require different treatments of their uncertainties.

When counting how many objects belong to a particular category, the uncertainty can often be estimated using Poisson statistics. The fractional uncertainty is large when only a few objects are counted and becomes smaller as the number of objects increases.

A fraction instead describes how many objects belong to a category relative to the total sample. For example, 20 planets of a particular type in a sample of 100 planets correspond to a fraction of 20 %. Binomial statistics are generally appropriate for estimating the uncertainty of such fractions. This uncertainty depends both on the total number of objects in the sample and on the measured fraction.

Consequently, the distinction between counts and fractions is important when comparing the sizes of different planetary populations.

### 4.5.4 Monte Carlo sampling

Measured planetary properties have uncertainties, which can affect the results of a population analysis. Monte Carlo (MC) sampling provides a means to estimate this effect by repeating the analysis many times with slightly different input values.

For each repetition, new values are drawn randomly from the uncertainty distributions of the measurements and the analysis is repeated. The resulting range of outcomes shows how much the final result can vary due to the measurement uncertainties.

For example, if an analysis depends on measured planetary radii, a new radius can be drawn for each planet according to its measured value and uncertainty. Repeating the analysis many times with these different radii produces a distribution of results. The spread of this distribution quantifies how strongly the radius uncertainties contribute to the uncertainty in the final result.

### 4.5.5 Bootstrap and randomisation methods

Bootstrap resampling is useful for assessing the sensitivity of a result to the specific objects included in a sample. Many new samples are constructed from the original data by repeatedly selecting objects at random, with replacement. This means that an object can appear more than once in a resampled dataset, while another object may not appear at all. The statistical analysis performed on the original data is repeated for each new sample.

The spread of the resulting values reflects the extent to which the result may vary due to sampling variability. Bootstrap resampling can therefore be used to estimate uncertainties and confidence intervals, as well as to assess the robustness of a result to changes in sample composition.

Randomisation tests address a different question. They probe whether an observed association could arise by chance if the association did not actually exist. For example, planets can be randomly reassigned between groups while keeping the measured planetary properties unchanged. This removes the original connection between a planet and its group. The analysis of interest is subsequently performed on each of a large number of randomised datasets. If the result measured from the real data is rarely reproduced in the randomised datasets, this provides evidence that the observed association is not easily explained by chance under the assumed null hypothesis.

Monte Carlo sampling, bootstrap resampling, and randomisation therefore serve different purposes. Monte Carlo sampling tests the effect of measurement uncertainties on a result, while bootstrap resampling evaluates how sensitive a result is to the particular sample that was observed. Randomisation examines whether an observed association could arise by chance under the null hypothesis of no association between the variables.

## From planet population to planetary systems

The population-level properties discussed in this chapter describe how planetary masses, radii, compositions, and occurrence rates are distributed across the observed exoplanet population. However, planets do not form and evolve in isolation, and their properties must also be considered in the context of the planetary systems to which they belong. The following chapter therefore shifts the focus from individual planets and population-level distributions to the architectures of multi-planetary systems, including observed multiplicities, orbital spacings, and intra-system similarities.

# CHAPTER 5

## Multi-planetary systems

Planets in multi-planetary systems can be examined both individually and in relation to one another. This chapter examines the architectures of multi-planetary systems and the relationships between the physical and orbital properties of planets within the same system.

The architecture of a planetary system comprises its planet multiplicity, the distribution and ordering of physical properties, such as planetary masses and radii, as well as the planets' orbital periods, spacings, eccentricities, mutual inclinations, and dynamical features, including mean-motion resonances. Together, these properties can provide constraints on planet formation and the subsequent dynamical evolution of the system.

The Solar System is first introduced as a reference point for comparison with exoplanetary systems. This chapter then considers observed exoplanetary systems, including the distinction between observed and intrinsic multiplicity, orbital spacings, and intra-system similarities. Finally, some of the formation and dynamical processes that can produce and modify planetary system architectures are discussed.

| Planet | $P$ [days] | $a$ [au] | $e$ | $R/R_\oplus$ | $M/M_\oplus$ |
|---|---|---|---|---|---|
| Mercury | 87.969 | 0.387 | 0.2056 | 0.383 | 0.0553 |
| Venus | 224.701 | 0.723 | 0.0068 | 0.950 | 0.815 |
| Earth | 365.256 | 1.000 | 0.0167 | 1.000 | 1.000 |
| Mars | 686.980 | 1.524 | 0.0935 | 0.532 | 0.107 |
| Jupiter | 4332.589 | 5.204 | 0.0487 | 10.973 | 317.83 |
| Saturn | 10755.699 | 9.573 | 0.0520 | 9.140 | 95.16 |
| Uranus | 30685.400 | 19.165 | 0.0469 | 3.981 | 14.54 |
| Neptune | 60189.018 | 30.178 | 0.0097 | 3.865 | 17.15 |

**Table 5.1:** Orbital periods $P$, semi-major axes $a$, orbital eccentricities $e$, radii $R$, and masses $M$ of the eight planets in the Solar System. Planetary radii and masses are expressed in Earth units. Data from Williams (2025).

## 5.1 The Solar System as a reference planetary system

The Solar System provides the most completely characterised example of a planetary system and is therefore a useful reference when comparing the architectures of exoplanetary systems (e.g. Murray & Dermott 1999; Lissauer & De Pater 2019; Perryman 2018). It contains eight planets orbiting the Sun: the four inner terrestrial planets Mercury, Venus, Earth, and Mars, followed by the gas giants Jupiter and Saturn and the ice giants Uranus and Neptune. Selected physical and orbital properties of the planets are listed in Table 5.1 and illustrated in Fig. 5.1.

Several properties of the Solar System provide useful reference points for comparisons with exoplanetary systems. The planetary orbits are broadly coplanar and, with the exception of Mercury, have relatively low eccentricities. The system also exhibits a pronounced distinction in physical properties between the four inner terrestrial planets and the four outer giant planets, providing the basis for its conventional division into the inner and outer Solar System.

The orbital periods span a particularly broad range, from approximately 88 days for Mercury to nearly 165 years for Neptune. In contrast, the observed population of multi-planetary systems is strongly weighted towards much shorter orbital periods because short-period planets are generally easier to detect, particularly through transit photometry. For example, 90% of the planets in the catalogue of confirmed multi-planetary systems compiled by Muresan et al. (2024) have orbital periods shorter than 100 days.

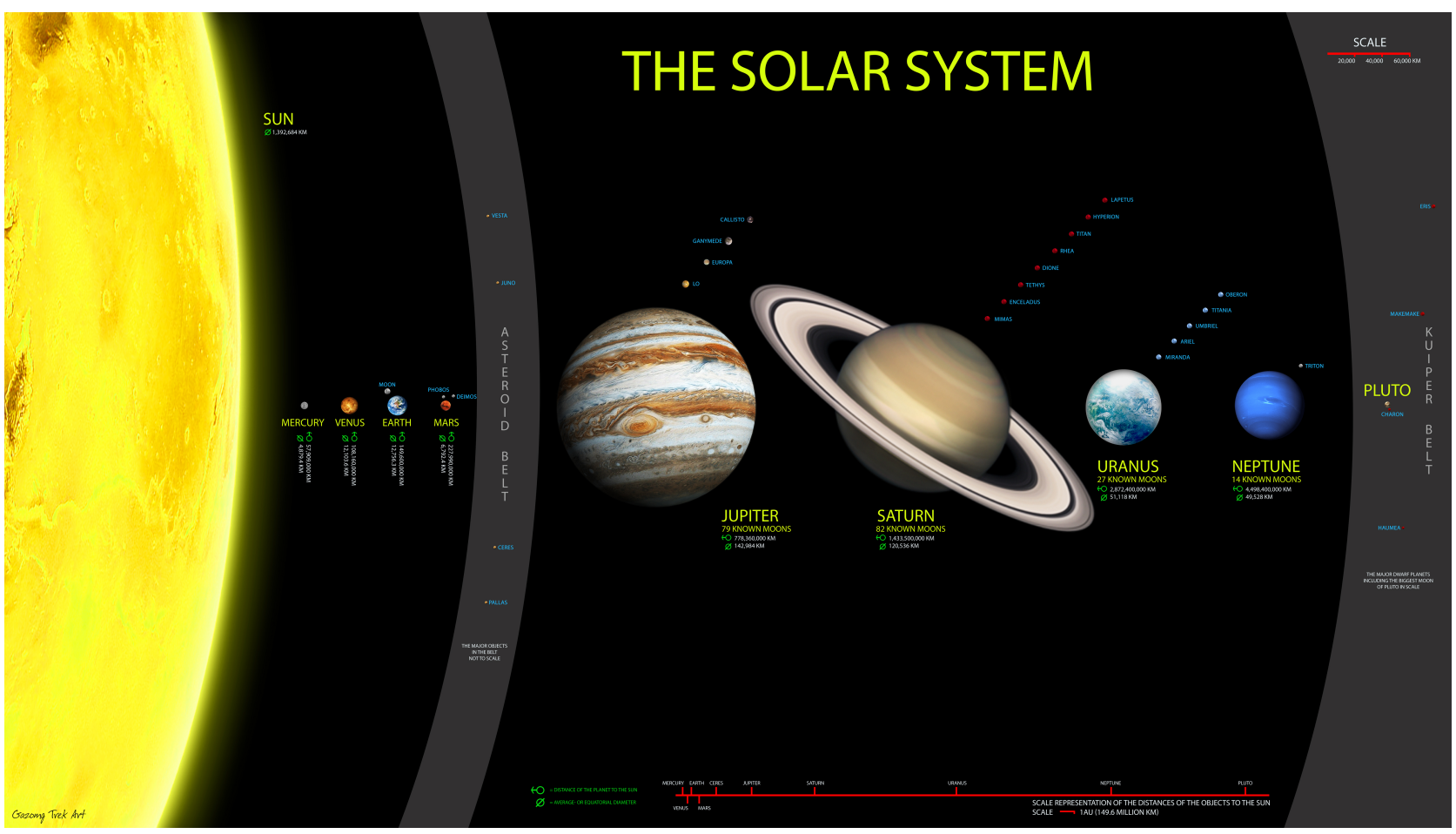

**Figure 5.1:** Illustration of the Solar System, including the Sun and the eight planets in order of increasing orbital distance. The asteroid belt, Kuiper belt, Pluto, and several smaller bodies are also visible for context. Illustration from Gazomg (2021).

Furthermore, the orbital spacings between adjacent planets in the Solar System are non-uniform. As introduced in Ch. 2, a convenient measure of the spacing between two adjacent planets is their orbital period ratio. The largest period ratio occurs between Mars and Jupiter,

$$\frac{P_{\mathrm{J}}}{P_{\mathrm{Ma}}} \simeq 6.3\,.$$

The wide orbital gap between these two planets contains the asteroid belt. The formation and subsequent dynamical influence of Jupiter, together with Saturn, strongly affected this region by exciting and removing material, thereby limiting further planetary growth. The other six adjacent planet pairs have period ratios ranging from approximately 1.6 for Venus–Earth to 2.8 for Saturn–Uranus. The Mars–Jupiter period ra-tio is therefore considerably larger than those of the other Solar-System planet pairs, while the remaining six ratios are typical of many observed exoplanet pairs (Malhotra 2015; Muresan et al. 2024).

The planetary masses and radii also vary considerably across the Solar System. Venus and Earth have relatively similar masses and radii, as do Uranus and Neptune, but the eight planets together span nearly four orders of magnitude in mass. The Solar System therefore contains local similarities without exhibiting uniform planetary properties or orbital spacings throughout the system.

The Solar System does not appear to be representative of the planetary architectures that dominate current exoplanet catalogues. Nevertheless, observational surveys preferentially detect systems containing short-period planets, while planets at Solar-System-like orbital distances are substantially more difficult to detect. Comparisons between the Solar System and observed exoplanetary systems must therefore account for observational biases.

## 5.2 Observed planetary system architectures

A substantial fraction of known exoplanets reside in systems containing at least two detected planets. Well-known examples of such multi-planetary systems include Kepler-90, with eight confirmed planets (Cabrera et al. 2014; Shallue & Vanderburg 2018), and TRAPPIST-1, which contains seven approximately Earth-sized planets (Gillon et al. 2016, 2017). Multi-planetary systems enable the study not only of the properties of individual planets, but also of their arrangement relative to one another.

An important distinction is that the architecture inferred from observations is not necessarily the complete architecture of the planetary system. The observed planet multiplicity of a system denotes the number of planets that have been detected in the system, whereas the intrinsic multiplicity is the actual number of planets in the system. Depending on the study, the term singles refers either to systems with only one detected planet or to the planets belonging to such systems. Analogously, the term multis refers either to systems with multiple detected planets or to the planets within such systems. For example, Weiss et al. (2018c) and Muresan et al. (2026) adopt the latter definition in both cases. These terms represent the observed planet multiplicity and do not necessarily reflect the intrinsic number of planets in a system.

The same distinction applies to neighbouring planets. In observational studies, two planets are generally considered adjacent if no additional detected planet has an orbital period between those of the two planets. They are not necessarily adjacent in the intrinsic system. For example, three consecutive planets with orbital periods of 5, 10, and 20 days have two adjacent period ratios of 2. If the 10-day planet remains undetected, the two detected planets instead appear to have a period ratio of 4. Missing an intermediate planet can therefore make the observed spacings substantially larger than the true adjacent spacings and, in systems with at least three detected planets, can make an intrinsically regular spacing pattern appear irregular (e.g. Muresan et al. 2024; Thomas et al. 2025).

Undetected inner or outer planets have a different effect: they can alter the intrinsic multiplicity and the range of orbital distances covered within a system without necessarily modifying the observed spacings between the already detected planets. The observational reasons why planets can remain undetected are discussed in Ch. 3. The important consequence here is that measured multiplicities, adjacency, and spacings describe the observed architecture and may not completely represent the underlying system.

### 5.2.1 Observed multiplicity and the *Kepler* dichotomy

Transit surveys provide a particularly clear illustration of the difference between observed and intrinsic planet multiplicity. In addition to systems hosting several transiting planets, *Kepler* detected a large number of systems containing only one detected transiting planet. Models in which planetary systems are drawn from common underlying distributions of intrinsic planet multiplicities and mutual inclinations can reproduce much of the observed population of systems with multiple transiting planets. However, such models tend to underpredict the large number of systems that contain only one detected transiting planet (e.g. Lissauer et al. 2011). The excess of apparently single-planet systems relative to the predicted number is commonly referred to as the *Kepler* dichotomy.

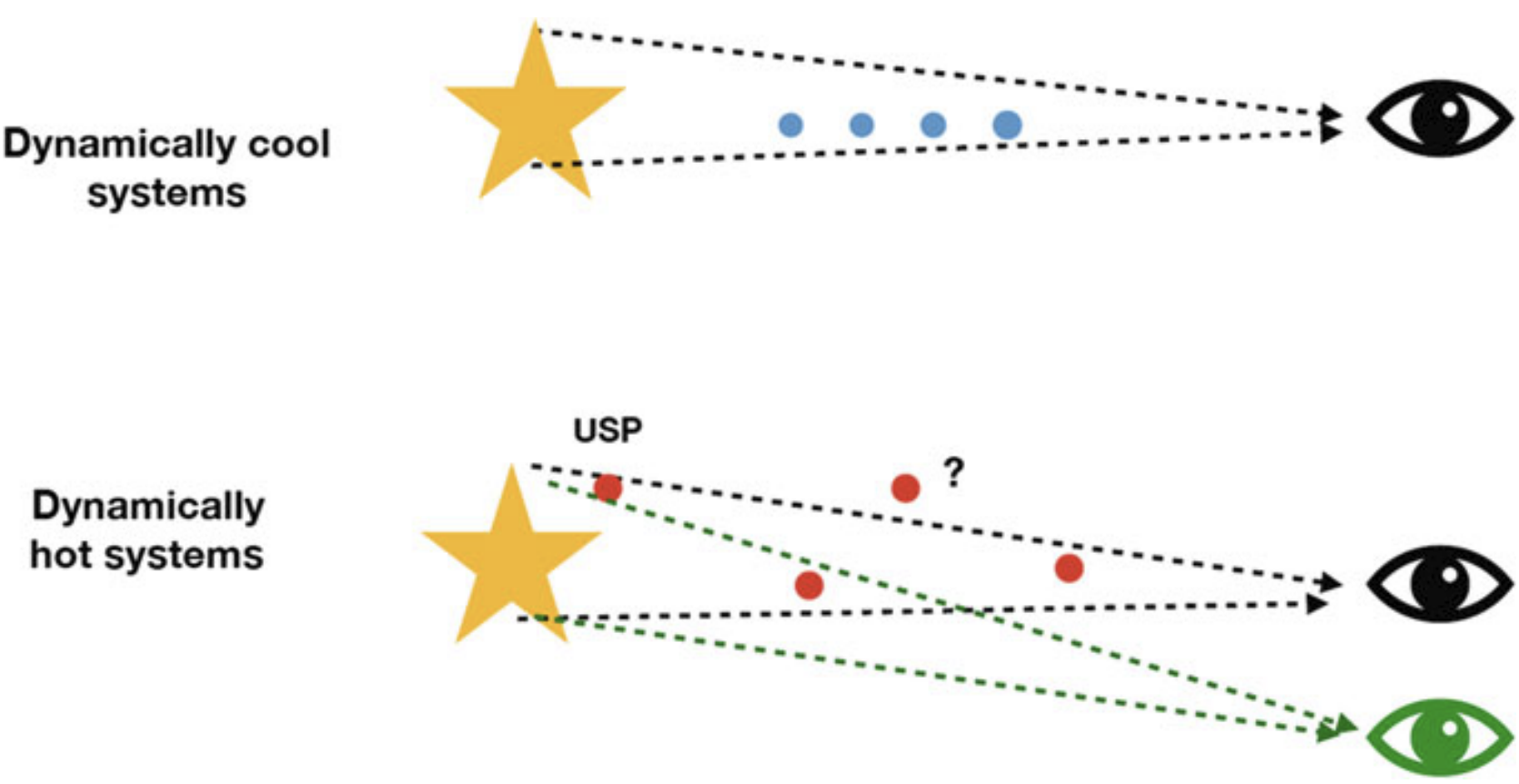

**Figure 5.2:** Schematic examples of how mutual inclinations and viewing geometry affect the observed transit multiplicity of a planetary system. Upper panel: A dynamically cold multi-planetary system with four planets exhibiting low mutual inclinations, all transiting along the indicated line of sight. Lower panel: A dynamically hot multi-planetary system hosting four planets with larger mutual inclinations, resulting in different numbers of planets to transit for different viewing directions. An intrinsically multi-planetary system can therefore be observed as a lower-multiplicity system (black eye) or as a single-planet system (green eye). Illustration from Biazzo et al. (2022).

Several explanations have been proposed for this dichotomy, including the existence of two distinct populations of planetary systems. One population may consist of compact, dynamically cold multi-planetary systems, in which planets occupy nearly circular and approximately coplanar orbits, whereas the other population may comprise intrinsically single-planetary systems and/or dynamically hot multi-planetary systems that host planets with larger orbital eccentricities and/or mutual inclinations (e.g. Ballard & Johnson 2016; Mulders et al. 2018; He et al. 2019).

Figure 5.2 illustrates an example in which an intrinsically multi-planetary system appears to contain fewer planets. In such a system, only planets with suitable orbital geometries transit the host star. Moreover, some planets may simply remain undetected because they are too small or have orbital periods that are too long for the survey to detect them reliably.

Additionally, the *Kepler* pipeline exhibits multiplicity-dependent detection efficiencies, whereby additional planets in systems with previously identified transiting planets are progressively more difficult to detect. Consequently, intrinsically multi-planetary systems may be mistakenly classified as systems hosting only a single planet (Zink et al. 2019).

The *Kepler* dichotomy should therefore not be interpreted as demonstrating the existence of two physically distinct populations of planetary systems. Rather, it describes the observed excess of single-transiting systems. Variations in intrinsic multiplicity, mutual inclination, and eccentricity, together with observational selection effects, can contribute to the observed multiplicity distribution (e.g. Ballard & Johnson 2016; Mulders et al. 2018; He et al. 2019).

*Kepler* systems with higher observed transit multiplicities are inferred to have lower eccentricities and mutual inclinations, on average, than lower-multiplicity systems (Xie et al. 2016; He et al. 2020). Observed multiplicity therefore contains information about the architecture of a planetary system, although it cannot be interpreted independently of transit geometry and detection completeness.

### 5.2.2 Orbital spacing and dynamical separation

As introduced in Ch. 2, the relative spacing between adjacent planets can be described by their orbital period ratio,

$$\mathrm{PR}_i = \frac{P_{i+1}}{P_i},$$

where the planets are indexed by increasing semi-major axis. A period ratio close to unity corresponds to comparatively closely spaced orbits, whereas a larger period ratio represents a wider relative orbital separation.

A period ratio is not itself a physical distance. However, for planets orbiting the same star and with planetary masses much smaller than the stellar mass, Kepler's third law (Eq. 2.1) gives approximately:

$$\frac{P_{i+1}}{P_i} \simeq \left(\frac{a_{i+1}}{a_i}\right)^{3/2}.$$

The period ratio therefore provides a convenient measure of relative orbital spacing and has the practical advantage that orbital periods can usually be measured very precisely.

Equal period ratios also correspond to equal separations in logarithmic orbital period,

$$\log(P_{i+1}) - \log(P_i) = \log\left(\frac{P_{i+1}}{P_i}\right).$$

Consequently, a sequence of planets with similar period ratios appears approximately evenly spaced when orbital period is plotted on a logarithmic scale. This is the sense in which a planetary system can exhibit regular orbital spacings (e.g. Weiss et al. 2018b; Muresan et al. 2024).

A compact system contains planets on comparatively short-period orbits. Compactness does not necessarily imply regular spacing: a compact system can have very different period ratios, while a more extended system can have larger but nearly equal ratios (Muresan et al. 2024).

Observed period-ratio distributions also show features near simple integer ratios such as 2:1 and 3:2 (e.g. Fabrycky et al. 2014; Steffen & Hwang 2015). Their possible connection to mean-motion resonances and orbital migration is discussed further in Sect. 5.3.

Period ratio describes the relative orbital spacing without accounting for planetary mass. A second measure expresses the separation in units of the mutual Hill radius, which provides a characteristic gravitational length scale for a pair of neighbouring planets and depends on the masses of both planets and the host star. The mutual Hill radius and the orbital separation in units of mutual Hill radii, $\Delta$, of two adjacent planets are given by Eqs. 2 and 2.2, respectively. These definitions follow those employed by Muresan et al. (2024).

In contrast to the period ratio, $\Delta$ depends on the planetary masses as well as on their orbital distances. Two planet pairs can therefore have the same period ratio but different separations in mutual Hill radii. For otherwise identical orbital configurations around the same star, a pair of lower-mass planets has a larger value of $\Delta$ than a pair of more massive planets (Muresan et al. 2024).

Larger values of $\Delta$ mean that the planets are more widely separated when their masses are taken into account, whereas smaller values mean that they are more closely spaced in this dynamical sense. However, $\Delta$ alone does not determine the long-term stability of a multi-planetary system. Stability also depends on properties such as eccentricities, inclinations, resonant configurations, and the presence of additional planets (e.g. Pu & Wu 2015).

A practical difference between the two spacing measures is the greater observational availability of orbital periods compared with precise planetary mass measurements. Mutual Hill radius separations can therefore be determined for substantially fewer observed planet pairs.

As discussed above for the orbital period ratios, an apparently large separation in units of mutual Hill radii can also result from an undetected intermediate planet. Observed orbital spacings must therefore be interpreted in the context of detection completeness.

### 5.2.3 Intra-system similarities and ordering

The large diversity observed from one planetary system to another contrasts with the degree of uniformity found within many individual systems. Previous studies have reported intra-system similarities in planetary radii, masses, and derived bulk densities, although the strength of these trends depends on the sample and analysis (e.g. Millholland et al. 2017; Weiss et al. 2018b; Otegi et al. 2022; Mamonova et al. 2024). Regular patterns are also observed in the orbital properties: many compact systems contain planets with similar adjacent period ratios, relatively low eccentricities, and small mutual inclinations (e.g. Weiss et al. 2018b; Jiang et al. 2020; Muresan et al. 2024; Van Eylen & Albrecht 2015; Xie et al. 2016; Fabrycky et al. 2014).

This tendency for planets within the same system to resemble one another has become known as the peas-in-a-pod pattern (Weiss et al. 2018b). It does not, however, reflect a single property, and a system may for instance be uniform in planet radius but not in mass, bulk density, or orbital spacing.

The apparent strength of intra-system similarities can also be affected by observational biases, since planet detectability depends on properties such as planetary radius and orbital period (Weiss & Petigura 2020; Mishra et al. 2021). The observed patterns should therefore be interpreted in the context of the selection effects described in Ch. 3.

Planetary system architectures can be analysed at both the population and system levels. Population-level analyses combine planets or adjacent planet pairs from many systems and are useful for identifying common correlations and trends. System-level analyses instead examine each planetary system separately. The two approaches provide complementary information, and a statistically significant population-level correlation does not imply that every individual system follows the same pattern (Muresan et al. 2024).

The observed intra-system similarities represent an important characteristic of the architectures of planetary systems and can be investigated at the system level. Several metrics have been employed to quantify intra-system similarities, providing a measure of the dispersion or similarity among planets within individual systems and enabling comparisons of the resulting measures across different systems (Millholland et al. 2017; Gilbert & Fabrycky 2020; Goyal & Wang 2022; Weiss et al. 2023; Mishra et al. 2023).

One of such metrics is the coefficient of variation (CV), which is a standard descriptive statistic for measuring the magnitude of variation in a set of numbers, defined as follows (Brown 1998):

$$\mathrm{CV}(q) = \frac{\sigma(q)}{\overline{q}} \ . \tag{5.1}$$

In the context of planetary systems, $\sigma(q)$ denotes the standard deviation of the values of a quantity $q$ within the system, such as the planetary radius, while $\overline{q}$ represents the mean of the values. The coefficient of variation is a positive number, and a small CV corresponds to similar values of the quantity $q$ among the planets, whereas a large CV indicates greater diversity.

Figure 5.3 illustrates several important architectural properties of all 26 systems listed in the NASA Exoplanet Archive that contain at least three confirmed planets, all of which have measurements of both mass and radius. The orbital periods, masses, and radii of the planets in this sample are presented for each system individually. The systems in Fig. 5.3 are shown from top to bottom in order of decreasing intra-system similarity in planetary radius, quantified by the coefficient of variation $\mathrm{CV}(R_\mathrm{p})$ using Eq. 5.1 with $q = R_\mathrm{p}$. Notably, the corresponding CV values for planetary masses do not necessarily follow the same ordering, and the systems would therefore generally appear in a different order in an analogous figure based on $\mathrm{CV}(M_\mathrm{p})$.

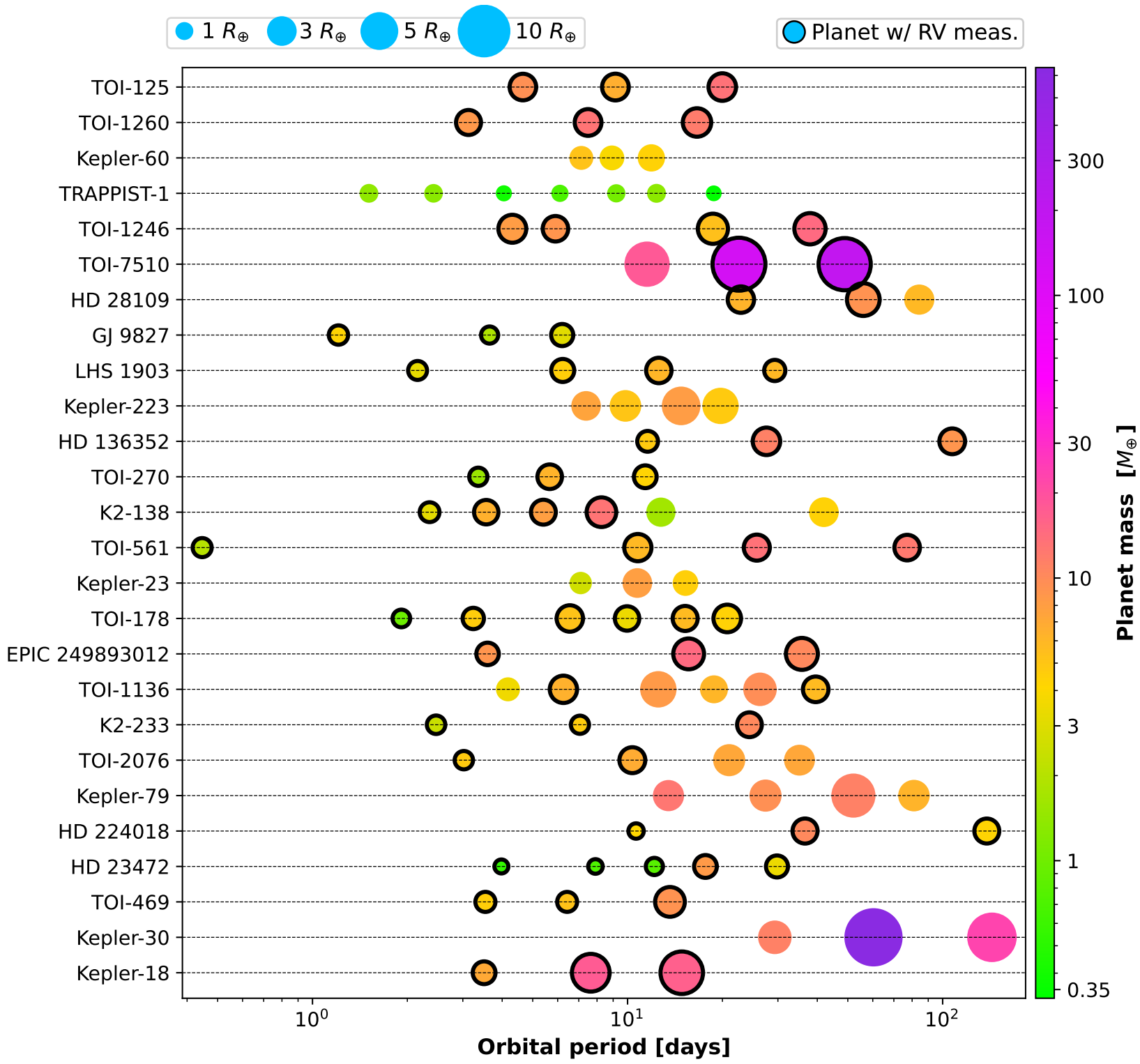

**Figure 5.3:** Architectures of 26 planetary systems that host at least three confirmed planets, all of which have measurements of both mass and radius. Each row represents an individual system, with orbital period on the horizontal axis. The marker size is proportional to the planetary radius, while the colour indicates the planetary mass, as shown in the figure legend. Planets with RV measurements are identified as indicated in the legend. The systems are ordered from top to bottom by decreasing intra-system similarity in planetary radius.

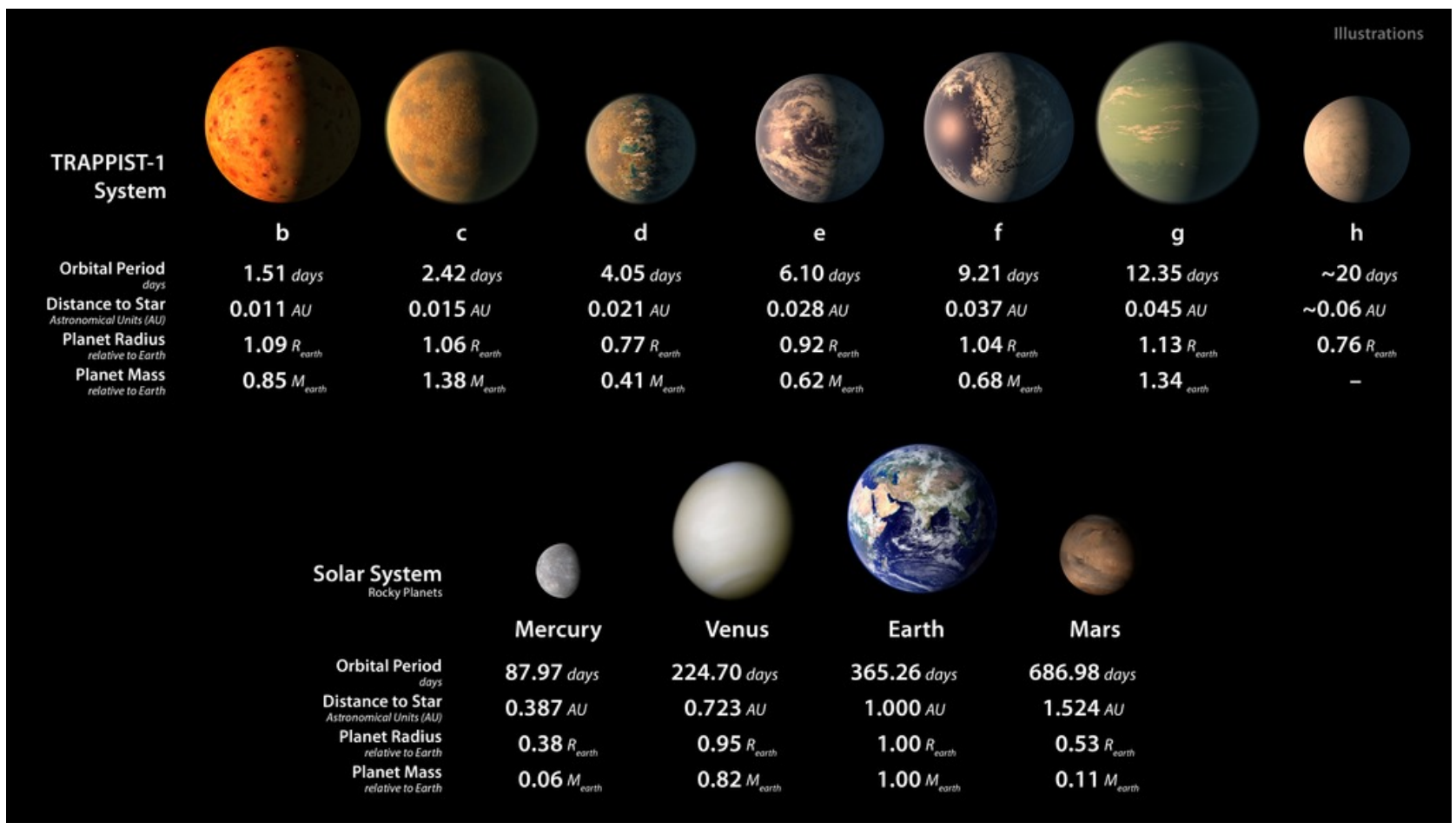

**Figure 5.4:** Seven planets detected in the TRAPPIST-1 system compared to the inner four planets in the Solar System. The TRAPPIST-1 planets have similar masses, radii, and orbital spacings (expressed as the orbital period ratios between adjacent planets). Overall, the intra-system similarities of these planets are much stronger than those exhibited by the inner Solar System planets. Illustration credit: NASA/JPL-Caltech 2017.

Different metrics for measuring intra-system similarities weight variations differently, and numerical values should therefore only be compared when the same metric is used. For instance, Muresan et al. (2024) employed a logarithmic measure of the fractional dispersion as the principal system-level metric. The coefficient of variation was evaluated independently as a consistency check, and the two metrics yielded consistent trends.

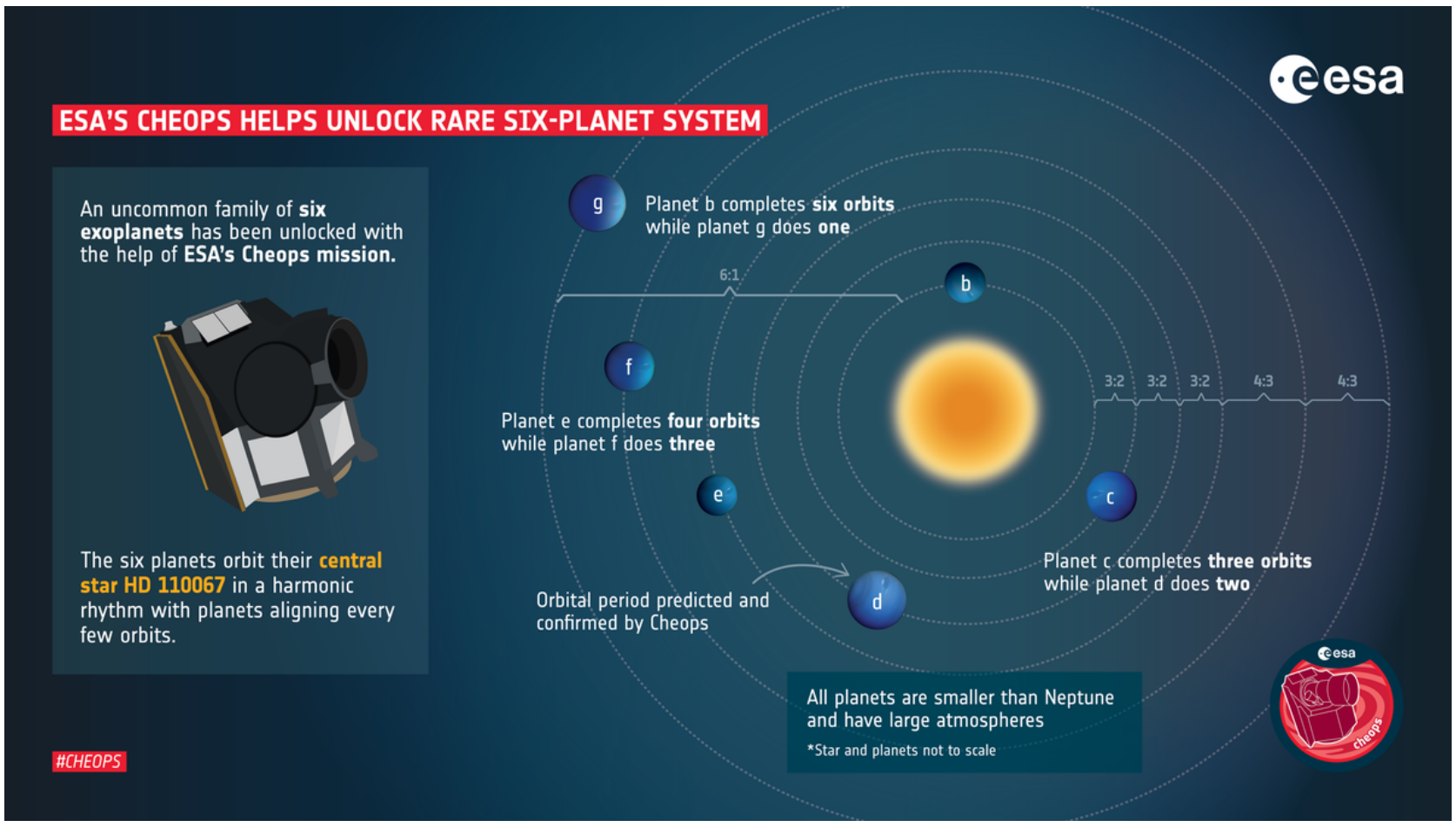

**Figure 5.5:** The six planets detected in the HD 110067 system exhibit similar masses and radii, while also forming a resonant chain in a compact orbital configuration (Luque et al. 2023). Illustration credit: ESA 2023.

The TRAPPIST-1 system represents a notable example of a compact system containing planets of broadly similar terrestrial sizes (Gillon et al. 2016, 2017). As illustrated in Figs. 5.3 and 5.4, the seven planets detected in the TRAPPIST-1 system exhibit much stronger intra-system similarities in planetary mass, radius, and orbital spacing compared to the four inner Solar System planets. The HD 110067 system is another example of a compact system hosting planets with similar radii (Luque et al. 2023), as shown in Fig. 5.5. Additionally, both exoplanetary systems are in resonant orbital configurations with small eccentricities and low mutual inclinations, as further discussed in Sect. 5.3.

Similarity and ordering with respect to orbital period describe different properties. For example, three planets with radii of $2.0\,R_\oplus$, $2.1\,R_\oplus$, and $2.2\,R_\oplus$, in order of increasing orbital period, are both similarly sized and monotonically ordered. On the other hand, if the radii of the two outer planets are interchanged, the three planets remain similar in size but no longer exhibit a monotonic ordering. Conversely, planets can display a clear increase or decrease in radius with orbital period even if their radii differ substantially. Additionally, a system may host plan-ets with regular orbital spacings but different masses or radii, whereas another system may contain planets with irregular spacings but similar sizes. Different architectural properties should therefore be examined independently before being considered jointly.

## 5.3 Formation and dynamical evolution

Planetary system architecture continues to evolve after the planets have formed. In particular, orbital spacings, multiplicity, eccentricities, and inclinations can change substantially during and after the dispersal of the protoplanetary gas disc. The observed present-day architecture therefore reflects both planet formation and subsequent dynamical evolution.

Planets within the same system form in the same protoplanetary disc and therefore share a common overall reservoir of gas and solid material and evolve in the same dynamical environment. This characteristic provides one possible origin for some of the intra-system similarities discussed in Sect. 5.2.3. However, temperature, chemical composition, and the availability of solid material vary with distance from the host star. A common formation environment therefore does not imply that the planets should be identical.

Orbital migration can substantially alter the arrangement established during planet formation. Interactions between a forming planet and the surrounding protoplanetary disc can move the planet inward or outward. In standard core-accretion models, in which planets begin forming through the growth of solid material, giant-planet formation can be favoured at larger orbital distances, particularly near or beyond the water snow line, the distance from the star beyond which water can condense into ice and thereby increase the amount of solid material available for planet formation. Hot Jupiters are therefore generally thought to have reached their present close-in orbits through migration or later dynamical evolution (Dawson & Johnson 2018).

Migration can also change the relative separations between adjacent planets. When neighbouring planets migrate at different rates such that their orbital separation decreases, the migration is described as convergent. This type of migration can bring their orbital periods close to ratios of small integers and allow the planets to become dynamically locked in a mean-motion resonance (e.g. Tamayo et al. 2017; Papaloizou 2021; Wong & Lee 2024).

A period ratio close to a ratio of small integers is a signature of a possible mean-motion resonance, but the period ratio alone is not sufficient to demonstrate that the planets are dynamically resonant. Common simple period ratios associated with possible resonances include 2:1, 3:2, 4:3, and 5:4. If several adjacent planet pairs become resonantly linked, the system can form a resonant chain (e.g. Siegel & Fabrycky 2021; Papaloizou 2021).

The seven planets detected in the TRAPPIST-1 system represent a clear example, as illustrated in Fig. 5.4. The planets form a resonant chain and exhibit measurable transit-timing variations caused by their mutual gravitational interactions (Gillon et al. 2017, see also Sect. 3.4). The HD 110067 system is another well-known resonant system, as illustrated in Fig. 5.5. The six detected planets orbit in a resonant chain and are also similar in mass and radius (Luque et al. 2023).

Resonant chains are not the dominant configuration among observed compact multi-planetary systems, although planet pairs near low-order mean-motion commensurabilities are common (e.g. Fabrycky et al. 2014; Steffen & Hwang 2015; Muresan et al. 2024). Resonant configurations established while the protoplanetary gas disc is still present can also be modified later. As the protoplanetary gas disc dissipates, its ability to keep planetary eccentricities and inclinations small is reduced. Resonant chains can then be disrupted through subsequent dynamical evolution (Izidoro et al. 2017).

Gravitational interactions can further reshape a system. Strong gravitational encounters between planets, commonly referred to as planet–planet scattering, can increase orbital eccentricities and mutual inclinations, while collisions and mergers can reduce the number of planets and increase the masses of the surviving planets. In more extreme cases, planets can also be ejected from the system. These processes can also increase the orbital spacings between the surviving planets (e.g. Mishra et al. 2021; Burn & Mordasini 2024).

Numerical simulations indicate that some of the intra-system patterns need not be fully established during the earliest stages of planet formation. Systems can develop similarities in planet size or orbital spacing through later dynamical interactions and collisions (e.g. Mishra et al. 2021; Lammers et al. 2023). Dynamical evolution can therefore both create and erase architectural regularities.

Massive outer planets can also influence the architecture of an inner planetary system. In particular, long-term gravitational perturbations from an inclined outer companion can increase the mutual inclinations of inner planets, potentially reducing the number that transit from a given viewing direction (Lai & Pu 2017). More generally, interactions involving giant planets can also modify eccentricities and orbital spacings during the dynamical evolution of a system (e.g. Burn & Mordasini 2024).

An observed association between an outer giant planet and a particular inner architecture does not, however, by itself demonstrate that the giant planet caused that architecture. Both may instead reflect common formation conditions, and observational selection can also influence which outer planets are detected.

Convergent migration followed by resonance disruption is not the only pathway capable of producing compact, non-resonant planetary systems. Models in which planets assemble largely in situ, meaning near their present-day orbital locations, or undergo limited migration can also produce compact architectures (e.g. Chiang & Laughlin 2013; Chatterjee & Tan 2014; MacDonald et al. 2020).

Population-synthesis models simulate the formation and evolution of large numbers of planetary systems by combining processes such as planetary growth, migration, gravitational interactions, as well as gas accretion and loss in order to test whether they can reproduce the observed diversity of planets and planetary system architectures (Burn & Mordasini 2024).

The diversity of observed planetary architectures therefore likely reflects a combination of formation location, planetary growth, migration, disc dispersal, and later dynamical evolution. Present-day architectures contain information about these processes, but they do not provide a unique record of the evolutionary path followed by an individual system.

The physical and demographic framework introduced in Ch. 4 together with the system-architecture concepts developed in this chapter provide the basis for the analyses summarised in Ch. 6. These analyses address orbital spacings and intra-system similarities, the properties of apparent singles and multis, and whether similarities in planetary bulk properties extend to inferred interior compositions.

# CHAPTER 6

# Paper summaries

This chapter summarises the aims, data sets, analyses, and principal results of Papers A–C, followed by a synthesis of how the three studies together address the research questions posed in Sect. 1.2.

## 6.1 Paper A

### Diversities and similarities exhibited by multi-planetary systems and their architectures I. Orbital spacings

Paper A examines a large sample of confirmed multi-planetary systems and their architectures, with particular emphasis on the orbital spacings between adjacent planets. We focused primarily on quantifying the intra-system similarities in planet masses, radii, and orbital spacings, as well as on investigating potential relationships between the orbital spacings and the sizes or masses of planets within the same system. In this work, the orbital spacing is quantified by the period ratio (PR) and, where attainable, also by the separation expressed in units of the mutual Hill radius, as introduced in Ch. 2.

## Data catalogues

The largest sample analysed in this work consists of all single-star systems that host at least three confirmed planets with measured orbital periods[1]. This period catalogue comprises 282 systems and 991 planets. We further constructed two additional samples, referred to as the radius catalogue and the mass catalogue, by selecting from the period catalogue all the systems containing at least three adjacent planets with measured radii and masses, respectively. The included masses were derived either via transit-timing variations or from radial-velocity observations combined with orbital inclinations measured from transits. The mass catalogue therefore excludes the inclination-dependent minimum masses obtained from radial-velocity measurements alone.

Approximately 90 % of the planets in the period catalogue have orbital periods shorter than 100 days, and most of them were discovered through transit photometry, particularly by the *Kepler* space telescope.

## Similarities and correlations

First, we performed population-level studies by examining all the pairs of adjacent planets in the three catalogues. For the analyses of adjacent orbital spacings, we compared the period ratio of the outer pair, $P_{i+2}/P_{i+1}$, with that of the inner pair, $P_{i+1}/P_i$. The full sample shows a significant positive Pearson correlation, meaning that adjacent orbital spacings tend to be similar. The result therefore extends the observed spacing-similarity pattern beyond the smaller and predominantly compact samples employed in previous studies.

We also examined how orbital spacing is related to planetary size and mass. In the radius catalogue, we identified a weak positive Pearson correlation between the orbital period ratio and the mean radius of every two adjacent planets. This correlation, however, disappears when adjacent planets with mean radii below $1\,R_\oplus$ are excluded.

[1] The catalogue was last updated using data from the NASA Exoplanet Archive on 2024-06-28.

Moreover, we identified a strong positive correlation between the orbital period ratios and the mean masses of adjacent planets in the mass catalogue. Such a mass–spacing relation among observed planets had not been previously reported. For the mass catalogue, Paper A also finds that spacing similarity is more pronounced when spacing is expressed as period ratios than in units of mutual Hill radii.

We additionally performed system-level analyses by examining each planetary system individually, facilitating both intra- and inter-system comparisons. We used the metric given in Eq. 2 in the paper to quantify the degree of intra-system similarity in a given planetary parameter among all planets within a system. This metric was independently evaluated for the planetary radii, masses, and orbital spacings, revealing that many of the observed systems exhibit strong intra-system similarities, particularly with respect to the orbital spacings.

Furthermore, we identified no significant correlation between the intra-system similarities in orbital spacings and those in planetary masses or radii. This result implies that planets orbiting the same star may be similarly spaced even if they do not have similar sizes or masses.

The full Solar System has an intermediate dispersion in its period ratios but very large radius and mass dispersions compared to the systems in the corresponding catalogues. When the inner and outer Solar System are considered separately, their dispersions in radius, mass, period ratio, and separation in units of the mutual Hill radius range from moderate to large.

Paper A also shows how orbital regularity can be used as a practical diagnostic of individual systems. If most adjacent pairs in a system have similar period ratios but one gap is much larger, the regular pattern identifies a natural region in which to search for an additional planet. Several such systems are highlighted in the paper.

Additionally, we found that the inner planets in all five systems that also host outer giants have relatively large spacing dispersions despite exhibiting more typical size and mass dispersions, thereby linking the orbital architecture of the inner system to the presence of outer giants.

## 6.2 Paper B

### Diversities and similarities exhibited by multi-planetary systems and their architectures

### II. Radii of singles and multis

Paper B investigates whether the observed planets in apparently single-planetary systems (henceforth singles) and those in confirmed multi-planetary systems (hereafter multis) are consistent with originating from the same underlying population, based on the observed distributions of planetary types and radii.

### FGK, early-M, and late-M samples

We constructed a catalogue of all the confirmed planets with measured radii and orbital periods hosted by single main-sequence stars with spectral classes ranging from late-M to late-F[2] After excluding planets with relative radius uncertainties larger than 15%, the catalogue contained a total of 1730 singles and 1522 multis in 677 multi-planetary systems.

To investigate the dependence on host-star spectral type, the catalogue was initially divided into five stellar samples: F, G, K, early M (M0–M2), and late M (M3–M9). The median and mean planetary radii of both the singles and multis increase on average with host-star temperature, from late-M to F-type hosts.

Hot Jupiters, defined here as planets with orbital periods $P < 10$ days and radii of $R > 6\,R_{\oplus}$, were found predominantly as singles orbiting stars with effective temperatures above 5000 K. After excluding the hot Jupiters, the F-, G-, and K-type stars appeared to host planets with similar radius distributions, and we therefore combined them into a single sample, hereafter denoted as the FGK sample. In contrast, the planets orbiting M dwarfs are smaller on average and were consequently not included in the FGK sample. Furthermore, the planets orbiting late-type M dwarfs, particularly those in multi-planetary systems, were found to have smaller median radii than the planets hosted by early-type M dwarfs, and these two M samples were therefore analysed separately.

[2] The catalogue was last updated from the NASA Exoplanet Archive on 2025-08-15.

## Planet types and radius distributions

The singles and multis were classified into six planet types based on radius and orbital period: ultra-short-period planets, small planets, hot giants, hot Jupiters, warm giants, and cold giants, as defined in Table 2 in the paper. After removing the hot Jupiters, the singles and multis contain similar relative fractions of the five remaining planet types. Therefore, the strong excess of hot Jupiters among singles represents a substantial difference between the full single- and multi-planetary systems with FGK host stars.

Further analysis revealed that after removing the hot Jupiters, the singles and multis in the FGK sample exhibit overall statistically indistinguishable radius distributions, particularly for planets with $R < 4\,R_{\oplus}$. These results are based on the three statistical tests employed in this work: the Kolmogorov–Smirnov, the Anderson–Darling, and the Mann–Whitney U tests.

Paper B therefore concludes that the observed singles and multis around FGK stars are overall consistent with originating from the same underlying population based on their planet types and radii, particularly for the smaller planets. These findings support the hypothesis that a substantial fraction of the apparent single-planet systems with FGK host stars contain additional planets that have not been detected due to high orbital eccentricities and mutual inclinations.

A particularly interesting exception occurs at radii of approximately $1.4 - 1.6\,R_{\oplus}$, where we identified a statistically significant local overabundance of multis relative to singles. The paper discusses few possible explanations for this newly identified feature.

For M-dwarf systems, the results are less conclusive due to the smaller sample sizes. Early- and late-type M dwarfs host planets with systematically smaller radii and shorter orbital periods compared to the planets orbiting FGK stars. According to the three independent comparison tests performed in this work, the radii of the singles and multis in neither the early- nor the late-type M samples can be drawn from the same distribution. In these two samples, singles tend to be larger on average than multis, and only the singles and multis with $R < 4\,R_{\oplus}$ orbiting early-M stars have statistically indistinguishable radius distributions. A tentative overabundance of multis is observed for planets hosted by M dwarfs as well, albeit at smaller radii of approximately $1.0 - 1.4\,R_\oplus$.

## Robustness of the FGK result

Paper B tested whether the principal FGK result depends on the heterogeneous origin of the catalogue or on plausible misclassification of some singles. Three smaller and more homogeneous *Kepler* subsets, as well as additional bootstrap-resampled tests, reproduced results consistent with our initial findings. The strongest and most consistent conclusion is therefore that the observed singles and multis, excluding hot Jupiters, with FGK host stars exhibit similar radius distributions, especially those with $R < 4\,R_\oplus$.

## 6.3 Paper C

### Diversities and similarities exhibited by multi-planetary systems and their architectures

### III. Trends in planetary interior compositions

Paper C examines whether the observed intra-system similarities in the bulk planetary properties also extend to the inferred planetary interiors, and whether systems with and without detected gas giants differ in the physical properties of their non-gas-giant planets.

### Data catalogue and samples

We constructed a catalogue, hereafter denoted as the mass–radius (M–R) catalogue, of all the single-star systems that host at least three confirmed planets with measured masses, radii, and orbital periods, yielding a total of 44 systems and 184 planets, all of which are confirmed[3].

The M–R catalog was divided into the following two samples, based on the presence of gas giants, which are defined here as planets with $R \geq 6\,R_{\oplus}$: sample 1 contains all the systems with no detected gas giants (a total of 30 systems and 125 planets), and sample 2 comprises the remaining systems that host at least one gas giant (14 systems and 59 planets). We examined the two samples separately and compared them in order to identify potential differences in the planetary properties and system architectures.

### Intra-system similarities

Consistent with previous findings, we identified significant positive Pearson correlations between the corresponding radii, masses, and bulk densities of adjacent planets. These similarities were found to be most prominent among planets with $R < 6\,R_{\oplus}$ and $M < 20\,M_{\oplus}$, regardless of the presence of a detected gas giant in the system. Furthermore, these similarities do not exhibit any apparent dependence on the host-star effective temperature.

[3]The catalogue was last updated from the NASA Exoplanet Archive on 2026-07-10.

We tested whether the observed positive correlations persist when planets from the M–R catalogue are randomly redistributed among the 44 systems. Our bootstrap-resampled catalogues revealed no significant correlations in the radii, masses, or bulk densities of adjacent planets, thereby suggesting that the observed intra-system similarities reflect intrinsic planetary system properties rather than random associations.

Our comparison between samples 1 and 2 revealed that the non-gas-giant planets in systems hosting gas giants are on average larger and more massive, while having more irregular orbital spacings than the planets in systems without detected gas giants. In the discussion, the paper considers possible explanation for these observed differences, including gravitational interactions with giant planets, differences in the host-star types of the two samples, and potential observational biases.

## Interior-composition inference

To investigate whether the observed intra-system similarities in planetary radius, mass, and bulk density imply similar planetary interiors, we modelled the composition of planets in sample 1. We employed the forward models of Aguichine et al. (2021), in which a planet consists of an iron-rich core, a silicate mantle, and a possible water-rich outer layer, together with the Bayesian MCMC framework described in Ch. 4. The principal inferred composition parameters are the core mass fraction (CMF) and water mass fraction (WMF), while the mantle mass fraction (MMF) follows from the requirement that the three fractions sum to unity.

Planets whose retrieved masses and radii were consistent with the corresponding observed values within one standard deviation were retained, yielding a final compositional subset of 28 planetary systems and 88 planets, all of which have $R < 4\,R_\oplus$. Within this subset, planets with $R \leq 1.5\,R_\oplus$ were inferred to be water-poor with WMFs $< 0.03$ while spanning a broad range of core mass fractions, whereas larger planets generally exhibited increasing WMFs and decreasing CMFs with increasing planetary radius.

This work also showed that the identified intra-system similarities in observable planetary properties do not necessarily translate into similarities in interior compositions. While adjacent planets display strong correlations in radius, mass, and bulk density, the inferred CMFs and WMFs of adjacent planets exhibit substantial scatter and only weak correlations. This finding indicates that planets within the same system can possess markedly different interior compositions despite having similar bulk properties.

Additionally, the compositional subset revealed that strictly monotonically increasing or decreasing planetary radii, CMFs, or WMFs with respect to orbital period within a system is uncommon. In particular, approximately half of the modelled systems exhibit mixed ordering trends simultaneously in planetary radius, CMF, and WMF. Paper C discusses this mixed ordering as a possible consequence of the combined effects of formation location, accretion history, orbital migration, atmospheric loss, giant impacts, and later dynamical evolution.

## 6.4 Overall conclusions

Papers A–C address different aspects of planetary system properties: orbital architecture, observed planet populations, planetary bulk properties, and inferred interior composition.

A central result of this thesis is that regularity in one aspect does not require regularity in another. Paper A shows that planets within the same system can have similar orbital spacings even when their radii or masses differ substantially. Paper C extends this result by indicating that adjacent planets with similar radii, masses, and bulk densities can have different inferred interior compositions. A planetary system can simultaneously display a highly regular orbital architecture and substantial physical or compositional diversity.

The three planets orbiting Kepler-60 provide a particularly clear example linking Papers A and C. In Paper A, they are found to exhibit low dispersions in both radius, mass, and orbital spacing. In Paper C, the same three planets are shown to have similar radii, masses, and bulk densities but different inferred core, mantle, and water mass fractions. Kepler-60 therefore shows directly that planets within the same system can appear uniform in their orbital and observable bulk properties while being substantially more diverse in their inferred interiors.

Consequently, papers A–C provide new insights into the properties and diversity of the observed exoplanets as well as the architectures of planetary systems. Together, these studies examine planetary systems on several complementary levels, including (I) the orbital configuration and the distribution of the physical properties of planets within multi-planetary systems, (II) differences between the observed singles and multis on a population level, (III) the extent to which the observed intra-system similarities reflect similarities in the inferred planetary composition, and (IV) differences in the properties of non-gas-giant planets in systems with and without detected gas giants.

Taken together, the results of these three works offer provisional answers to the research questions posed in Sect. 1.2. Papers A and C reveal that the confirmed planets in multi-planetary systems frequently exhibit intra-system similarities in their measured radii, masses, bulk densities, and orbital spacings. Paper C further reveals that these observed similarities of planets within the same system do generally not extend to their inferred interior compositions in terms of their core and water mass fractions. Moreover, the findings of Paper B indicate that the observed singles and multis orbiting FGK host stars, after excluding hot Jupiters, are consistent with originating from a common underlying planetary population, suggesting that many apparent single-planetary systems may host additional, currently undetected planets.

Collectively, these results show that similarities in observable planetary properties do not necessarily imply compositional similarity, while the apparent distinction between singles and multis may arise substantially due to detection biases. Papers A–C therefore demonstrate that planet multiplicity, bulk properties, interior compositions, as well as physical and orbital system architectures provide important, complementary characteristics of planetary systems.

# CHAPTER 7

## Outlook

The results of Papers A–C identify several areas in which further observations can extend the work presented in this thesis: detecting additional planets in known systems, increasing the number of systems with precise planetary masses and radii, and adding constraints on planetary composition beyond mass and radius alone. Future surveys and follow-up observations can address these needs in complementary ways.

The PLAnetary Transits and Oscillations of stars (PLATO) mission, planned for launch in March 2027, will observe more than 200 000 stars with high-precision photometry and is designed to detect terrestrial planets in orbits extending up to the habitable zones of Sun-like stars (European Space Agency 2026a). Transit measurements combined with accurately determined stellar radii will yield precise planetary radii, while radial-velocity follow-up of suitable targets will provide planetary masses. Asteroseismology of suitable host stars will additionally provide precise stellar masses, radii, and ages, thereby improving the derived planetary properties (Rauer et al. 2025).

These capabilities will help address several limitations of the current samples. Longer photometric time series and the detection of additional planets can improve estimates of observed planet multiplicities and provide more complete system architectures. A large PLATO transit sample would provide an opportunity to expand the observed samples of singles and multis and to perform analyses analogous to those presented in Paper B. In particular for FGK host stars, the origin of the overabundance of multis at radii of approximately $1.4 - 1.6\, R_{\oplus}$ remains unresolved, and an additional large survey can test whether this feature persists. Larger samples of confirmed planets around M dwarfs are likewise required before firm conclusions can be drawn about differences between singles and multis orbiting these stars. Precise radii, radial-velocity masses, and stellar ages will also increase the number of multi-planetary systems for which planetary bulk properties and interior compositions can be studied in detail.

Continued long-baseline radial-velocity monitoring and astrometry will improve the characterisation of the outer regions of known planetary systems. This is important as long-period massive planets can remain undetected in present catalogues. Gaia Data Release 4, planned for December 2026 (European Space Agency 2026b), is expected to include non-single-star orbital solutions and companion masses (European Space Agency 2026c). For suitable systems, Gaia astrometry could complement long-baseline radial-velocity measurements in identifying and characterising massive companions on wider orbits. Such detections will provide a more complete picture of planetary system architectures, particularly at orbital separations that are difficult to probe with transit surveys.

The Nancy Grace Roman Space Telescope (Roman), launched in August 2026, will extend exoplanet demographics to orbital separations that are poorly represented in transit samples (NASA 2026a). Its Galactic Bulge Time-Domain Survey is expected to detect more than a thousand wide-orbit planets through gravitational microlensing and more than 100 000 transiting planets in the same survey fields (NASA 2026b). Microlensing surveys primarily probe planets at wider orbital separations than transit surveys, thereby providing complementary demographic information beyond the predominantly short-period planets analysed in this thesis.

Improving the inference of planetary interior compositions requires both more precise bulk properties and additional constraints. Mea-surement uncertainties in mass and radius propagate directly into inferred composition parameters (Otegi et al. 2020b), while different interior structures can reproduce the same bulk properties even when mass and radius are measured accurately (Rogers & Seager 2010). Host-star chemical composition, atmospheric characterisation, age information, as well as temperature and pressure profiles may therefore provide useful supplementary constraints. Interior models that include H/He-rich envelopes are also required for planets whose measured masses and radii cannot be reproduced by core–mantle–water models alone. For rocky planets, smaller mass uncertainties are particularly important, while sub-Neptunes and Neptune-sized planets can additionally necessitate more precise radii and, in some cases, age or atmospheric information to reduce degeneracies between thermal structure and envelope composition.

Atmospheric spectroscopy provides one such source of additional information, and JWST is already characterising individual exoplanet atmospheres (e.g. Ahrer et al. 2025). The forthcoming Atmospheric Remote-sensing Infrared Exoplanet Large-survey (Ariel), currently planned for launch in 2031, is designed to characterise the atmospheres of about 1000 exoplanets in a large spectroscopic survey (European Space Agency 2026d). Such measurements can help distinguish between planets with different types of low-density outer layers and thereby improve the interpretation of interior models that are otherwise constrained mainly by mass and radius.

Host-star abundances provide another constraint on rocky planetary interiors. Dorn et al. (2015) showed that stellar abundances of Fe, Si, and Mg can reduce degeneracies in rocky-planet interior models and improve the inference of mantle composition. Combining such abundance measurements with precise planetary masses and radii may therefore provide stronger constraints on the internal structure of rocky planets.

Together, these developments will yield more complete planetary system catalogues, larger samples with precise bulk properties, and additional constraints on planetary interiors. They will provide valuable insights into the properties and demographics of exoplanets, as well as the architectures, formation and evolution of planetary systems, allowing the patterns identified in this thesis to be tested with larger and better characterised samples.